\documentclass[12pt]{article}
\usepackage{amsfonts,amsthm,amsmath,amssymb,upgreek,bm,mathtools}
\usepackage[bookmarks = false, linktocpage=true]{hyperref}
\usepackage[paper=letterpaper,margin=.85in]{geometry}
\usepackage{graphicx}
\usepackage{units}
\usepackage{float}
\usepackage[export]{adjustbox}
\usepackage{bbold}
\usepackage{xcolor}
\usepackage[shortlabels]{enumitem}
\usepackage{dsfont}
\usepackage[mathscr]{euscript}
\usepackage[normalem]{ulem}
\usepackage{braket}
\usepackage{slashed}
\usepackage{comment}
\usepackage{multicol}
\allowdisplaybreaks
\numberwithin{equation}{section}

\usepackage{tikz-cd}
\usepackage{tikz}
\usetikzlibrary{decorations,calc,shapes,patterns, math, shadings, fadings, arrows}
\tikzset{snake it/.style={decorate, decoration=snake}}
\tikzset{cross/.style={cross out, draw=black, minimum size=2*(#1-\pgflinewidth), inner sep=0pt, outer sep=0pt},
cross/.default={1pt}}
\usetikzlibrary{shapes.geometric}
\tikzset{
    partial ellipse/.style args={#1:#2:#3}{
        insert path={+ (#1:#3) arc (#1:#2:#3)}
    }
}
\usetikzlibrary{decorations.pathmorphing}
\definecolor{vert}{rgb}{0.1367 0.543 0.1367}

\newcommand{\be}{\begin{equation}}
\newcommand{\ee}{\end{equation}}

\newcommand{\SL}{\ensuremath \mathrm{SL}}
\newcommand{\Tr}{\ensuremath \mathrm{Tr}}

\newcommand{\Aut}{\ensuremath \mathrm{Aut}}
\newcommand{\Hom}{\ensuremath \mathrm{Hom}}
\newcommand{\Mod}{\ensuremath \mathrm{Mod}}
\newcommand{\Homeo}{\ensuremath \mathrm{Homeo}}
\newcommand{\baseRing}[1]{\ensuremath{\mathbb{#1}}}
\newcommand{\K}{\baseRing{K}}
\newcommand{\Z}{\baseRing{Z}}
\newcommand{\R}{\baseRing{R}}

\newcommand{\one}{\baseRing{1}}
\newcommand{\bF}{\mathbb{F}}
\newcommand{\bS}{\mathbb{S}}
\newcommand{\bB}{\mathbb{B}}
\newcommand{\mH}{\mathcal{H}}
\newcommand{\Sch}{\ensuremath \mathrm{Sch}}

\begin{document}
\thispagestyle{empty}

\vspace*{2.5cm}
\begin{center}

{\bf {\LARGE Puzzles in 3D Off-Shell Geometries via VTQFT}}

\begin{center}

\vspace{1cm}

{\bf Cynthia Yan}\\
  \bigskip \rm

\,\vspace{0.5cm}\\
{Stanford Institute for Theoretical Physics,\\ Stanford University, Stanford, CA 94305, USA}
       
\vspace{0.9cm}

\rm
  \end{center}

\vspace{2.5cm}
{\bf Abstract}
\end{center}
\begin{quotation}
\noindent

We point out a difficulty with a naive application of Virasoro TQFT methods to compute path integrals for two types of off-shell 3-dimensional geometries. Maxfield-Turiaci \cite{Maxfield:2020ale} proposed solving the negativity problem of pure 3d gravity by summing over off-shell geometries known as Seifert manifolds. We attempt to compute Seifert manifolds using Virasoro TQFT. Our results don't match completely with Maxfield-Turiaci. We trace the discrepancies to not including the mapping class group properly. We also compute a 3-boundary torus-wormhole by extrapolating from an on-shell geometry. We encounter challenges similar to those observed in the comparison between the genuine off-shell computation of a torus-wormhole by Cotler-Jensen \cite{Cotler:2020ugk} and the extrapolation from an on-shell configuration.

\end{quotation}

\setcounter{page}{0}
\setcounter{tocdepth}{3}
\setcounter{footnote}{0}
\newpage

\setcounter{page}{2}
\tableofcontents
\pagebreak

\section{Introduction}

Finding the dual of AdS$_3$ gravity \cite{Collier:2019weq, Belin:2020hea, Chandra:2022bqq, Collier:2023fwi, Belin:2023efa, Collier:2024mgv, Jafferis:2024jkb} analogous to the duality between Jackiw-Teitelboim (JT) gravity \cite{Teitelboim, Jackiw, AlmheiriPolchinski} and random matrix theory \cite{Saad:2019lba, Stanford:2019vob} has been of great interest recently. Proposed dualities are statistics of OPE coefficients of an ensemble of formal CFT$_2$s with large central charge $c$ and a sparse low-energy spectrum  \footnote{This is not a true microscopic ensemble of CFT$_2$s. See \cite{Belin:2023efa} for a discussion of the microscopic ensemble. For a free toy model of CFT$_2$ ensemble microscopically defined see \cite{Maloney:2020nni, Afkhami-Jeddi:2020ezh}.}. These dualities are different from earlier examples of AdS/CFT and their boundary ensembles satisfy the Eigenstate Thermalization Hypothesis (ETH) \cite{ethSrednicki, ethDeutsch}. Virasoro TQFT \cite{Collier:2023fwi, Collier:2024mgv} gives an algorithmic procedure of computing the partition function of a fixed on-shell, i.e. hyperbolic, geometry. It is natural to ask the question whether one can get an answer for the partition function of an off-shell geometry by extrapolating an on-shell geometry computed using VTQFT. It turns out we do not get our expected answers. Nevertheless, we can still learn from these computations features of the off-shell geometries. 

One motivation of studying off-shell geometries is that existing on-shell calculations only teach us about statistics of OPE coefficients, but we also want to learn about spectral statistics of the density of states $\rho$. In this paper, we explore three different off-shell geometries using VTQFT by extrapolating on-shell geometries to off-shell geometries
\begin{itemize}
    \item $\overline{\rho\rho}$ is probed by torus-wormhole. A Maldacena-Maoz wormhole \cite{Maldacena:2004rf} $\Sigma\times I$ has two asymptotic boundaries with identical topology and constant moduli. The boundaries are Riemann surfaces. One simple option is for both boundaries to be tori and we get a torus-wormhole. This off-shell geometry was computed by Cotler and Jensen \cite{Cotler:2020ugk}. And it was further studied in \cite{Collier:2023fwi} and \cite{Yan:2023rjh}. We review these in Appendix \ref{toruswormhole};
    \be
    \includegraphics[valign=c,width=0.2\textwidth]{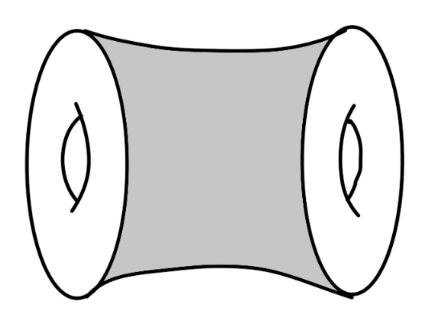}
    \ee
    \item Non-perturbative correction to $\overline{\rho}$ is probed by Seifert manifolds, which we try to compute in section \ref{seifert}. A Seifert manifold that we consider looks like a solid torus with $n$ smaller solid tori inside carved-out and then filled-in but twisted by $SL(2,\Z)$ actions;
    \be
    \includegraphics[valign=c,width=0.25\textwidth]{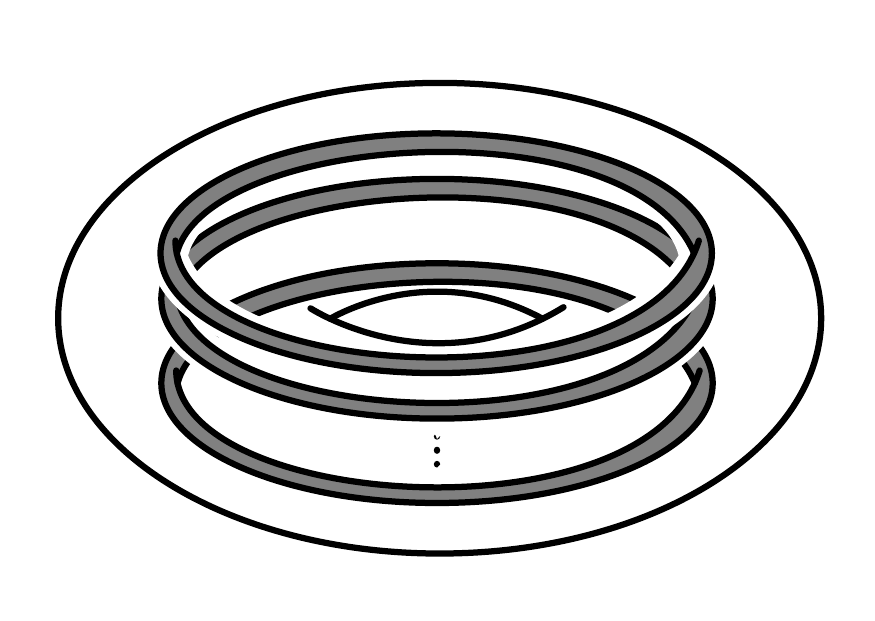}
    \ee
    \item $\overline{\rho\rho\rho}$ is probed by a 3-boundary torus-wormhole which has topology pair-of-pants times $S^1$, which we try to compute in section \ref{3toruswormhole}.
    \be
    \includegraphics[width=0.25\textwidth, valign=c]{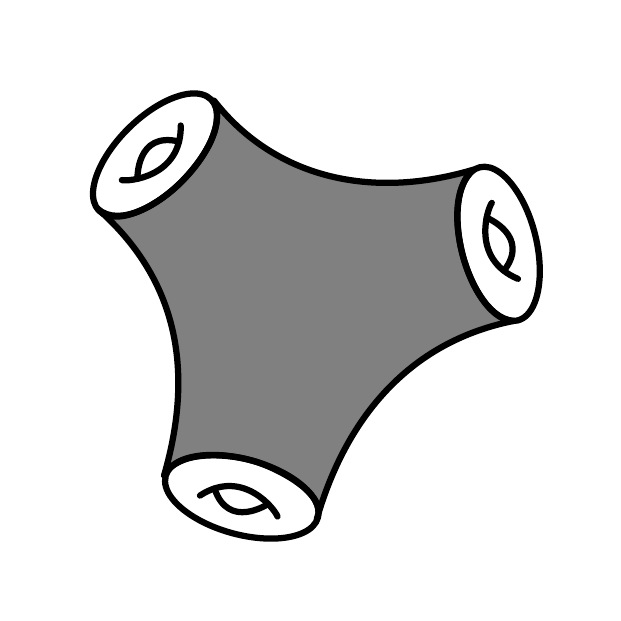}
    \ee
\end{itemize}
In addition to probing spectral statistics, there is another important motivation to study Seifert manifolds. Pure 3d gravity was studied by \cite{Maloney:2007ud, Keller:2014xba}. In the near extremal limit, there is the problem with negative density of states \cite{Benjamin:2019stq}. Maxfield and Turiaci made a proposal of solving this negativity problem by summing over off-shell geometries called Seifert manifolds \cite{Maxfield:2020ale}. They calculated these geometries by relating them to disks with defects in JT gravity which we review in Appendix \ref{negativity}. 
\be
\includegraphics[valign=c,width=0.25\textwidth]{figures/seifertn}\quad\mapsto\quad\includegraphics[valign=c,width=0.2\textwidth]{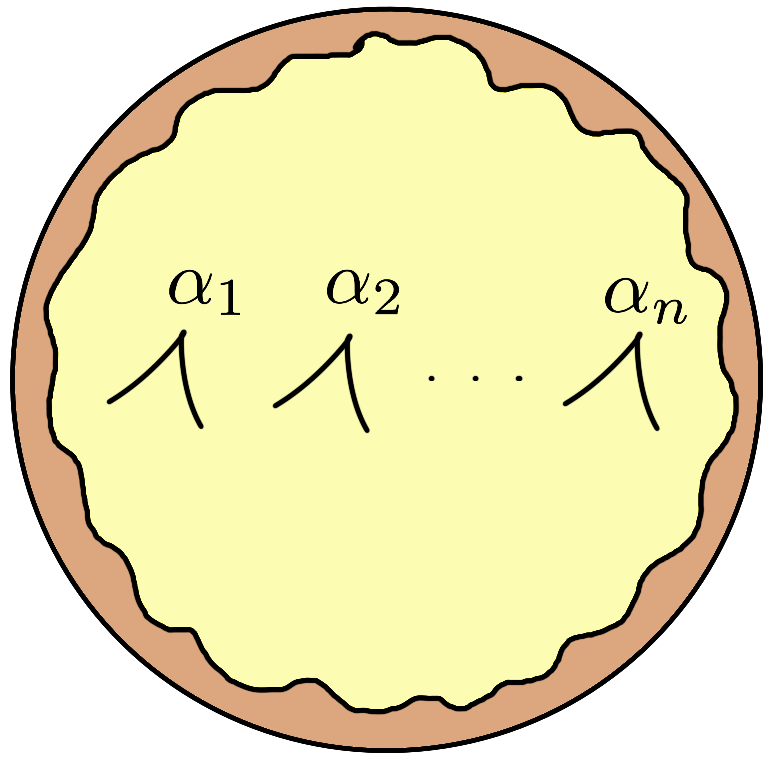}
\ee
But there still lacks a genuine 3d computation of these Seifert manifolds. In section \ref{seifert}, we make an attempt to do such a computation in the framework of VTQFT. We find that the results don't match completely with Maxfield-Turiaci \cite{Maxfield:2020ale}, but we can identify the problem of the computation to mapping class group by looking into a 2-dimensional analog. Our VTQFT computation exactly agrees with a wrong computation in JT gravity of a disk with $n$-defects while ignoring the mapping class group, where we first cut a disk with n defects like a watermelon, then glue the individual pieces together.
\be
\includegraphics[valign=c,width=0.28\textwidth]{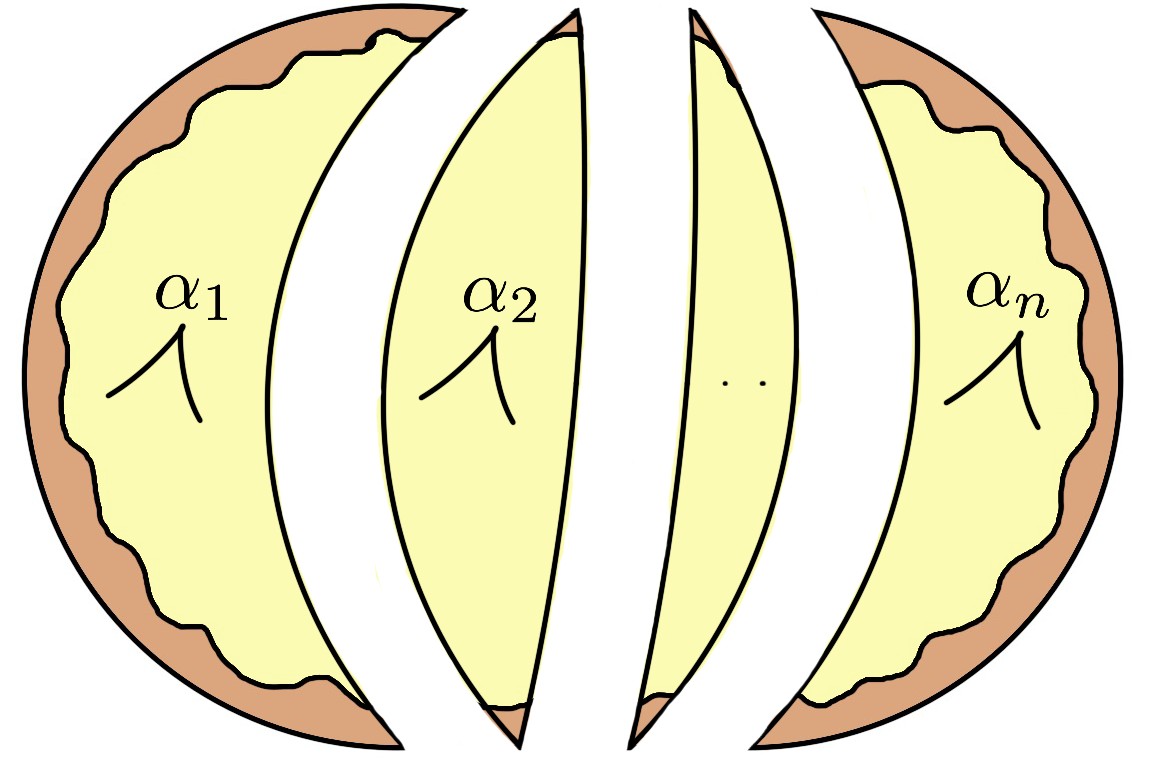}
\ee
Using VTQFT, we could compute torus wormhole and 3-boundary torus wormhole by first adding matter insertions of mass $\Delta$ and then taking the $\Delta\rightarrow0$ limit
\be
\includegraphics[width=0.2\textwidth, valign=c]{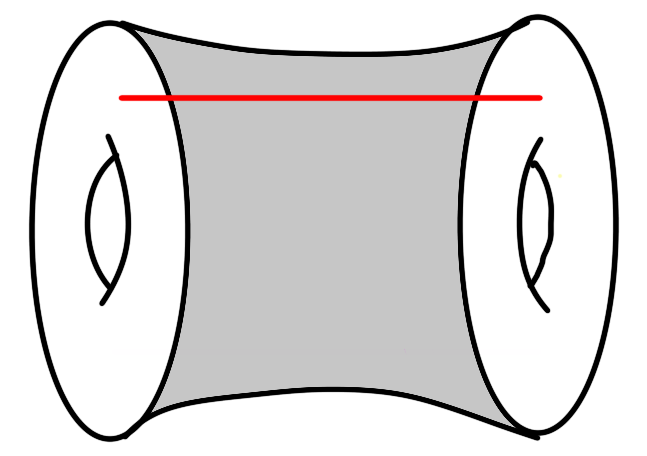}\quad\quad \includegraphics[width=0.25\textwidth, valign=c]{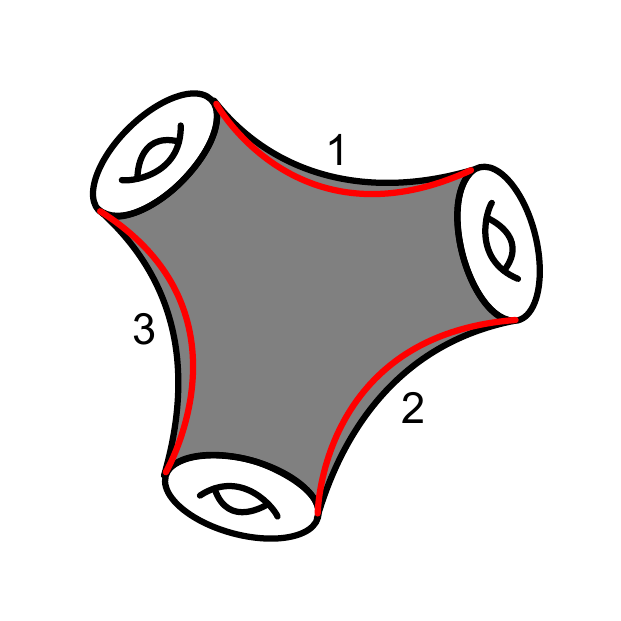}
\ee
The torus-wormhole computation using VTQFT do not match our expectation from Cotler-Jensen \cite{Cotler:2020ugk}. However, this is what we expect to happen because we can understand the difference by dimensional reduction to 2 dimensions. 
\be
\includegraphics[width=0.18\textwidth,valign=c]{figures/toruswormhole.png}\mapsto\includegraphics[width=0.25\textwidth,valign=c]{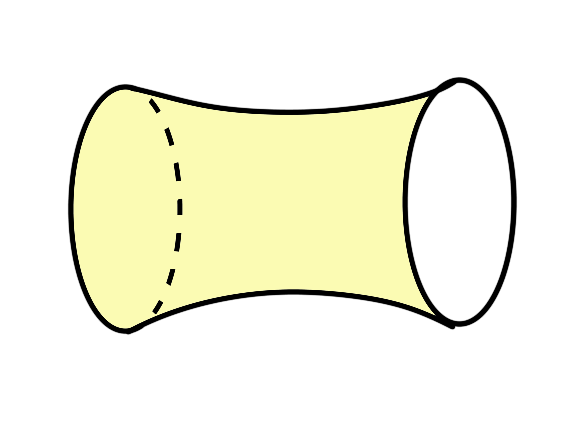}\quad\quad \includegraphics[width=0.25\textwidth, valign=c]{figures/3To2}\mapsto \includegraphics[valign=c,width=0.2\textwidth]{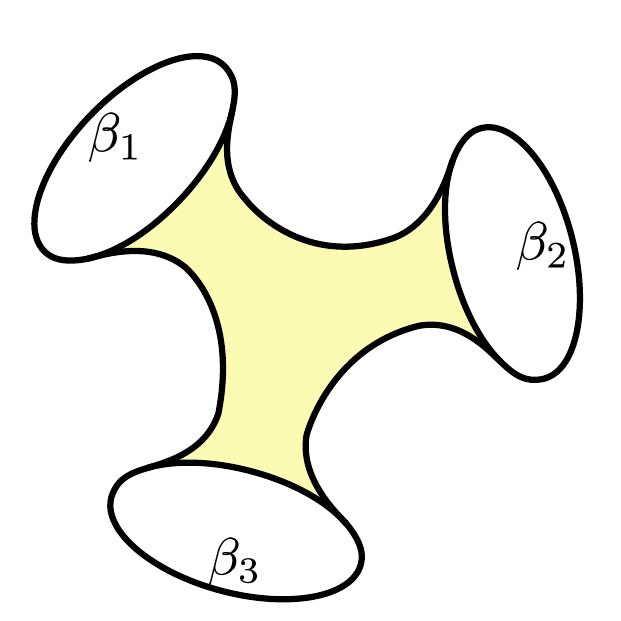}
\ee
In JT gravity, we know definitively how to compute both off-shell geometries and on-shell geometries.
\be
\includegraphics[valign=c,width=0.2\textwidth]{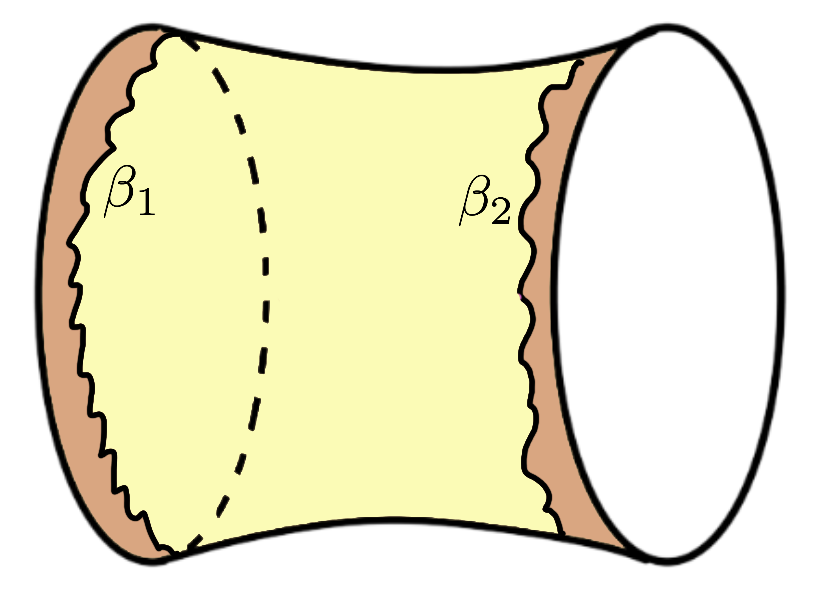} \quad\text{v.s.}\quad \includegraphics[valign=c,width=0.2\textwidth]{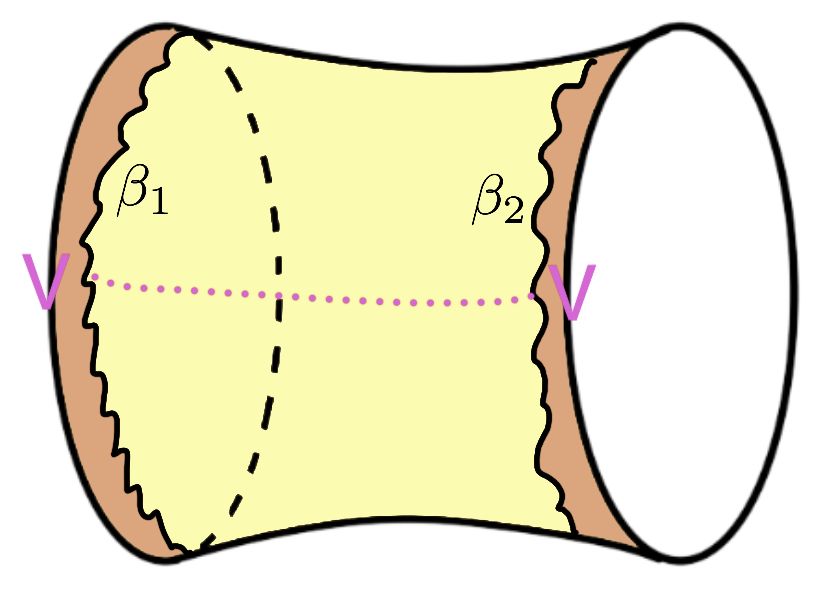}\quad\quad \includegraphics[valign=c,width=0.2\textwidth]{figures/3holenomatter}\quad\text{v.s.}\quad
\includegraphics[valign=c,width=0.2\textwidth]{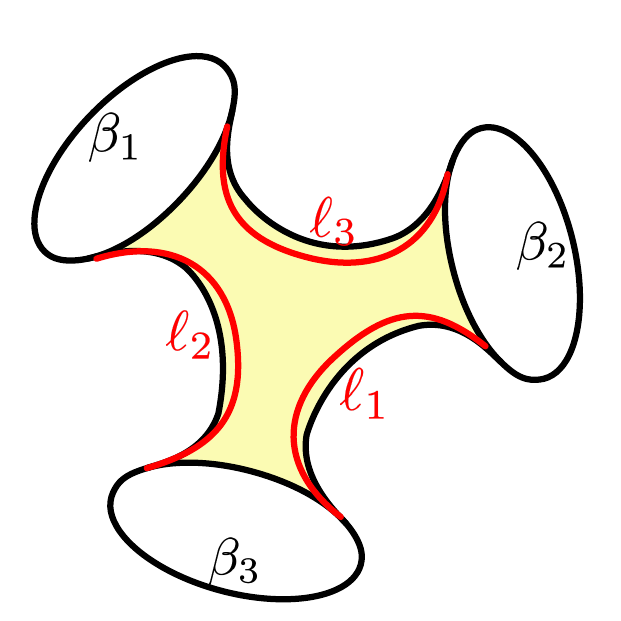}
\ee
In the $\Delta\rightarrow0$ limit, we can recover an on-shell answer from an off-shell answer by including winding terms
\be
\text{on-shell}=\text{off-shell}+\text{winding}
\ee
The answers we get in 3d for torus wormhole and 3-boundary torus wormhole look intuitively very similar to the above situation. But in the 3d case, we want to recover an off-shell answer from an on-shell computation. Even in 2d, this is the opposite direction than what we know how to do. However, we can make a guess on the answer of an off-shell 3-boundary torus-wormhole by referring to the differences between on-shell and off-shell computations of a torus-wormhole. 

In section \ref{3toruswormhole}, we give two methods of computing 3-boundary torus-wormhole using VTQFT, and arrive at the same answer. These computations also provide a sanity check of VTQFT for geometries with torus boundaries. 

\noindent\textbf{Note (v3)} There is a simultaneous work on similar subjects by Jan de Boer, Joshua Kames-King, and Boris Post \cite{deBoer:2025rct}. 

\section{Seifert manifolds in Virasoro TQFT}
\label{seifert}

The partition function of pure 3d gravity was computed by Maloney-Witten \cite{Maloney:2007ud} and Keller-Maloney \cite{Keller:2014xba}. It was pointed out by \cite{Benjamin:2019stq} that the density of states becomes negative in the near-extremal limit \cite{Benjamin:2019stq}. Maxfield-Turiaci had a proposal of solving the negativity problem by adding Seifert manifolds \cite{Maxfield:2020ale}. They calculated these geometries by relating them to disks with defects in JT gravity. (See appendix \ref{negativity} for a review of their computation.) But there still lacks a genuine 3d computation of these Seifert manifolds. In section \ref{seifert:VTQFT}, we make an attempt to do such a computation in the framework of VTQFT. We will find that the results don't match completely with Maxfield-Turiaci. In section \ref{seifert2d}, we identify the problem of the computation to mapping class group by doing dimensional reduction to 2 dimensions.

\subsection{VTQFT Computation}
\label{seifert:VTQFT}

In this subsection, we give a computation of Seifert manifolds using VTQFT \footnote{The method we show here was first pointed out by Boris Post.}. Before computing Seifert manifolds, let's warmup by first introducing alternative ways of obtaining SL(2,$\Z$) blackholes. We parameterize a torus boundary by modular parameter $\tau$. The spatial cycle is shown in red, and we fill in the Euclidean time cycle to obtain 
a BTZ blackhole. If we first act on $\tau$ by a an SL(2,$\Z$) transform $\gamma=\begin{pmatrix}a&b\\r&d\end{pmatrix}$ to get $\gamma\cdot\tau=\frac{a\tau+b}{r\tau+d}$ and then fill in the Euclidean time cycle, we get an $SL(2,\Z)$ blackhole (see Appendix \ref{seifert:review} for a detailed review).
\be
\includegraphics[valign=c,width=0.25\textwidth]{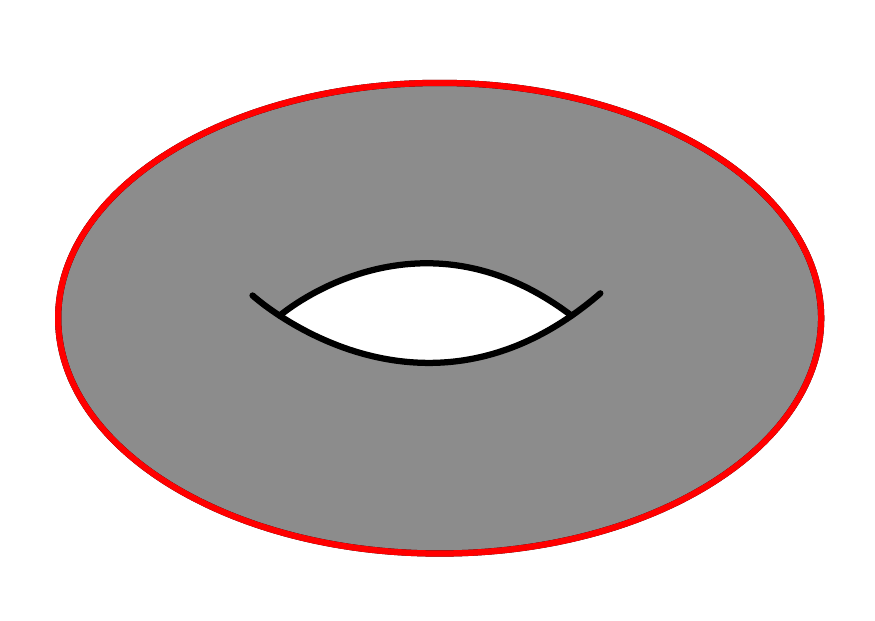}\quad\xrightarrow{\gamma}\quad\includegraphics[valign=c,width=0.25\textwidth]{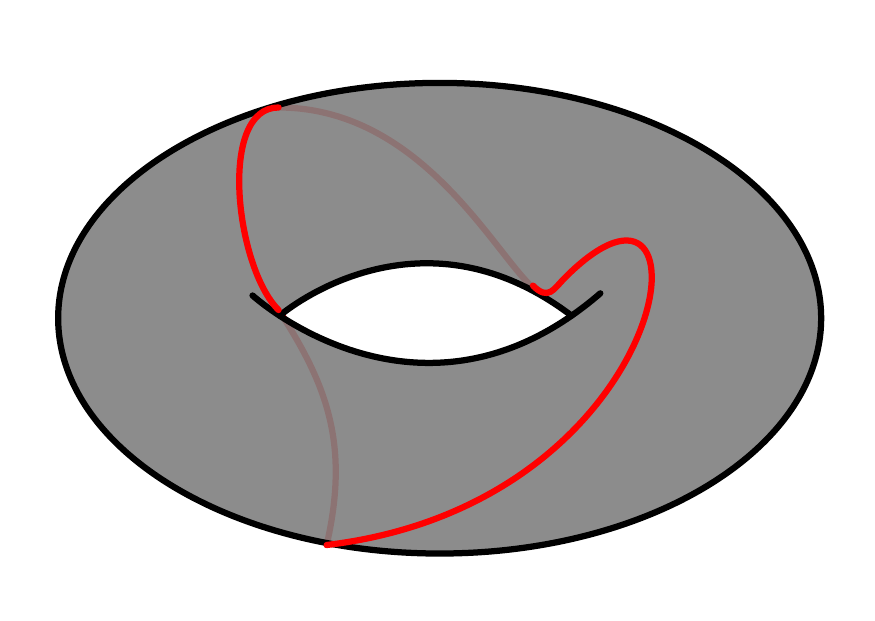}
\ee

\paragraph{An alternative way of obtaining an $SL(2,\Z)$ blackhole} 
Alternatively, we can obtain an $SL(2,\Z)$ blackhole by first carving out a smaller solid torus inside a solid torus, and then gluing back a solid torus but twisted by $\gamma$. 
\be
\includegraphics[valign=c,width=0.25\textwidth]{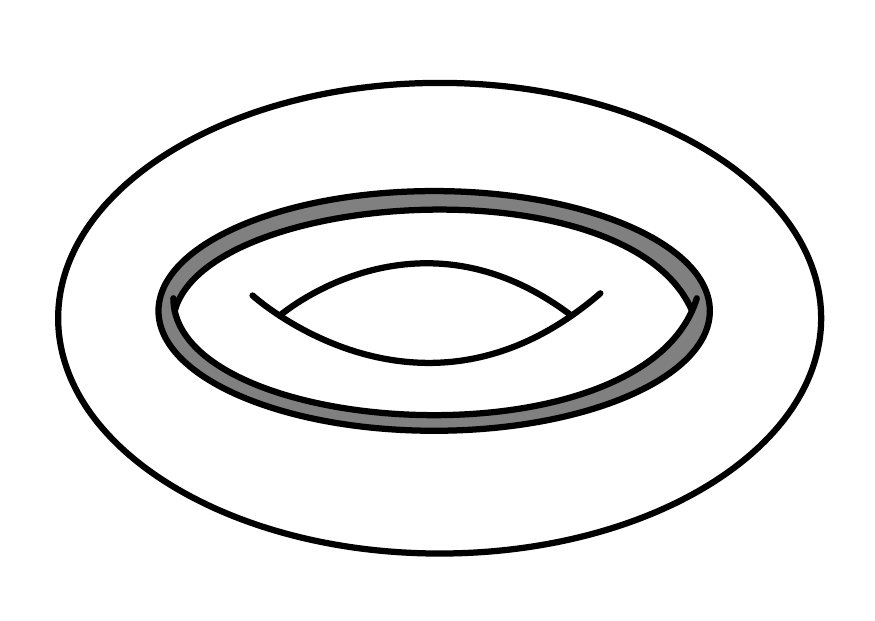}
\ee
More concretely, we start with a bulk torus-wormhole, and glue the outside to an asymptotic 3d trumpet and fill in the inside with a solid torus but twisted by $\gamma$. There are three components to this computation: (1) A 3d trumpet is a solid torus with a Wilson loop labeled by $(P,\bar{P})$ running through the middle \cite{Mertens:2022ujr, Jafferis:2024jkb} which is given by 
\be
Z^{3d}_{\text{trumpet}}=\chi_P(-1/\tau)\chi_{\bar{P}}(-1/\bar{\tau})
\ee 
Here conformal weight is given by $h=\frac{c-1}{24}+P^2=\frac{1}{4}(b+b^{-1})^2+P^2$ in terms of the Liouville parameter $P$.
(2) A bulk torus-wormhole is given by 
\be
\delta(P-P_1)\delta(\bar{P}-\bar{P}_1)
\ee
labeled by two sets of Liouville parameters. (See (\ref{tw1bd}) in appendix \ref{onshell32} for more details.)
(3) Let's describe how to fill in the inner torus. The spatial circle of the to-be-glued small solid torus of slope 0 is glued to a circle of slope $-d/r$ on the carved-out boundary. This gluing process is called a Dehn-filling, and is represented by integrating the modular crossing kernels $\mathbb{K}^{(d,-r)}_{\one P_1}[\one]\mathbb{K}^{(d,-r)}_{\one \bar{P}_1}[\one]$ over $P_1$ and $\bar{P}_1$. The modular transform of the character $\chi_{P}(\tau)$ is implemented via the modular crossing kernel
\be
\chi_P(\gamma\cdot \tau)=\int_0^\infty dP'\,\K_{P'P}^{(r,d)}[\one]\chi_{P'}(\tau)
\ee
See (\ref{kkernel}) for the expression of $\mathbb{K}^{(d,-r)}_{\one P_1}[\one]$ and appendix \ref{slbh} for more details.

We notice that all the above quantities as in any VTQFT computations factorize into a chiral part and an anti-chiral part multiplied together. Let's just focus on the chiral part from now on. The chiral part of the SL(2,$Z$) blackhole partition function is given by
\be
Z^{vir}_{(r,d)}=\int dP\,dP_1\,\underbrace{\chi_P(-\frac{1}{\tau})}_{\substack{\text{glue outer bdy to}\\\text{asymptotic 3d trumpet}}}\underbrace{\delta(P-P_1)}_{\substack{\text{bulk}\\\text{torus-wormhole}}}\underbrace{\mathbb{K}^{(d,-r)}_{\one P_1}[\one]}_{\substack{\text{Dehn-filling}\\\text{on inner bdy}}}=\chi_\one(\gamma\cdot\tau)
\ee
as expected. 

If we do this carve-out and fill-in procedure on more than one tori inside a solid torus, we get Seifert manifolds with torus boundary. After we carve out the $i$th solid torus, we glue back a solid torus with spatial circle of slope $-d_i/r_i$
\be
\includegraphics[valign=c,width=0.25\textwidth]{figures/seifertn}
\ee
But we are still not ready to compute Seifert manifolds in VTQFT yet, we need to give another way of calculating SL(2,$Z$) blackhole, and then we can generalize from there.

\paragraph{An alternative way to the alternative way} Now instead of gluing the outside torus to an asymptotic 3d trumpet directly, we do an S-transform on the outside torus and then glue it to the S-transformed asymptotic 3d trumpet.
\be
\includegraphics[valign=c,width=0.25\textwidth]{figures/2tp}\xrightarrow{\text{S-transform}} \includegraphics[valign=c,width=0.25\textwidth]{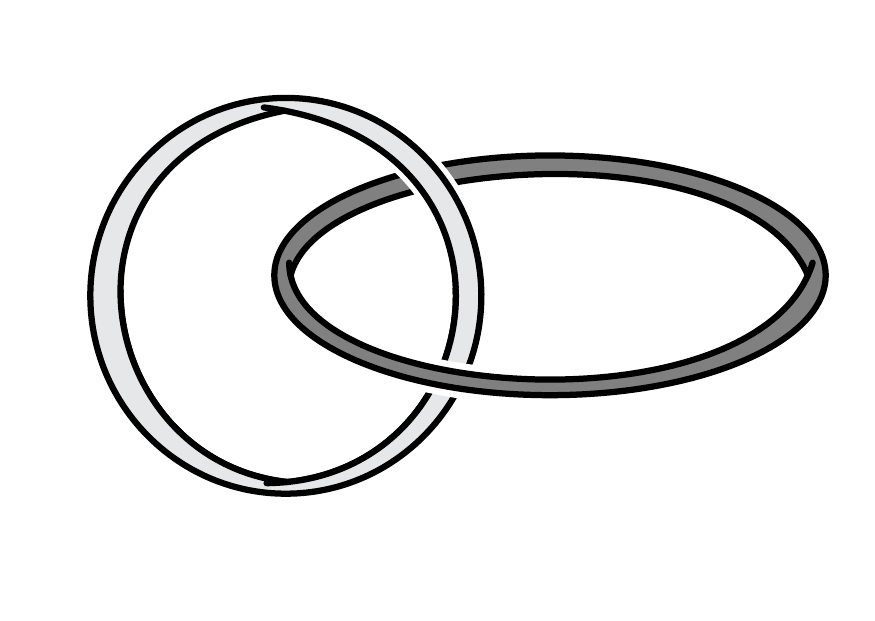}
\ee
Here the outside torus becomes the vertical torus after S-transform. In other words, there are still three components to this computation but two of them have changed: (1) An S-transformed 3d trumpet. (2) A Hopf link 
\be
\includegraphics[valign=c,width=0.25\textwidth]{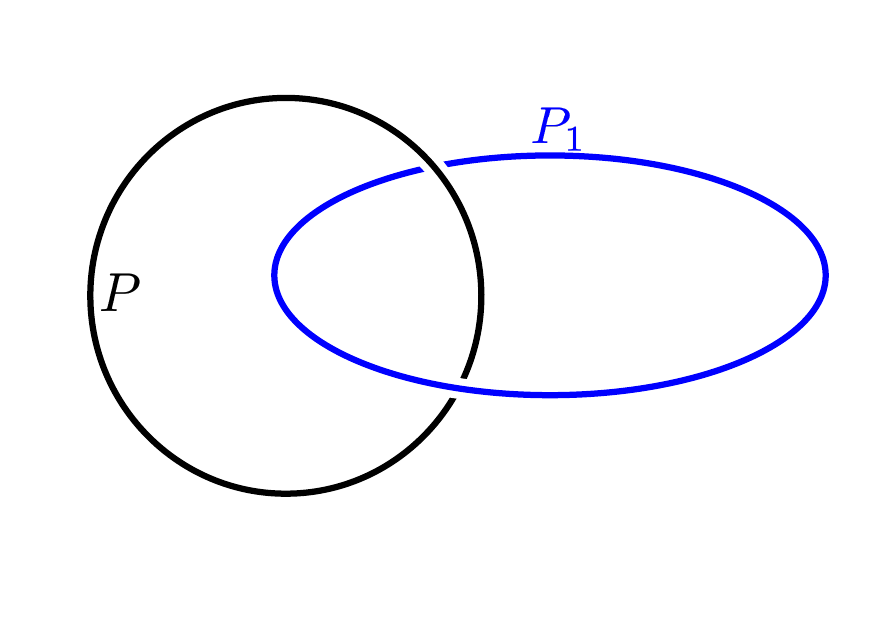}=\bS_{P_1P}[\one]\label{1link}
\ee
(See (\ref{thlink1bd}) in appendix \ref{onshell32} for one derivation and see appendix \ref{linkthlink} for another derivation of this formula.) (3) Dehn filling on the inner torus.

We then calculate the chiral part of the partition function of an SL(2,$Z$) blackhole as follows
\begin{align}
Z^{vir}_{(r,d)}&=\int dP'dP\,dP_1\,\underbrace{\chi_{P'}(-\frac{1}{\tau})\bS_{P'P}[\one]}_{\substack{\text{glue to S-transformed}\\\text{asymptotic 3d trumpet}}}\includegraphics[valign=c,width=0.2\textwidth]{figures/linkp1}\underbrace{\mathbb{K}^{(d,-r)}_{\one P_1}[\one]}_{\substack{\text{Dehn-filling}\\\text{on inner bdy}}}\label{altalt}\\
&=\int dP'dP\,dP_1\,\chi_{P'}(-\frac{1}{\tau})\bS_{P'P}[\one]\bS_{P_1P}[\one]\mathbb{K}^{(d,-r)}_{P_1,\one}[\one]\\
&=\int dP\,\chi_{P}(\tau)\mathbb{K}^{(r,d)}_{P,\one}[\one]=\chi_\one(\gamma\cdot\tau)
\end{align}
so we again get our expected answer. 

\paragraph{Seifert Manifolds} We use similar method to calculate Seifert manifolds. This method of computing Seifert manifolds was pointed out by Boris Post and the related links were computed in Post-Tsiares \cite{Post:2024itb}. We do an S-transform on the outside torus and then glue it to the S-transformed asymptotic 3d trumpet. 
\be
\includegraphics[valign=c,width=0.25\textwidth]{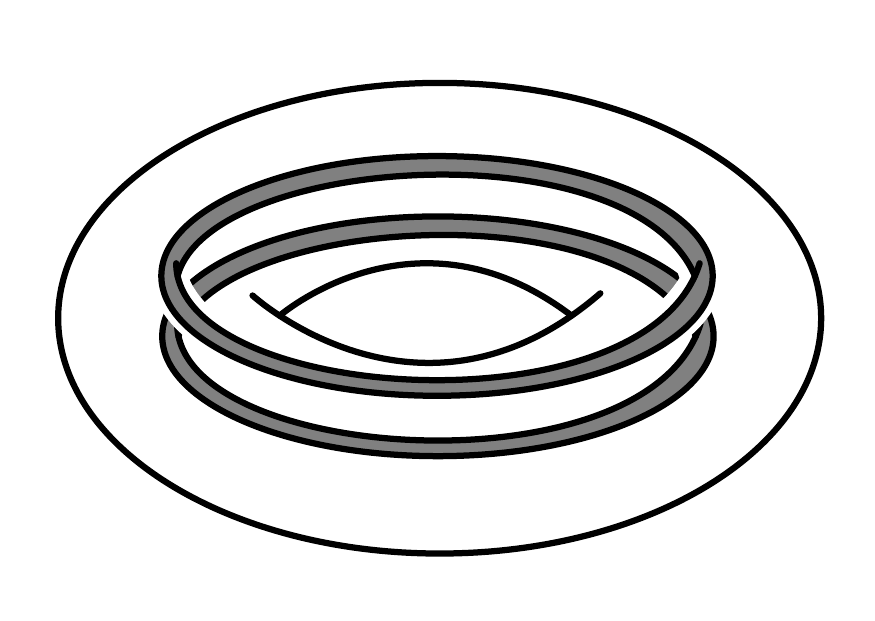}\xrightarrow{\text{S-transform}} \includegraphics[valign=c,width=0.25\textwidth]{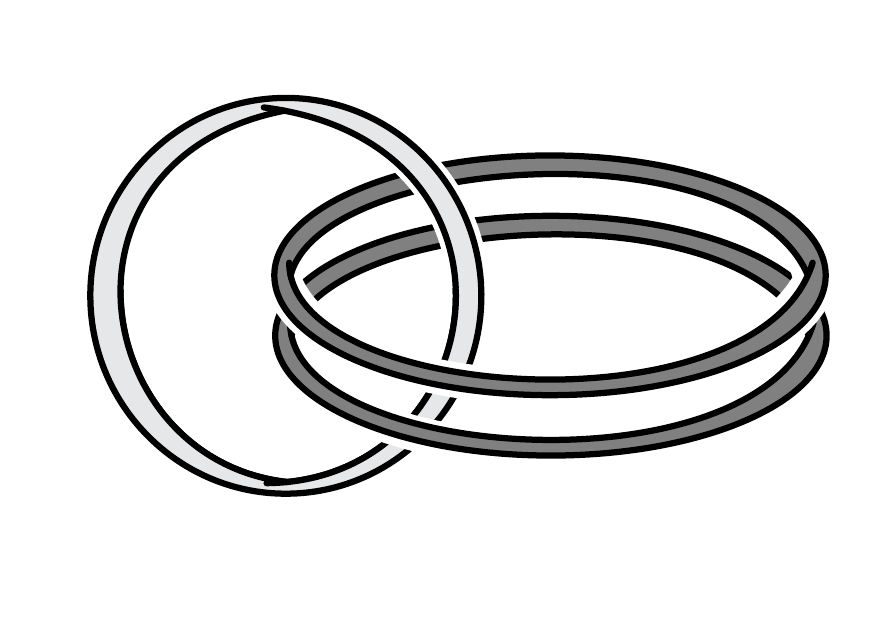}
\ee
Here the outside torus becomes the vertical torus after S-transform. There are four components to this computation: (1) An S-transformed 3d trumpet. (2) A link with 3 loops
\be
\includegraphics[valign=c,width=0.25\textwidth]{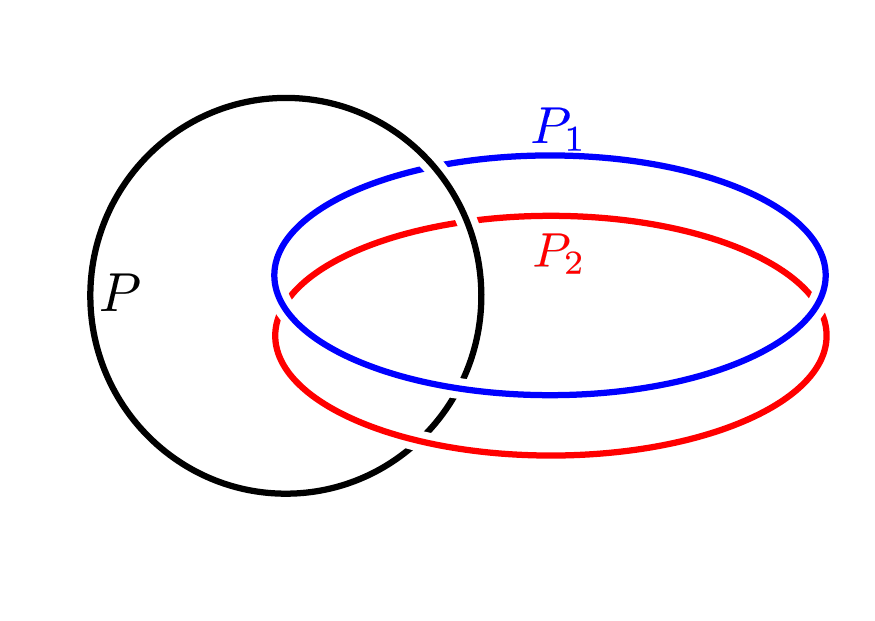}=\frac{\bS_{P_1P}[\one]\bS_{P_2P}[\one]}{\bS_{P\one}[\one]}
\ee
(See appendix \ref{link3more} for a derivation of the above formula.) (3) Dehn-filling on first inner torus. (4) Dehn-filling on second inner torus. The chiral part of the partition function is then given by
\begin{align}
Z^{vir}(\includegraphics[valign=c,width=0.2\textwidth]{figures/3To1})&=\int dP'dP\,dP_1dP_2\,\underbrace{\chi_{P'}(-\frac{1}{\tau})\bS_{P'P}[\one]}_{\substack{\text{glue to S-transformed}\\\text{asymptotic bdy}}}\includegraphics[valign=c,width=0.2\textwidth]{figures/linkp1p2}\underbrace{\mathbb{K}^{(d_1,-r_1)}_{P_1,\one}[\one]}_{\substack{\text{Dehn-filling}\\\text{on torus 1}}}\underbrace{\mathbb{K}^{(d_2,-r_2)}_{P_2,\one}[\one]}_{\substack{\text{Dehn-filling}\\\text{on torus 2}}}\\
&=\int dP'dP\,dP_1dP_2\,\chi_{P'}(-\frac{1}{\tau})\bS_{P'P}[\one]\frac{\bS_{P_1P}[\one]\bS_{P_2P}[\one]}{\bS_{P\one}[\one]}\mathbb{K}^{(d_1,-r_1)}_{P_1,\one}[\one]\mathbb{K}^{(d_2,-r_2)}_{P_2,\one}[\one]\\
&=\int dP\,\chi_{P}(\tau)\frac{\mathbb{K}^{(r_1,d_1)}_{P_1,\one}[\one]\mathbb{K}^{(r_2,d_2)}_{P_2,\one}[\one]}{\bS_{P\one}[\one]}\label{3bdySeifert}
\end{align}
In general, we can use the same method to calculate a solid torus with $n$ small tori carved-out and filled-in using the identity
\be
\includegraphics[valign=c,width=0.25\textwidth]{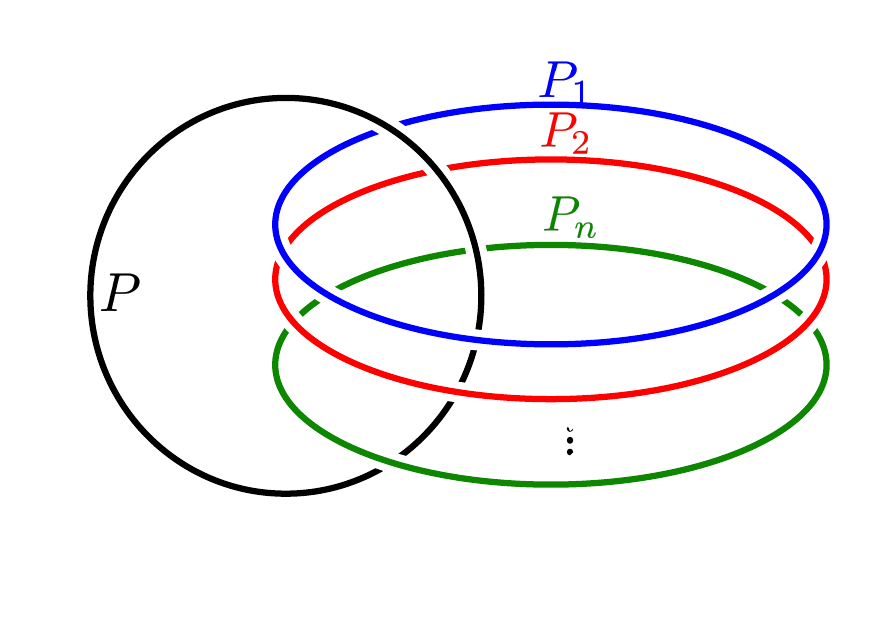}=\frac{\prod_{i=1}^n\bS_{P_iP}[\one]}{\bS_{P\one}[\one]^{n-1}}
\ee
(See appendix \ref{link3more} for a derivation of the above formula.) The chiral part of the partition function is given by
\begin{align}
Z^{vir}(\includegraphics[valign=c,width=0.2\textwidth]{figures/seifertn})&=\int dP'dP\,\prod_{i=1}^n dP_i\,\underbrace{\chi_{P'}(-\frac{1}{\tau})\bS_{P'P}[\one]}_{\substack{\text{glue to S-transformed}\\\text{asymptotic bdy}}}\includegraphics[valign=c,width=0.25\textwidth]{figures/linkpn}\underbrace{\prod_{i=1}^n\mathbb{K}^{(d_i,-r_i)}_{\one P_i}[\one]}_{\text{Dehn-filling}}\\
&=\int dP\,\chi_{P}(\tau)\frac{\prod_{i=1}^n\mathbb{K}^{(r_i,d_i)}_{\one P}[\one]}{\bS_{\one P}[\one]^{n-1}}\label{seifertbefore}
\end{align}

\paragraph{Modular Invariance} To ensure modular invariance, we need to sum over all possible Dehn-filling AND we also need to sum over all possible modular transform when we glue to the asymptotic 3d trumpet. Thus (\ref{seifertbefore}) becomes 
\begin{align}
&\int dP''dP'dP\,\prod_{i=1}^n dP_i\,\underbrace{\chi_{P''}(-\frac{1}{\tau})\bS_{P''P'}[\one]}_{\substack{\text{glue to S-transformed}\\\text{asymptotic bdy}}}\underbrace{\color{orange}\K_{P',P}^{(r,d)}[\one]\color{black}}_{\substack{\text{SL(2,$\Z$) transform}\\\text{on asymptotic boundary}}}\includegraphics[valign=c,width=0.25\textwidth]{figures/linkpn}\underbrace{\prod_{i=1}^n\mathbb{K}^{(d_i,-r_i)}_{\one P_i}[\one]}_{\text{Dehn-filling}}\\
=&\int dP'dP\,\K_{P',P}^{(r,d)}[\one]\chi_{P'}(\tau)\frac{\prod_{i=1}^n\mathbb{K}^{(r_i,d_i)}_{\one P}[\one]}{\bS_{\one P}[\one]^{n-1}}
\end{align}
The corresponding chiral part of density of states is given by
\be
\tilde{\rho}_n^{(r,d),(r_i,d_i)}(P')=\int dP\,\K_{P',P}^{(r,d)}[\one](\tau)\frac{\prod_{i=1}^n\mathbb{K}^{(r_i,d_i)}_{\one P}[\one]}{\bS_{\one P}[\one]^{n-1}}
\ee
The total density of states is then given by summing over all possible $n$, coprime $(r,d)$, and all coprime $(r_i,d_i)$ with $r_i>d_i$
\begin{align}
    \tilde{\rho}(P,\bar{P})&=\rho_{MWK}(P,\bar{P})+\sum_{n=2}^\infty\tilde{\rho}_n(P,\bar{P})\\  \tilde{\rho}_n(P,\bar{P})&=\sum_{(r,d)}\sum_{\substack{(r_i,d_i)\\r_i>d_i}}\tilde{\rho}_n^{(r,d),(r_i,d_i)}(P)\tilde{\rho}_n^{(r,d),(r_i,d_i)}(\bar{P})
\end{align}
where $\rho_{MWK}$ is the density of states of BTZ blackhole plus SL(2,$Z$) blackholes. This density of states is manifestly modular invariant because of the sum over $(r,d)$.

\paragraph{Near-extremal limit} In the near-extremal limit, for $n$ tori being carved-out and filled-in inside a solid torus, the leading order contribution is given by $(r,d)=(0,1)$ and $(r_i,d_i)=(2,1)$. Thus we have the density of states is given by 
\be
\tilde{\rho_n}(P,\bar{P})\approx\frac{\left(\mathbb{K}^{(2,1)}_{\one P}[\one]\mathbb{K}^{(2,1)}_{\one\bar{P}}[\one]\right)^n}{\left(\bS_{\one P}[\one]\bS_{\one\bar{P}}[\one]\right)^{n-1}}\approx 32\pi^3e^{S_0(J)}\lambda^n \frac{1}{\pi P^{2n-2}}
\ee
where
\be
\lambda=\frac{(-1)^J}{8\pi^2}e^{-\frac{S_0(J)}{2}}
\ee
In the following table we compare our result with result of Maxfield-Turiaci \cite{Maxfield:2020ale} (see appendix \ref{MTs} for review) in the near-extremal limit.
\be
\begin{tabular}{c|c}
    \\
    Maxfield-Turiaci & VTQFT\\
    \\
    \hline
    \\
    $\rho_2(P,\bar{P})\approx-\frac{1}{2\pi^2P^2}$&$\tilde{\rho_2}(P,\bar{P})\approx\frac{1}{2\pi^2P^2}$\\
    \\
    \hline
    \\
    $\rho_n(P,\bar{P})\approx 2^n\binom{\frac{1}{2}}{n}32\pi^3e^{S_0(J)}\lambda^n \frac{1}{\pi P^{2n-2}}$ & $\tilde{\rho_n}(P,\bar{P})\approx 32\pi^3e^{S_0(J)}\lambda^n \frac{1}{\pi P^{2n-2}}$\\
    \\
\end{tabular}\label{tablecompare}
\ee
We notice that $e^{\# S_0(J)}$ and $P^\#$ match, but we miss a combinatorial factor $2^n \binom{\frac{1}{2}}{n}$. The contrast is easiest to see when $n=2$, where the two results differ by a minus sign. In the next section, we try to trace down the problem by doing dimensional reduction to 2 dimensions.

\subsection{Explaining our problem by going to 2d} 
\label{seifert2d}

Recall that a seifert manifold which looks like a solid torus with $n$ smaller solid tori carved-out and filled-in can be dimensionally reduced to a disk with $n$ defects in JT gravity \cite{Maxfield:2020ale}
\be
\includegraphics[valign=c,width=0.25\textwidth]{figures/seifertn}\quad\mapsto\quad\includegraphics[valign=c,width=0.2\textwidth]{figures/ndefects.png}
\ee
and in the near-extremal limit we are interested in, it's the same to consider a disk with $n$ holes, where the perimeters of the holes are related to the angles of the defects $b_i=2\pi i \alpha_i$. Now let's start by doing a wrong calculation of disk with two holes, where we just cut the disk like cutting a watermelon. Compared to the correct way of cutting into a trumpet and a 3-hole sphere (see (\ref{trumpetandvolume}) in appendix \ref{MTs}), this wrong method ignores the fact that a lot of the ways of doing the cut are equivalent because they are related by large diffeomorphisms (the mapping class group) and counts each one as if they were distinct.
\begin{align}
Z^{\text{wrong}}_2(\beta,b_1,b_2)&=\includegraphics[valign=c,width=0.2\textwidth]{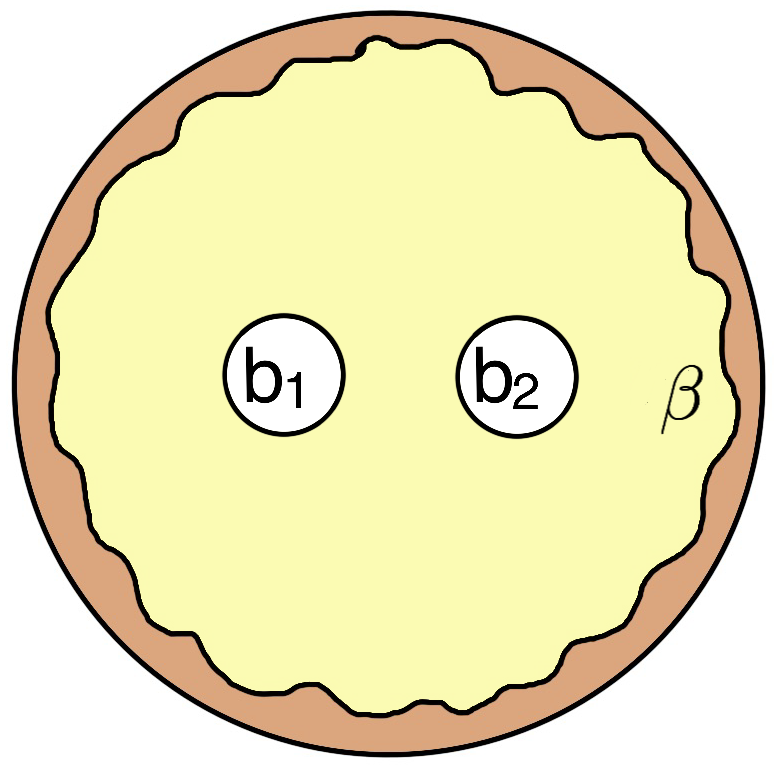}\\
&=\int\,e^\ell d\ell\includegraphics[valign=c,width=0.2\textwidth]{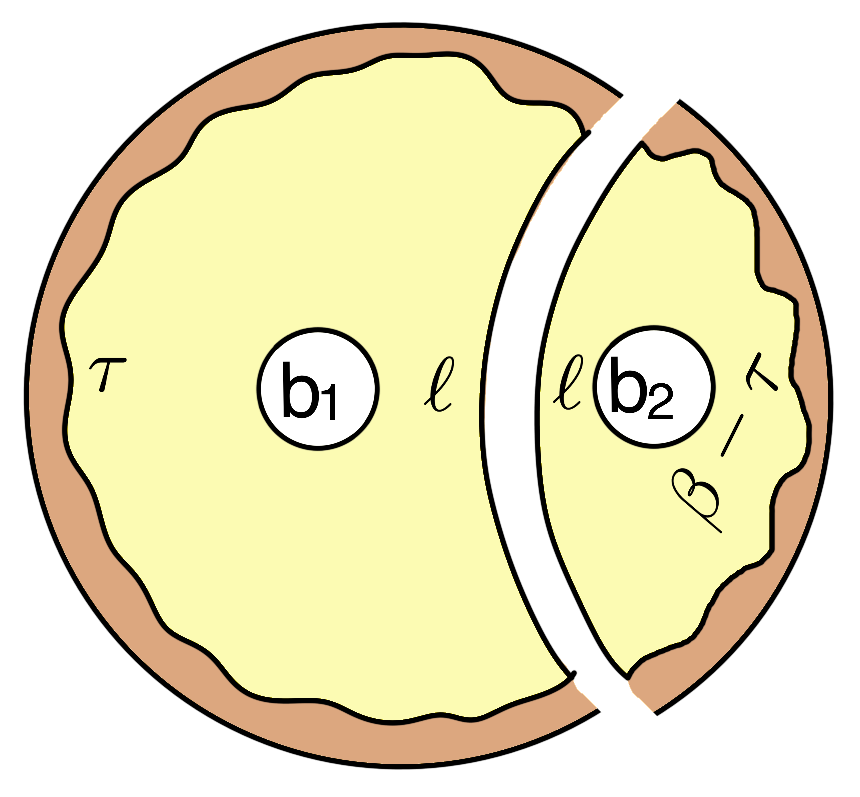}\\
&=e^{-S_0}\int\,e^\ell d\ell\,\varphi_{\text{Trumpet},\tau}(\ell,b_1)\varphi_{\text{Trumpet},\beta-\tau}(\ell,b_2)\\
&=e^{-S_0}\int_0^\infty ds\,\frac{\rho_1(s,b_1)\rho_1(s,b_2)}{\rho_0(s)}e^{-\beta s^2/2}
\end{align}
The corresponding density of states is given by 
\begin{align}
\tilde{\rho}_2(s)&=\frac{\rho_1(s,b_1)\rho_1(s,b_2)}{\rho_0(s)}\\
&\approx \frac{1}{\pi s^2}
\end{align}
This is exactly minus of the correct answer. Here $s$ is related to the energy by $E=\frac{1}{2}s^2$, and $\rho_0$ and $\rho_1$ are the disk and trumpet density of states respectively. (See (\ref{diskrho}) and (\ref{trumpetrho}) in appendix \ref{seifert:2d} for their formulas. And see (\ref{2hole}) in appendix \ref{MTs} to review how to get the correct answer.) This is the same situation as in our VTQFT computation where the $\tilde{\rho}^{3d}_2(P,\bar{P})$ differ from Maxfield-Turiaci by a minus sign as was shown in table \ref{tablecompare}. 

Now we argue that the dimensional reduction of our VTQFT computation of Seifert manifolds with torus boundary is calculating disk with $n$ defect while ignoring the mapping class group. Now instead of computing a disk with $n$ holes the correct way by cutting it into a trumpet and a ($n+1$)-hole sphere (see (\ref{trumpetandvolumen}) in appendix \ref{MTs}), let's calculate disk with $n$ holes but ignoring the mapping class group by cutting it like cutting a watermelon.
\begin{align}
Z^{wrong}_n(\beta,b_1,\cdots, b_n)&=\includegraphics[valign=c,width=0.2\textwidth]{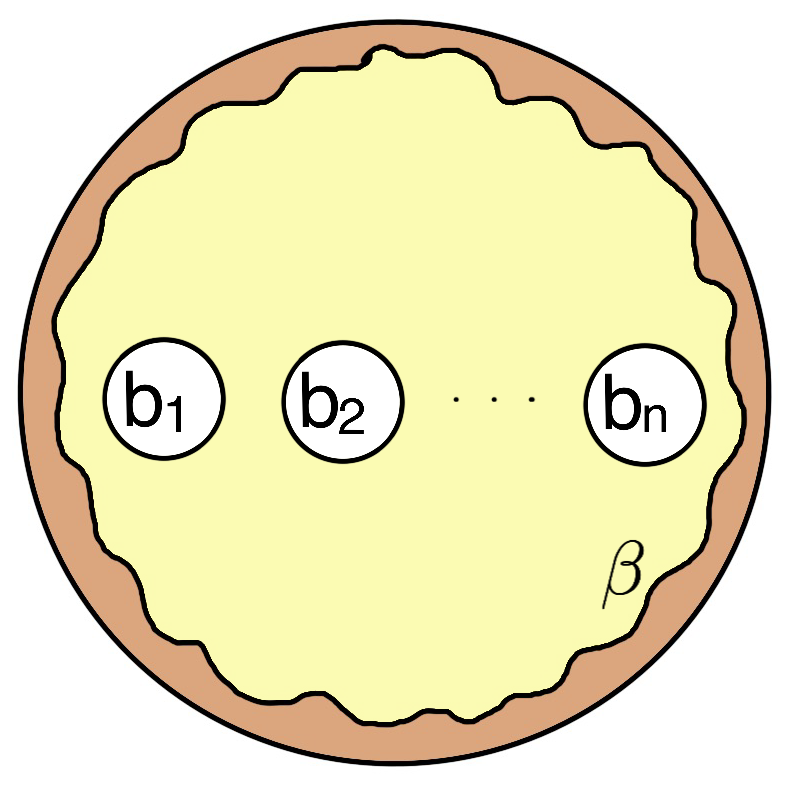}\\
&=\int\prod_{i=1}^{n-1}e^{\ell_i}d\ell_i\includegraphics[valign=c,width=0.25\textwidth]{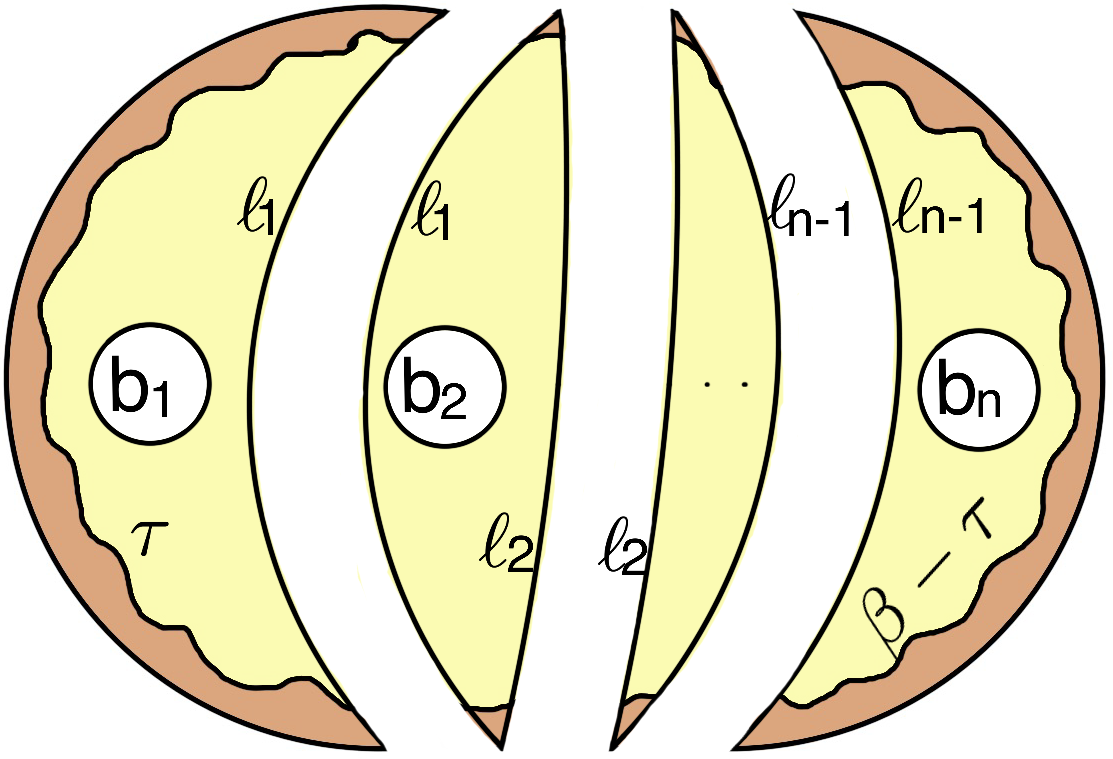}\\
&=e^{-(n-1)S_0}\int \prod_{i=1}^{n-1}e^{\ell_i}d\ell_i\, \varphi_{\text{Trumpet},\tau}(\ell_1,b_1)\left(\prod_{i=2}^{n-1}\braket{\ell_{i-1},b_i|\ell_i}\right)\varphi_{\text{Trumpet},\beta-\tau}(\ell_{n-1},b_n)\\
&=e^{-(n-1)S_0}\int_0^\infty ds\,\frac{\prod_{i=1}^n\rho_1(s,b_i)}{\rho_0(s)^{n-1}}e^{-\beta s^2/2}
\end{align}
(For a review of how to do this JT computation see appendix \ref{MTs} in particular (\ref{wedgeeq}).) The corresponding density of states is give by 
\begin{align}
\tilde{\rho}_n(s)&=\frac{\prod_{i=1}^n\rho_1(s,2\pi i\alpha_i)}{\rho_0(s)^{n-1}}\\
&\approx \frac{1}{\pi s^{2n-2}}
\end{align}
where in the second line we have taken the limit $s\rightarrow0$. This agrees with VTQFT computation in near-extremal limit 
\be
\tilde{\rho}_n^{3d}(P,\bar{P})\approx 32\pi^3e^{S_0(J)}\lambda^n\tilde{\rho}_n^{2d}(P)
\ee
Therefore, we conclude that our VTQFT computation disagree with Maxfield-Turiaci because we didn't take into account the mapping class group. (See Apendix \ref{MCG} for mapping class groups of some 3d objects.) Intuitively, this could be understood as in our VTQFT computation different carved-out and filled-in tori are independent of each other, and our 2d computation is in a BF theory where all the boundaries factorize. 

\section{Three-boundary torus wormhole in Virasoro TQFT}
\label{3toruswormhole}

In this section, we consider a 3-boundary torus-wormhole, which has topology a pair-of-pants times $S^1$, and has three asymptotic torus boundaries. 
\be
\includegraphics[width=0.25\textwidth, valign=c]{figures/3To2}=\includegraphics[width=0.25\textwidth, valign=c]{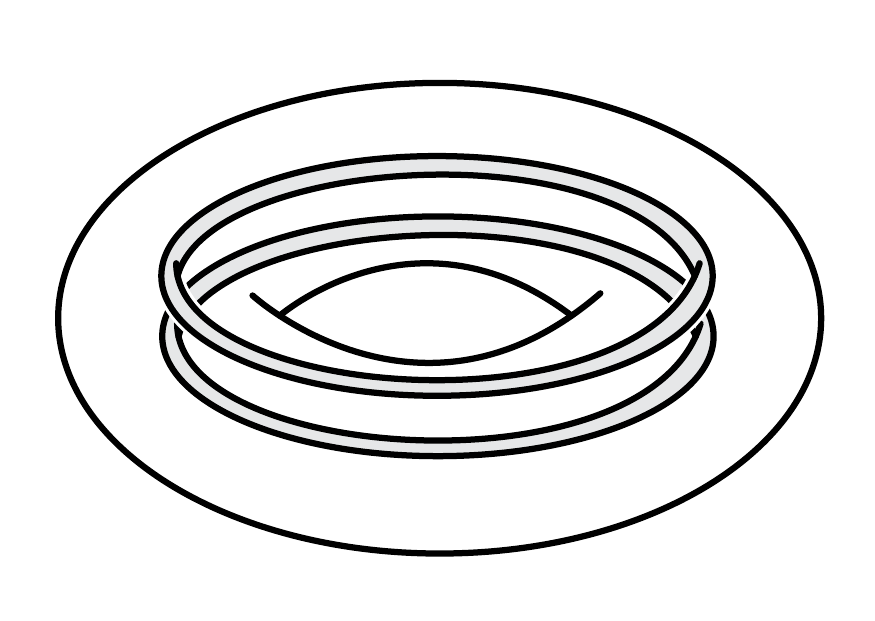}
\ee
The simplest and most-symmetric way of making this geometry on-shell is by inserting two operators of mass $\Delta$ on each boundary. 
\be
\includegraphics[width=0.25\textwidth, valign=c]{figures/3T}=\includegraphics[width=0.25\textwidth, valign=c]{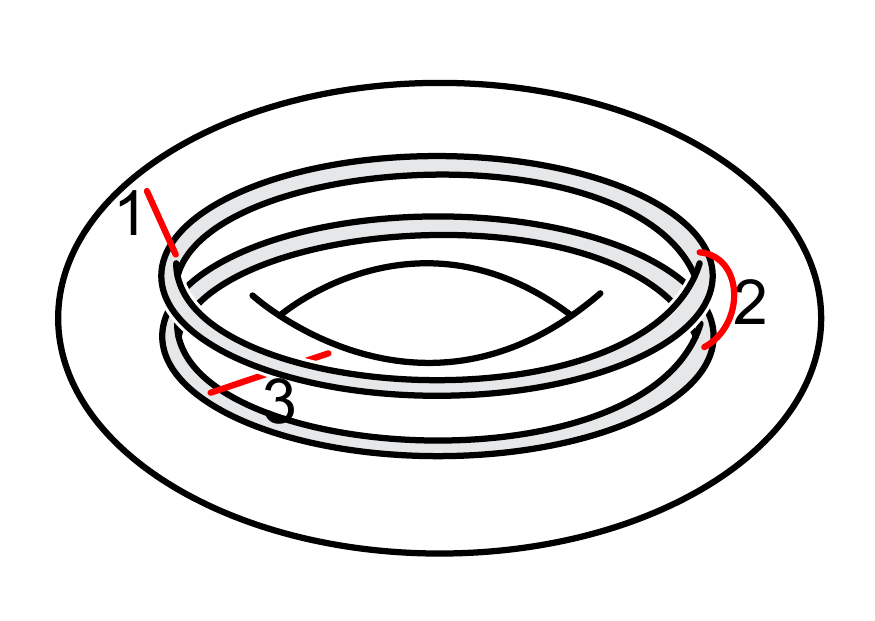}
\ee
We provide two ways of computing this geometry using VTQFT and arrive at the same answer. In the limit $\Delta\rightarrow0$, each single geometry, before summing over mapping class group, is given by
\be
Z^{\text{on-shell}}(\tau_1,\bar{\tau}_1,\tau_2,\bar{\tau}_2,\tau_3,\bar{\tau}_3)\propto
\frac{1}{\Delta^6}\prod_{i=1}^3\int dP_id\bar{P}_i\,\chi_{P_i}(\tau_i)\bar{\chi}_{\bar{P}_i}(\bar{\tau}_i)
\ee
We do not expect this to agree with a true off-shell 3-boundary torus-wormhole. But we can make a guess inspired by the difference between an on-shell torus-wormhole and an off-shell torus-wormhole (reviewed in section \ref{32}). Our guess is, before summing over mapping class group
\be
Z^{\text{off-shell}}(\tau_1,\bar{\tau}_1,\tau_2,\bar{\tau}_2,\tau_3,\bar{\tau}_3)\propto\prod_{i=1}^3f(\tau_i)\int dP_id\bar{P}_i\,P_i\bar{P}_i\chi_{P_i}(\tau_i)\bar{\chi}_{\bar{P}_i}(\bar{\tau}_i)
\ee
for some function $f$. In appendix \ref{23}, we give a detailed analysis of on-shell v.s. off-shell 2-dimensional 3-boundary wormhole in JT gravity. 

A 3-boundary torus wormhole was first computed in Post-Tsiares \cite{Post:2024itb} using VTQFT, but they arrived at a finite answer.

\subsection{First method}
\label{33}

Before we present our VTQFT computation of 3-boundary torus-wormhole. Let's first review Heegaard Splitting in VTQFT. For a closed orientable 3-manifold $M$, we can choose any 2d closed genus-g surface $\Sigma_g\subset M$ that splits $M$ into two compression bodies $M_1$ and $M_2$ glued along their outer boundary: the splitting surface $\Sigma_g$ and we denote
\be
M=M_1\cup_f M_2
\ee
where $f$ is an element of the mapping class group of $\Sigma_g$. The VTQFT partition function is then given by
\be
Z_{vir}(M)=\braket{Z_{vir}(M_1)|U(f)|Z_{vir}(M_2)}
\ee
where $U(f)$ is the representation of the modular transformation $f$ on the Hilbert space $\mathcal{H}_{\Sigma_g}$.

In our particular case, we have a 3-boundary torus wormhole with 6 insertions
\be
M=\includegraphics[width=0.25\textwidth, valign=c]{figures/3T}=\includegraphics[width=0.25\textwidth, valign=c]{figures/3T0}
\ee
We do Heegaard splitting along the blue torus and split $M$ into $M_1$ and $M_2$
\begin{multline}
   \includegraphics[width=0.25\textwidth, valign=c]{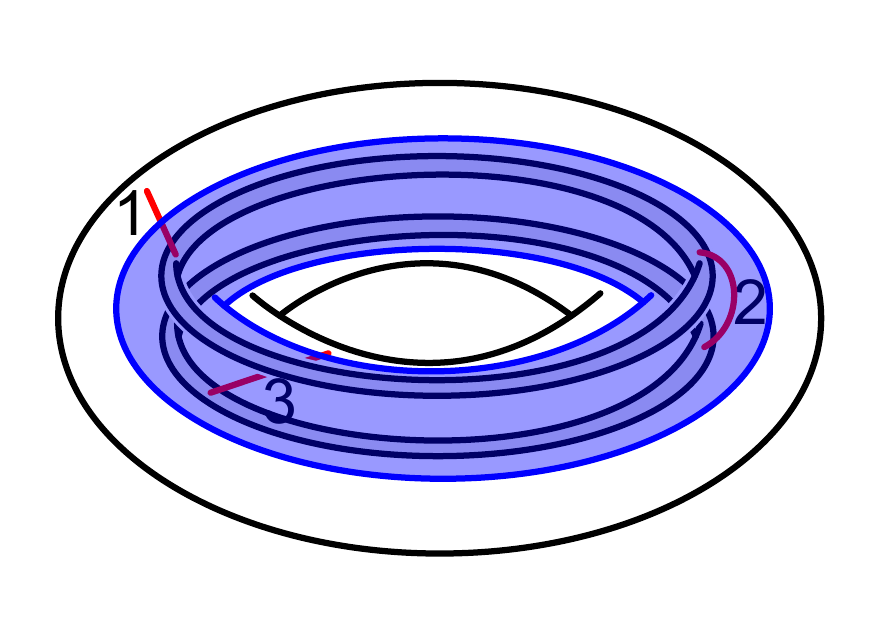}\\
   \rightarrow\quad  M_1=\includegraphics[width=0.25\textwidth, valign=c]{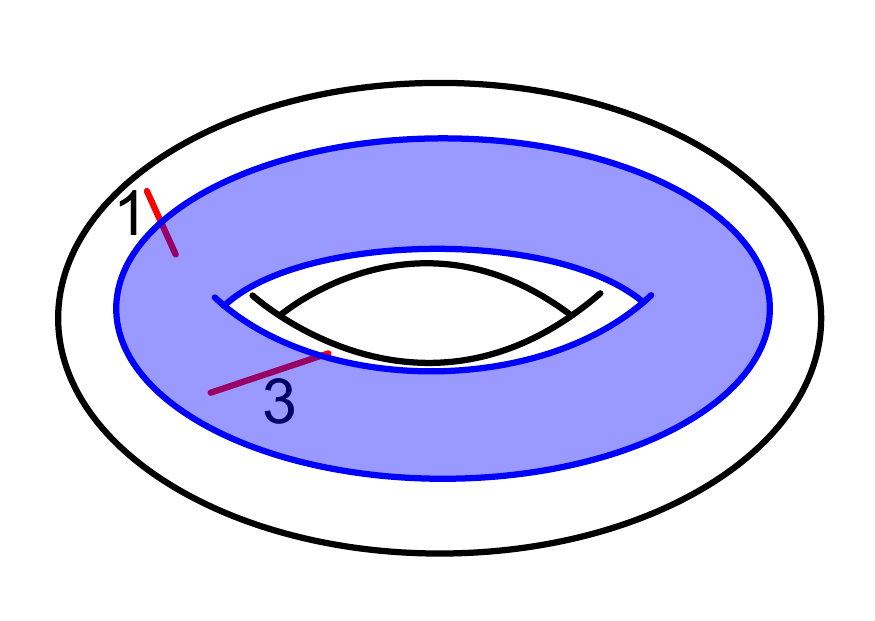} \quad \text{and}\quad M_2=\includegraphics[width=0.25\textwidth, valign=c]{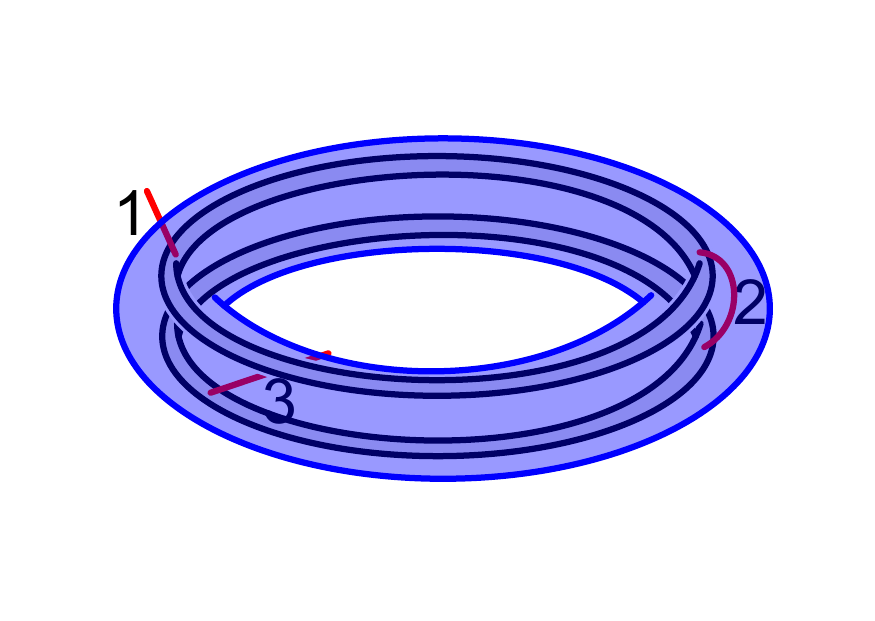} 
\end{multline}
then
\be
Z_{vir}(M)=\braket{Z_{vir}(M_1)|Z_{vir}(M_2)}
\ee
See appedix \ref{heegard1} for computation details. It turns out
\begin{multline}
Z_{vir}(M)=\int dP_p dP_q dP_r dP_s dP_t dP_u\,\rho_0(P_p)\rho_0(P_s)\rho_0(P_u)\\
\times \frac{C_0(P_1,P_p,P_q)C_0(P_1,P_t,P_u)C_0(P_2,P_p,P_q)C_0(P_2,P_r,P_s)C_0(P_3,P_r,P_s)C_0(P_3,P_t,P_u)} {C_0(P_p,P_r,P_t)C_0(P_q,P_r,P_u)C_0(P_t,P_q,P_s)}\\
\times\mathbb{F}_{2t}\begin{bmatrix}P_p&P_r\\P_q&P_s\end{bmatrix}\mathbb{F}_{1r}\begin{bmatrix}P_p&P_t\\P_q&P_u\end{bmatrix}\mathbb{F}_{3q}\begin{bmatrix}P_s&P_t\\P_r&P_u\end{bmatrix}\\
\times\underbrace{\includegraphics[width=0.2\textwidth, valign=c]{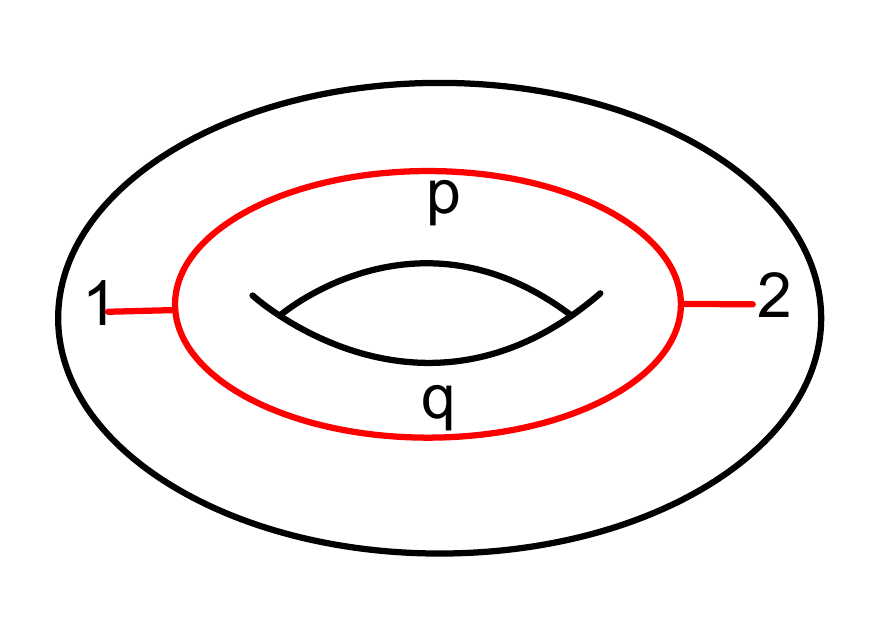}\quad \includegraphics[width=0.2\textwidth, valign=c]{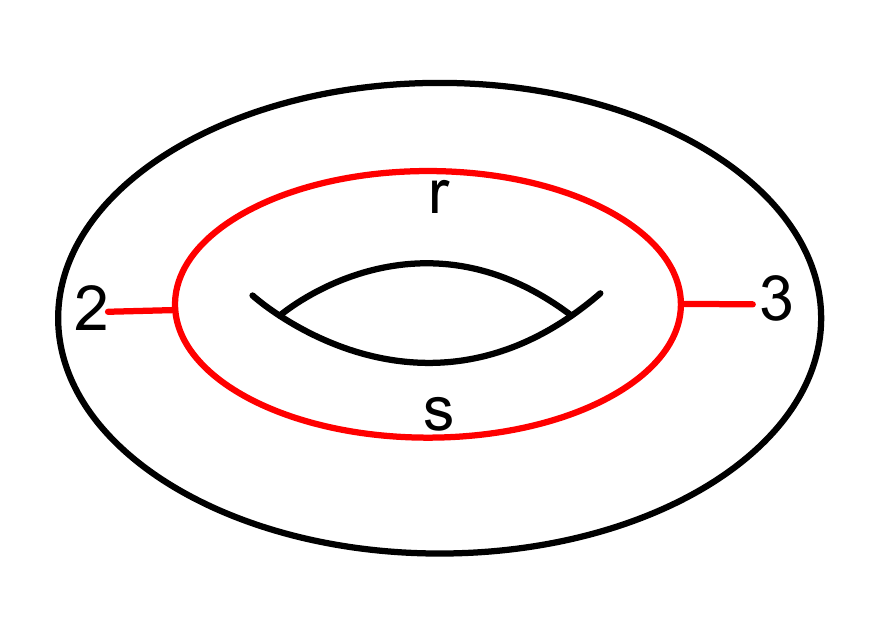}\quad \includegraphics[width=0.2\textwidth, valign=c]{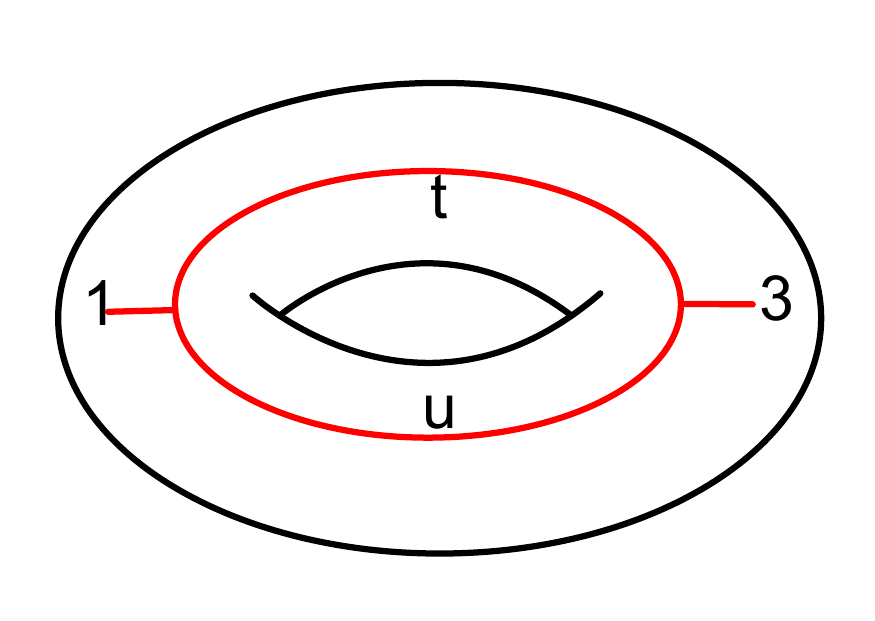}}_{\text{conformal blocks}}\label{method1}
\end{multline}
Now we want to take the limit that the mass $\Delta$ of insertions $1$, $2$, and $3$ goes to zero. Let us first use the identity
\be
\lim_{P_1\rightarrow\one}C_0(P_1,P_p,P_q)=\frac{\delta(P_p-P_q)}{\rho_0(P_p)}
\ee
and simplify $Z_{vir}(M)$ to
\begin{multline}
Z_{vir}(M)=\int dP_p dP_r  dP_t \,\frac{C_0(P_1,P_t,P_t)C_0(P_2,P_p,P_p)C_0(P_3,P_r,P_r)} {C_0(P_p,P_r,P_t)C_0(P_q,P_r,P_u)C_0(P_t,P_q,P_s)}\\
\times\mathbb{F}_{2t}\begin{bmatrix}P_p&P_r\\P_p&P_r\end{bmatrix}\mathbb{F}_{1r}\begin{bmatrix}P_p&P_t\\P_p&P_t\end{bmatrix}\mathbb{F}_{3q}\begin{bmatrix}P_r&P_t\\P_r&P_t\end{bmatrix}\\
\times\underbrace{\includegraphics[width=0.2\textwidth, valign=c]{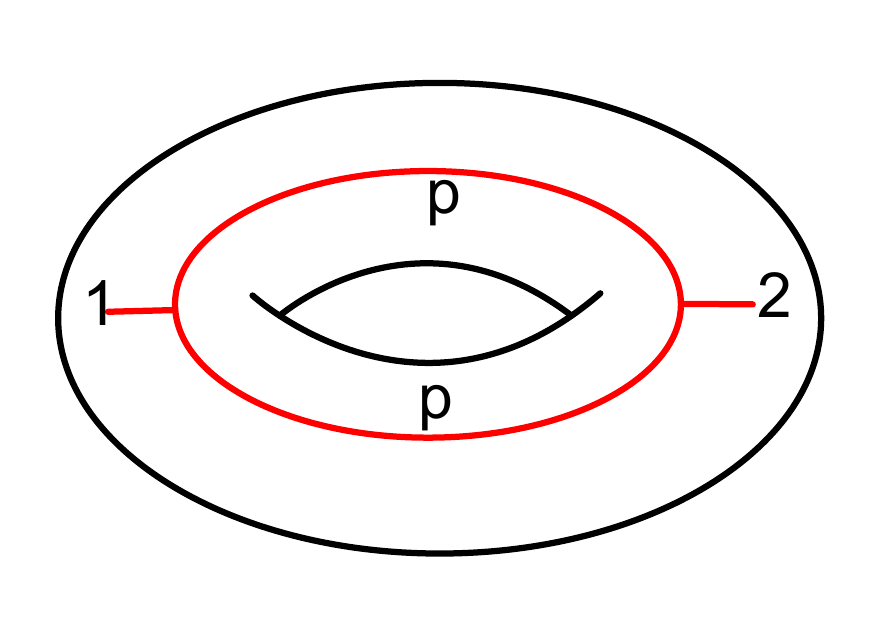}\quad \includegraphics[width=0.2\textwidth, valign=c]{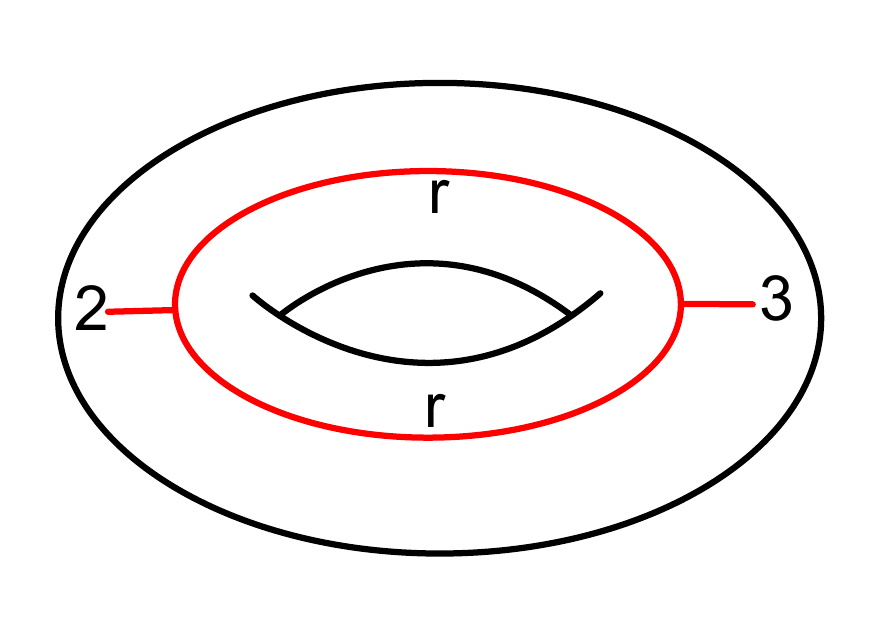}\quad \includegraphics[width=0.2\textwidth, valign=c]{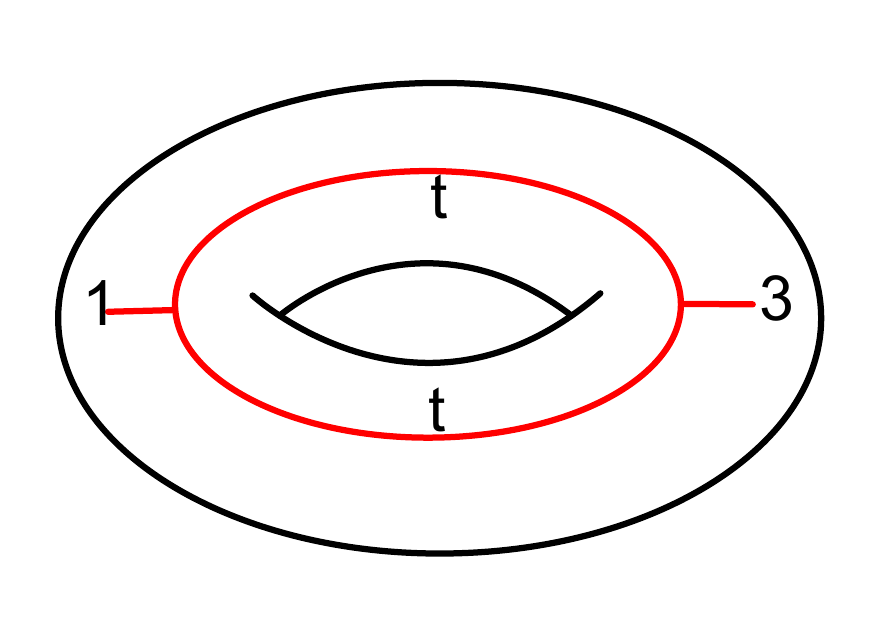}}_{\text{conformal blocks}}\label{method1a}
\end{multline}
Now let's use the following 3 identities: first
\be
\mathbb{F}_{\mathbb{1}t}\begin{bmatrix}P_p&P_r\\P_p&P_r\end{bmatrix}=\rho_0(P_t)C_0(P_p,P_r,P_t)
\ee
Then recall that in the limit $\Delta\rightarrow0$ (see appendix B.3 of \cite{Yan:2023rjh} for more detail) 
\be
C_0(P_1,P_t,P_t)\propto \frac{1}{\Delta\rho_0(P_t)}
\ee
And finally, in the limit $\Delta\rightarrow0$, the conformal blocks becomes characters
\be
\chi_{P_p}(\tau_3)\chi_{P_r}(\tau_1)\chi_{P_t}(\tau_2)
\ee
Therefore, in the limit $\Delta\rightarrow0$
\be
Z_{vir}(M)\propto\frac{1}{\Delta^3}\int dP_p dP_r dP_t \,\chi_{P_p}(\tau_3)\chi_{P_r}(\tau_1)\chi_{P_t}(\tau_2)\label{onshellchiral}
\ee
Notice that the above factorizes in terms of the three boundaries. And finally, the 3-boundary torus-wormhole partition function is given by a chiral part and an anti-chiral part. Therefore,
\begin{multline}
Z^{\text{on-shell}}(\tau_1,\bar{\tau}_1,\tau_2,\bar{\tau}_2,\tau_3,\bar{\tau}_3)\propto\\
\frac{1}{\Delta^6}\int dP_1d\bar{P}_1\,\chi_{P_1}(\tau_1)\bar{\chi}_{\bar{P}_1}(\bar{\tau}_1)\int dP_2d\bar{P}_2\,\chi_{P_2}(\tau_2)\bar{\chi}_{\bar{P}_2}(\bar{\tau}_2)\int dP_3d\bar{P}_3\,\chi_{P_3}(\tau_3)\bar{\chi}_{\bar{P}_3}(\bar{\tau}_3)
\end{multline}

\subsection{Second method}
\label{second}
In this section, we show an alternatively way of computing 3-boundary torus wormhole in the framework of VTQFT, and connect back to the links we mentioned in section \ref{seifert:VTQFT}. We show that they agree with what we get in \ref{33}.

We start with a 3-boundary torus-wormhole with 6 insertions
\be
M=\includegraphics[width=0.25\textwidth, valign=c]{figures/3T}=\includegraphics[width=0.25\textwidth, valign=c]{figures/3T0}
\ee
In this section, we use a different Heegaard splitting with a genus-2 surface
\be
\includegraphics[width=0.25\textwidth, valign=c]{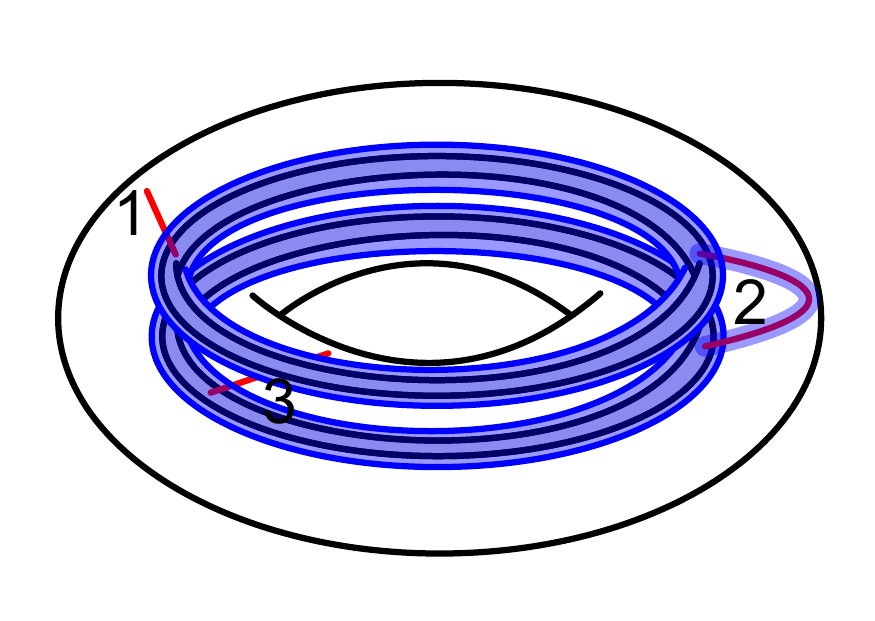}
\ee
and this splits $M$ into two parts
\be
M_1=\includegraphics[width=0.25\textwidth, valign=c]{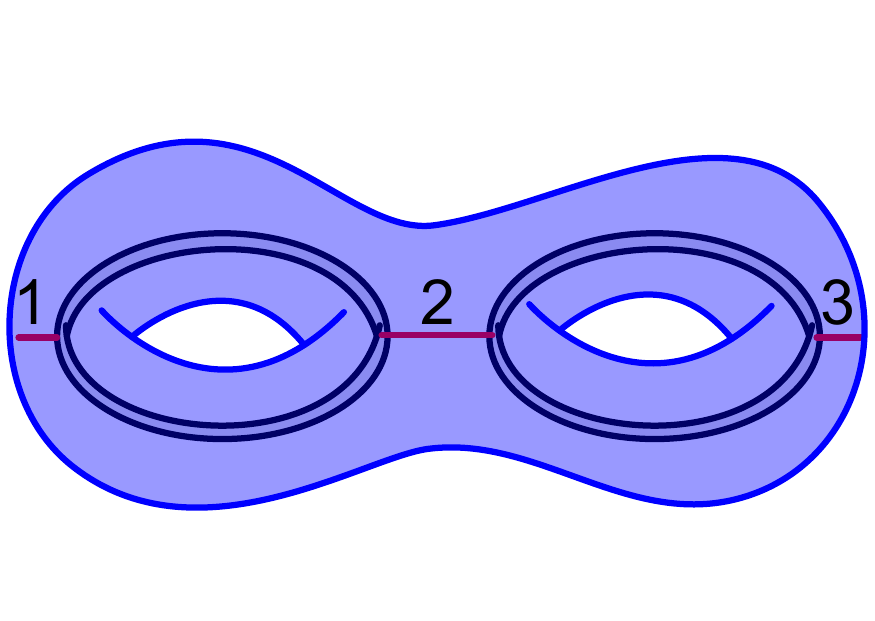}\quad\text{and}\quad M_2=\includegraphics[width=0.25\textwidth, valign=c]{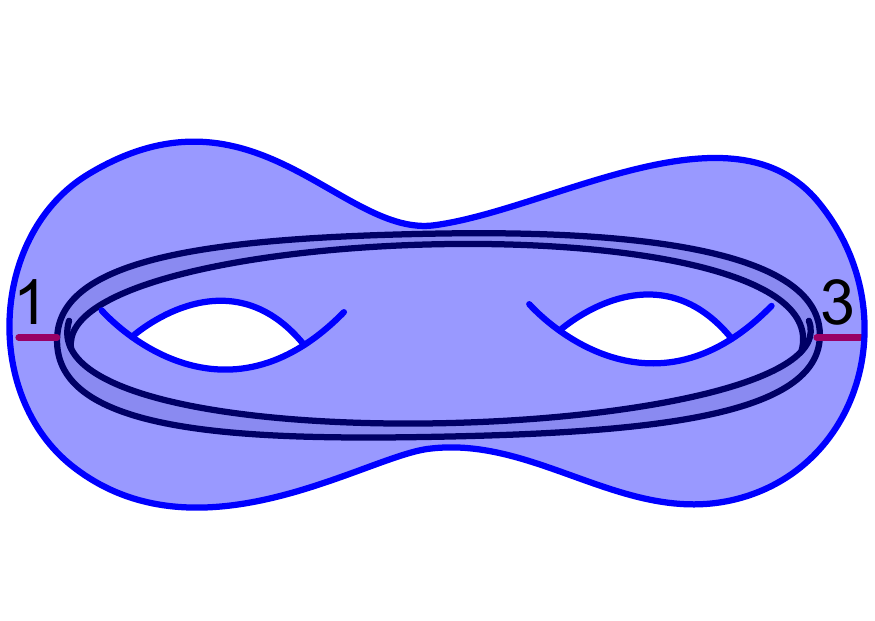}
\ee
See appendix \ref{heegard2} for computation details. It turns out that
\begin{multline}
Z_{vir}(M)=\int dP_u dP_t\,\rho_0(P_u)\rho_0(P_t)\, C_0(P_1,P_t,P_u)C_0(P_2,P_t,P_u)C_0(P_3,P_t,P_u)\\
\times\underbrace{\includegraphics[width=0.2\textwidth, valign=c]{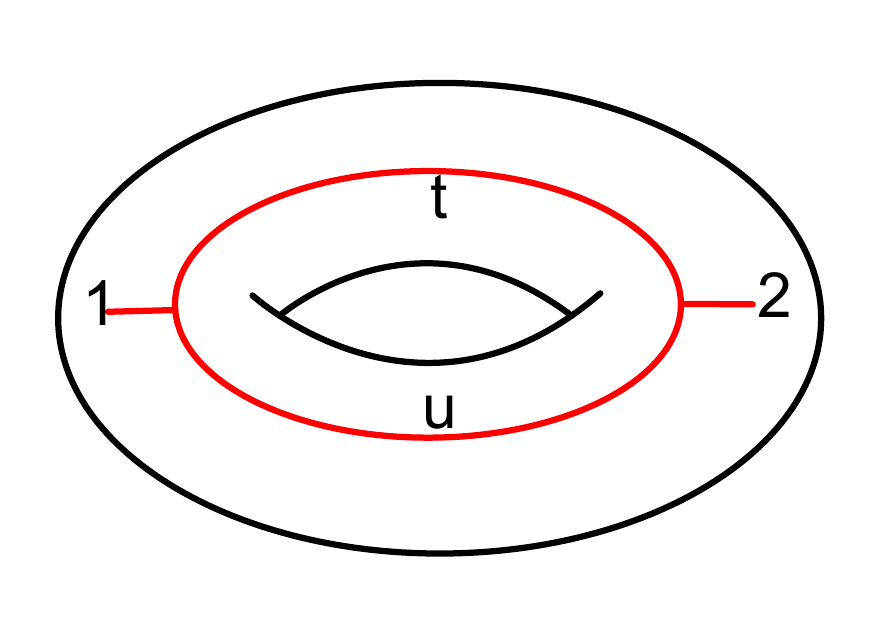}\quad \includegraphics[width=0.2\textwidth, valign=c]{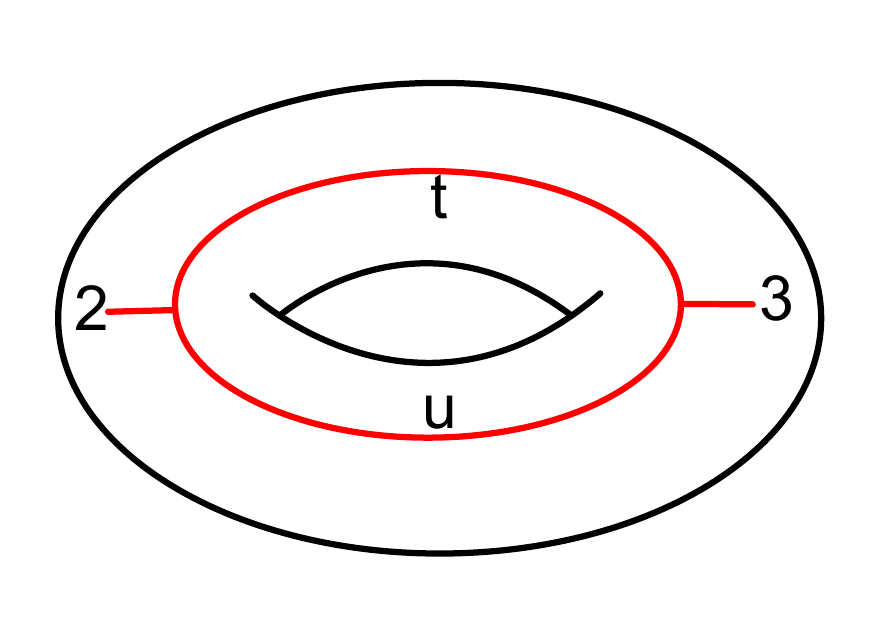}\includegraphics[width=0.2\textwidth, valign=c]{figures/diag9}}_{\text{conformal blocks}}\label{method2a}
\end{multline}
This result doesn't obviously agree with (\ref{method1a}) that we obtained using method 1. Nevertheless, method 2 is consistent with boundary-side CFT ensemble expectation, so one might be worried that the two methods do not agree. Also, we should arrive at the same answer from the same geometry even though we used different Heegard splittings. In what follows, we show that they agree in the limit $\Delta\rightarrow0$. 

Now, let's take the limit $\Delta\rightarrow0$ and first use the identity
\be
\lim_{P_1\rightarrow\one}C_0(P_1,P_t,P_u)=\frac{\delta(P_t-P_u)}{\rho_0(P_t)}
\ee
then $Z_{vir}(M)$ becomes
\begin{multline}
Z_{vir}(M)=\int dP_t\,\rho_0(P_t)\, C_0(P_2,P_t,P_t)C_0(P_3,P_t,P_t)\\
\times\underbrace{\includegraphics[width=0.2\textwidth, valign=c]{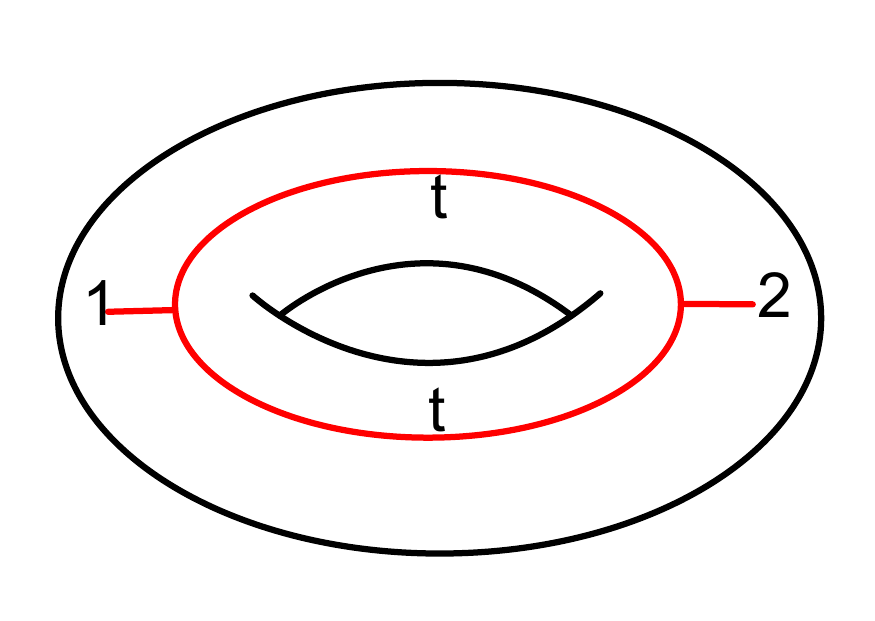}\quad \includegraphics[width=0.2\textwidth, valign=c]{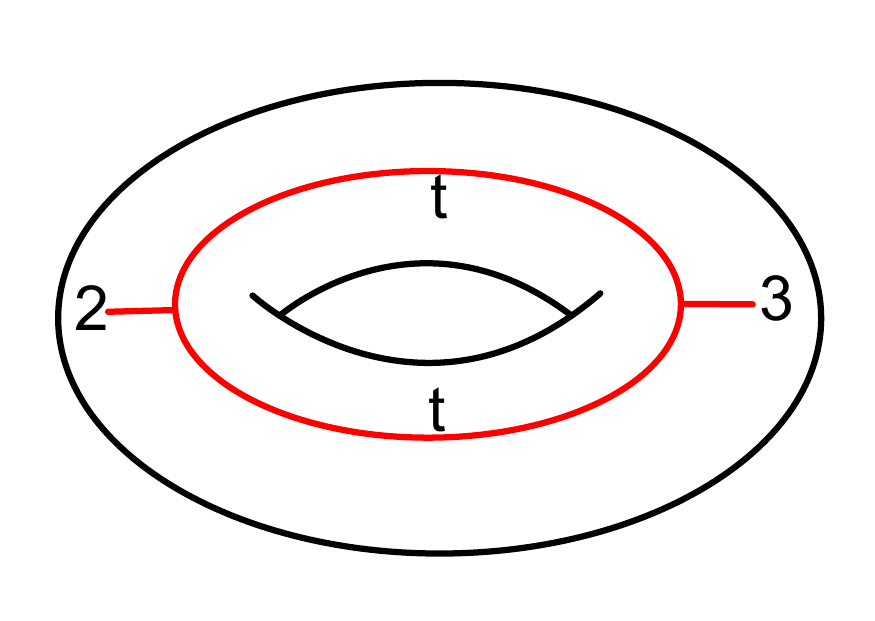}\includegraphics[width=0.2\textwidth, valign=c]{figures/diag9}}_{\text{conformal blocks}}
\end{multline}
Then recall that in the limit $\Delta\rightarrow0$  
\be
C_0(P_1,P_t,P_t)\propto \frac{1}{\Delta\rho_0(P_t)}
\ee
And finally, in the limit $\Delta\rightarrow0$, the conformal blocks becomes characters
\be
\chi_{P_t}(\tau_3)\chi_{P_t}(\tau_1)\chi_{P_t}(\tau_2)
\ee
Therefore, in the limit $\Delta\rightarrow0$
\be
Z_{vir}(M)\propto\frac{1}{\Delta^2}\int dP_t\,\frac{\chi_{P_t}(\tau_1)\chi_{P_t}(\tau_2)\chi_{P_t}(\tau_3)}{\rho_0(P_t)}\label{heegard22}
\ee
This naively looks very different from (\ref{onshellchiral}) in method 1. However, we can make the following manipulators to make them the same
\begin{align}
    Z_{vir}(M)&\propto\frac{1}{\Delta^2}\int dP_t\,\frac{\chi_{P_t}(\tau_1)\chi_{P_t}(\tau_2)\chi_{P_t}(\tau_3)}{\rho_0(P_t)}\\
    &\propto\frac{1}{\Delta^2}\int dP_1dP_2dP_3dP_t\,\frac{\bS_{P_1P_t}[\one]\bS_{P_2P_t}[\one]\bS_{P_3P_t}[\one]}{\rho_0(P_t)}\chi_{P_1}(-\frac{1}{\tau_1})\chi_{P_2}(-\frac{1}{\tau_2})\chi_{P_3}(-\frac{1}{\tau_3})\\
    &\propto\frac{1}{\Delta^3}\int dP_1dP_2dP_3\,\chi_{P_1}(-\frac{1}{\tau_1})\chi_{P_2}(-\frac{1}{\tau_2})\chi_{P_3}(-\frac{1}{\tau_3})
\end{align}
using the identity 
\be
\int dP\,\frac{\bS_{P_1P}[\one]\bS_{P_2P}[\one]\bS_{P_3P}[\one]}{\rho_0(P)}\propto\frac{1}{\Delta}
\ee
Now let's prove the above identity assuming the Virasoro-Verlinde formula propsed by Post-Tsiares in \cite{Post:2024itb}
\begin{multline}
\frac{1}{\rho_0(P_3)C_0(P_0,P_3,P_3)}\int dP\,\frac{\bS_{P_1P}[P_0]\bS_{P_2P}[\one]\bS^*_{PP_3}[P_0]}{\rho_0(P)}\\
=\frac{1}{\sqrt{C_0(P_0,P_1,P_1)C_0(P_0,P_3,P_3)}}\begin{Bmatrix}P_3&P_3&P_0\\P_1&P_1&P_2\end{Bmatrix}
=\frac{\bF_{P_0P_2}\begin{bmatrix}P_3&P_1\\P_3&P_1\end{bmatrix}}{\rho_0(P_2)C_0(P_1,P_3,P_2)}=\frac{\bF_{P_0P_2}\begin{bmatrix}P_3&P_1\\P_3&P_1\end{bmatrix}}{\bF_{\one P_2}\begin{bmatrix}P_3&P_1\\P_3&P_1\end{bmatrix}}
\end{multline}
where the Virasoro 6j-symbol is related to the F block (see (\ref{6jformula}) in appendix \ref{heegard1} for more details). In the limit $P_0\rightarrow\one$, the above equation is $1$. Thus, we have
\begin{align}
\int dP\,\frac{\bS_{P_1P}[\one]\bS_{P_2P}[\one]\bS_{P_3P}[\one]}{\rho_0(P)}&=\lim_{P_0\rightarrow\one}\int dP\,\frac{\bS_{P_1P}[P_0]\bS_{P_2P}[\one]\bS^*_{PP_3}[P_0]}{\rho_0(P)}\\
&=\rho_0(P_3)C_0(P_0,P_3,P_3)\\
&\propto \frac{1}{\Delta}
\end{align}
Also see appendix \ref{link3more} to see links in VTQFT related to this identity.

Finally as an aside, recall that in section \ref{seifert:VTQFT}, we introduced a way to compute a Seifert manifold which looks like a solid torus with two tori inside carved-out and then filled in (\ref{3bdySeifert}). Now if we instead of filling in the two inside tori, glue them to asymptotic boundaries, we get
\begin{align}
\includegraphics[width=0.25\textwidth, valign=c]{figures/3To2}
&\propto\frac{1}{\Delta^2}\int dP'dP\,dP_1dP_2\,\chi_{P'}(-\frac{1}{\tau_3})\bS_{P'P}[\one]\frac{\bS_{P_1P}[\one]\bS_{P_2P}[\one]}{\bS_{P\one}[\one]}\chi_{P_1}(-\frac{1}{\tau_1})\chi_{P_2}(-\frac{1}{\tau_2})\\
&\propto\frac{1}{\Delta^2}\int dP\,\frac{\chi_P(\tau_1)\chi_P(\tau_2)\chi_P(\tau_3)}{\rho_0(P)}\label{3bdyglue}
\end{align}
This is exactly the same as what we got using Heegard splitting (\ref{heegard22}). Note the $\frac{1}{\Delta^2}$ comes from the fact that each additional asymptotic boundary is associated with a $\frac{1}{\Delta}$ divergence in the chiral part (see appendix \ref{onshell32} for more details).

\section{Discussion}
\label{discussion}

In this paper, we showed that our VTQFT computations of Seifert manifolds and 3-boundary torus-wormholes do not fully align with the expectations from 3D gravity. We traced these discrepancies to the omission of contributions from the mapping class group. To correct this issue, one approach is to explicitly incorporate the mapping class group by hand into the Virasoro TQFT framework when considering off-shell objects. Additionally, alternative methods of extrapolating from on-shell to off-shell configurations, beyond taking the limit of vanishing matter insertions, could be explored.

\section*{Acknowledgements} 
We want to thank Scott Collier for initial collaboration and discussions. We want to give special thanks to Douglas Stanford. We also want to thank Eleny Ionel, Jordan Cotler, Lorenz Eberhardt, Kristan Jensen, Juan Maldacena, Boris Post, and Steven Kerckhoff for discussions. This work was made possible by Institut Pascal at Université Paris-Saclay with the support of the program “Investissements d’avenir” ANR-11-IDEX-0003-01. CY was supported in part by the Heising-Simons Foundation, the Simons Foundation, and grant no. PHY-2309135 to the Kavli Institute for Theoretical Physics (KITP).

\appendix

\section{Review: Pure 3d Gravity Negativity Problem}
\label{negativity}
In this section, we first give a review of the negativity problem of pure 3d gravity found by \cite{Benjamin:2019stq} in \ref{seifert:review}, and then review the dimensional reduction of BTZ blackhole and SL(2,$\Z$) blackhole to JT gravity in \ref{seifert:2d}, and finally we review the resolution of the negativity problem given by Maxfield-Turiaci \cite{Maxfield:2020ale} in \ref{MTs}.  Along the way, we also review some useful JT gravity tools. 

\subsection{Pure 3d gravity}
\label{seifert:review}
3d Einstein-Hilbert action is given by
\be
I_{EH}=-\frac{1}{16\pi G_N}\left(\int d^3x\,\sqrt{g_3}(R_3+\frac{2}{\ell_3^2})+2\int_{\partial}d^2x\,\sqrt{\gamma_2}(\kappa_3-\frac{1}{\ell_3})\right)\label{3daction}
\ee
where $g_3$ is the metric, $R_3$ is the curvature, $\ell_3$ is AdS length, and $\kappa_3$ is the  extrinsic curvature; all of them in 3d. And $\gamma_2$ is the induced metric on the boundary. 

Let us start with AdS$_3$ and impose boundary condition to be a flat torus parametrized by spatial angle $\phi$ and Euclidean time $t_E$ such that $(t_E,\phi)\sim(t_E,\phi+2\pi)\sim(t_E+\beta,\phi+\theta)$
\be
\includegraphics[valign=c,width=0.25\textwidth]{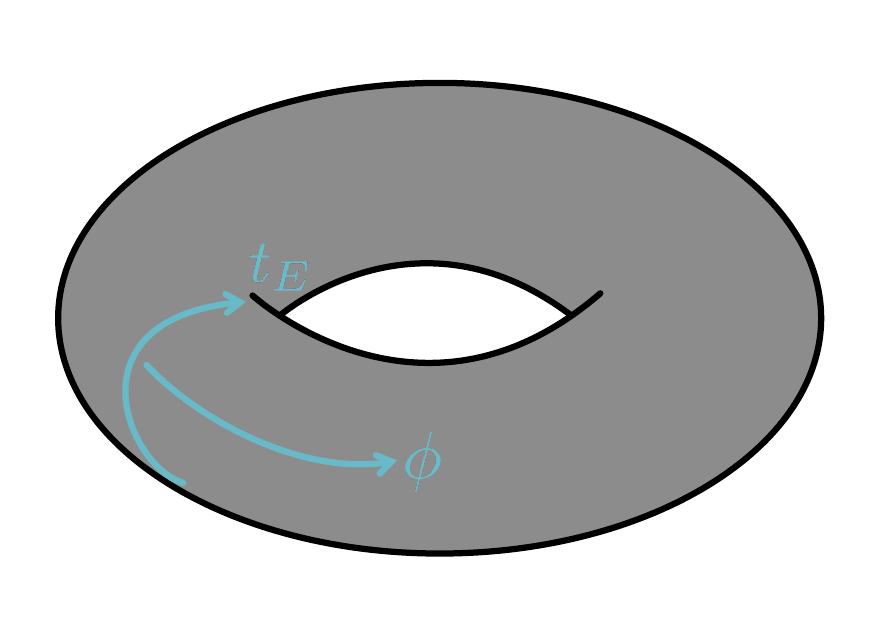}
\ee
so the metric of AdS$_3$ is
\be
\ell_3^{-2}ds^2=(1+r)^2dt_E^2+\frac{dr^2}{1+r^2}+r^2d\phi^2
\ee
and the induced metric on the boundary is
\be
\gamma_2=\epsilon^{-2}dt_Ed\phi
\ee
where $\epsilon$ is a holographic renormalization parameter which is taken to zero. If we define the modular parameters
\be
\tau=\frac{\theta+i\beta}{2\pi}\quad\quad\bar{\tau}=\frac{\theta-i\beta}{2\pi}
\ee
the partition function can be written as
\be
Z(\beta,\theta)=\Tr\left[e^{-\beta H-i\theta J}\right]=\Tr\left[e^{2\pi i\tau(L_0-\frac{c}{24})-2\pi i\bar{\tau}(\bar{L}_0-\frac{c}{24})}\right]
\ee
where $H$ is the Hamiltonian and $J$ is the angular momentum
\be
H=L_0+\bar{L}_0-\frac{c}{12}\quad\quad J=\bar{L}_0-L_0
\ee
Here $L_0$ is the left-moving conformal generator and $\bar{L}_0$ is the right-moving conformal generator.

\subsubsection{BTZ Blackhole} 
Filling in the Euclidean time cirlce, we get BTZ blackhole
\be
\includegraphics[valign=c,width=0.25\textwidth]{figures/btzcircle}
\ee
where the red circle is the spatial circle. Its partition function is given by
\be
Z_{BTZ}=\chi_\one(-\frac{1}{\tau})\chi_\one(-\frac{1}{\bar{\tau}})
\ee
where
\be
\chi_\one(\tau)=\frac{(1-q)q^{-\frac{c-1}{24}}}{\eta(\tau)}\quad\quad q=e^{2\pi i\tau}
\ee
In terms of the variables 
\be
P=\sqrt{\frac{E-J}{2}}\quad\quad \bar{P}=\sqrt{\frac{E+J}{2}}
\ee
the corresponding density of states is 
\be
\rho_{BTZ}(P,\bar{P})=\mathbb{S}_{P\mathbb{1}}[\mathbb{1}]\mathbb{S}_{\bar{P}\mathbb{1}}[\mathbb{1}]
\ee
A modular S-transform is implemented via a modular S kernel $\mathbb{S}_{P'P}[\mathbb{1}]$ \cite{Collier:2019weq}
\be
\chi_P(-1/\tau)=\int_0^\infty dP'\mathbb{S}_{P'P}[\mathbb{1}]\chi_{P'}(\tau)\quad\quad\mathbb{S}_{P'P}[\mathbb{1}]=2\sqrt{2}\cos(4\pi PP')
\ee
The identity S kernel is obtained from subtracting the $h=1$ kernel from the $h=0$ kernel
\be
\mathbb{S}_{P\mathbb{1}}[\mathbb{1}]=\mathbb{S}_{P,\frac{i}{2}(b^{-1}+b)}[\mathbb{1}]-\mathbb{S}_{P,\frac{i}{2}(b^{-1}-b)}[\mathbb{1}]=4\sqrt{2}\sinh(2\pi bP)\sinh(2\pi b^{-1}P)
\ee
In the near-extremal limit $E\approx J$, we have
\be
\rho_{BTZ}(P,\bar{P})\approx 32\pi^2e^{2\pi Q\bar{P}}P^2=32\pi^2e^{S_0(J)}P^2
\ee
where
\be
S_0(J)=2\pi\sqrt{\frac{\ell_3 J}{4G_N}}
\ee
is the extremal entropy of BTZ blackhole. 

\subsubsection{SL(2,$\Z$) Blackhole} 
\label{slbh}
We construct an SL(2,$\Z$) blackhole by acting on $\tau$ by an element of SL(2,$\Z$)
\be
\gamma=\begin{pmatrix}a&b\\r&d\end{pmatrix}\in SL(2,\Z)
\ee
and get
\be
\tilde{\tau}=\gamma\cdot\tau=\frac{a\tau+b}{r\tau+d}
\ee
We then fill in the new Euclidean time circle of this transformed torus boundary to get a SL(2,$\Z$) blackhole
\be
\includegraphics[valign=c,width=0.25\textwidth]{figures/sl2zcircle}
\ee
where the red circle is the spatial circle. 

To be more specific, if we parametrize the original torus boundary by $z$
\be
z=\frac{\phi+it_E}{2\pi}\quad\quad z\sim z+1\sim z+\tau
\ee
then after the trnasformation $\gamma$, the new coordinate is
\be
\tilde{z}=\frac{\tilde\phi+i\tilde{t}_E}{2\pi}=\frac{z}{r\tau+d}\quad\quad \tilde{z}\sim \tilde{z}+1\sim \tilde{z}+\tilde{\tau}
\ee
But it turns out that the only thing that matters for an SL(2,$Z$) blackhole is the ratio between $r$ and $d$, and we can label them by coprime pair $(r,d)$

The partition function of an SL(2,$Z$) blackhole is given by
\be
Z_{(r,d)}=\chi_\one(\gamma\cdot\tau)\chi_\one(\overline{\gamma\cdot\tau})
\ee
In particular
\be
Z_{BTZ}=Z_{(1,0)}
\ee
The density of states of an SL(2,$\Z$) blackhole is given by
\be
\rho_{\gamma}(P,\bar{P})=\mathbb{K}^{(r,d)}_{P\mathbb{1}}[\mathbb{1}]\mathbb{K}^{(r,d)}_{\bar{P}\mathbb{1}}[\mathbb{1}]
\ee
A general modular transform labeled by $\gamma\in\SL(2,\Z)$ is implemented via a modular crossing kernel $\mathbb{K}^{(r,d)}_{P' P}$ \cite{Benjamin:2020mfz}
\be
\chi_P(\gamma\cdot \tau)=\int_0^\infty dP'\,\K_{P'P}^{(r,d)}[\one]\chi_{P'}(\tau)\quad\quad
\mathbb{K}^{(r,d)}_{P' P}[\one] = \epsilon \sqrt{\frac{8}{r}}e^{\frac{2\pi i }{r}( a P'^2 + d P^2)}\cos\left(\frac{4\pi}{r}P P'\right)
\ee
Again identity modular transform kernel is obtained from subtracting the $h=1$ kernel from the $h=0$ kernel
\be
\mathbb{K}_{P\mathbb{1}}^{(r,d)}=\K_{P,\frac{i}{2}(b^{-1}+b)}^{(r,d)}-\K_{P,\frac{i}{2}(b^{-1}-b)}^{(r,d)}=\epsilon(\gamma)\sqrt{\frac{2}{r}}e^{-2\pi i\frac{a}{r}\frac{c-1}{24}}e^{2\pi i\frac{d}{r}P^2}\sinh(\frac{2\pi}{r}bP)\sinh(\frac{2\pi}{r}b^{-1}P)\label{kkernel}
\ee
In the near extremal limit
\be
\rho_{(r,d)}(P,\bar{P})\approx \frac{4}{r}e^{\frac{2\pi Q\bar{P}}{r}}e^{-\frac{2\pi i d}{r}\bar{P}^2}\left(1-e^{\frac{2\pi i(d^{-1})_r}{r}}\right)=\frac{4}{r}e^{\frac{S_0(J)}{r}}e^{-\frac{2\pi i d}{r}\bar{P}^2}\left(1-e^{\frac{2\pi i(d^{-1})_r}{r}}\right)
\ee
From the entropy term, we see that the leading contribution is when $(r,d)=(2,1)$
\be
\rho_{(2,1)}(P,\bar{P})\approx 4(-1)^{\bar{P}^2}e^{\pi Q\bar{P}}=4(-1)^J e^{\frac{S_0(J)}{2}}
\ee
Now let's summarize. In the near-extremal limit
\be
\rho_{BTZ}(P,\bar{P})\approx 32\pi^2e^{S_0(J)}P^2\quad\quad \rho_{(2,1)}(P,\bar{P})\approx 4(-1)^J e^{\frac{S_0(J)}{2}}
\ee
the second term can be larger than the first when $P$ is very small, for odd spin $\rho$ would become negative. This is the negativity problem.

\subsection{Dimensional reduction}
\label{seifert:2d}
In this section, we review the dimensional reduction of BTZ and SL(2,$Z$) blackholes following \cite{Ghosh:2019rcj}. 

The 3d action (\ref{3daction}) could be Kaluza-Klein reduced to a Einstein-dilaton theory and in the near-extremal limit it becomes Jackiw-Teitelboim (JT) gravity \cite{Teitelboim, Jackiw, AlmheiriPolchinski} with action
\be
I_{JT}=-S_0(J)\chi-\frac{1}{2}\int_M d^2x\,\sqrt{g_2}\Phi(R_2+\frac{2}{\ell_2^2})-\int_{\partial M} du\,\Phi(\kappa_2-\frac{1}{\ell_2})
\ee
where the first term is the 2d Einstein-Hilbert action which is purely topological and $\chi=2-2g-n$ is the Euler character for manifold $M$ with $g$ the genus and $n$ the number of boundaries, and $S_0(J)$ is the zero-temperatuer bulk entropy, and in this case, the extremal entropy of BTZ blackhole. Classically, the equation of motion fixes the bulk geometry to be AdS$_2$ with $R_2=-2/\ell_2^2$ and the action reduces to a Schwarzian action on the boundary \cite{MaldacenaStanfordYang}.

Two simplest geometries of Euclidean AdS are a hyperbolic disk which has one asymptotically boundary with renormalized length $\beta$, and a hyperbolic trumpet which has one asymptotic boundary with renormalized length $\beta$ and one geodesic boundary with length $b$. 
\be
\includegraphics[valign=c,width=0.5\textwidth]{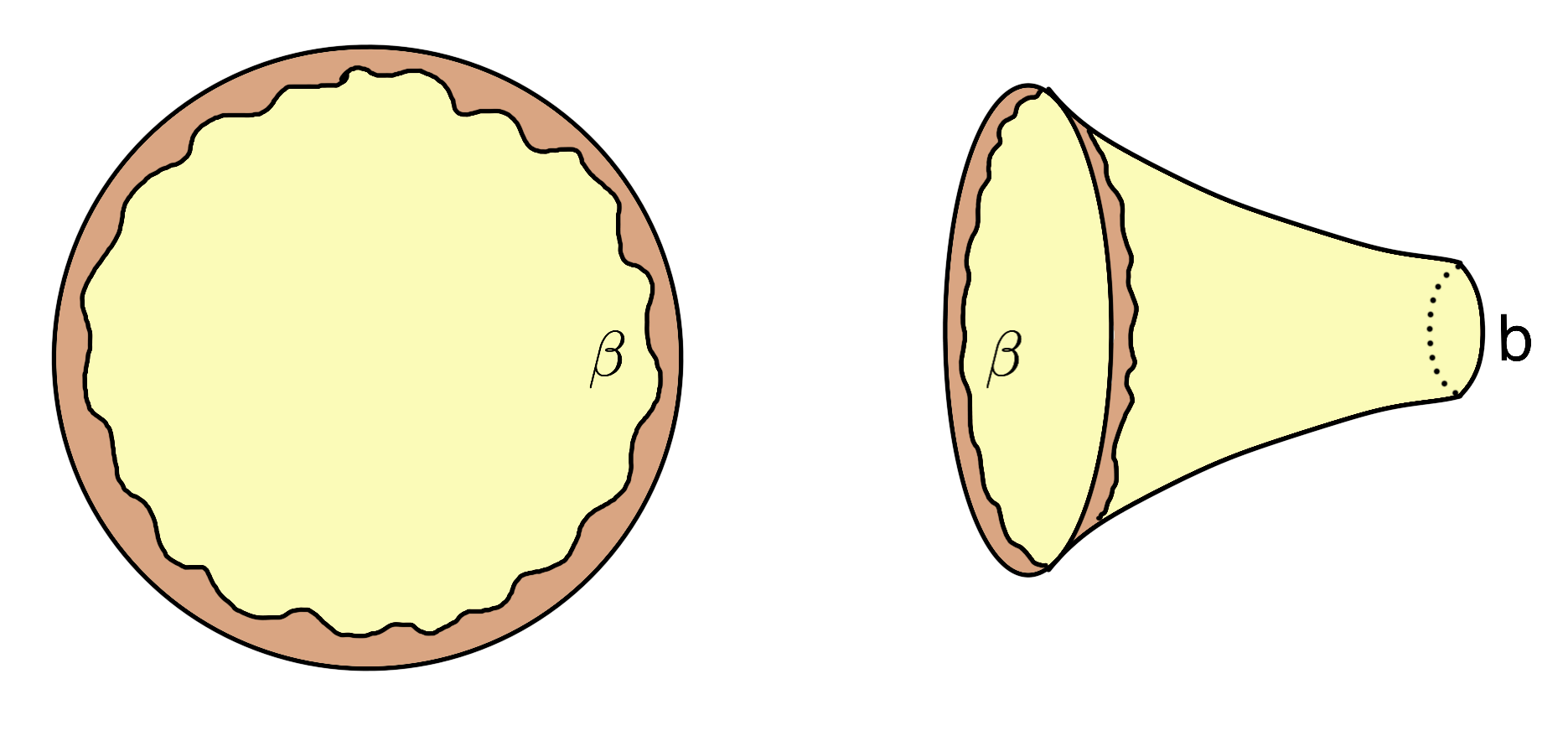}
\ee
JT path integrals without operator insertions can be computed directly by doing the path integral over the wiggly boundary of the disk and the trumpet explicitly. Disk \cite{BagretsAltlandKamenev16, 9authors, BagretsAltlandKamenev17, StanfordWitten17, SchwarzianBelokurovShavgulidze, SchwarzianMertensTuriaciVerlinde, KitaevSuhBH, Yangsingleauthor, IliesiuPufuVerlindeWang} and trumpet partition functions \cite{StanfordWitten17, Saadsingleauthor} are given respectively by
\begin{align}
Z_{\text{Disk}}(\beta)&=e^{S_0}\frac{e^{\frac{2\pi^2}{\beta}}}{\sqrt{2\pi\beta^3}}=e^{S_0}\int_0^\infty ds\,\underbrace{\frac{s \sinh(2\pi s)}{2\pi^2}}_{\rho_0(s)}e^{-\beta s^2/2}\label{diskrho}\\
Z_{\text{Trumpet}}(\beta,b)&=\frac{e^{-\frac{b^2}{2\beta}}}{\sqrt{2\pi \beta}}=\int_0^\infty ds\,\underbrace{\frac{\cos(b s)}{\pi}}_{\rho_1(s,b)}e^{-\beta s^2/2}\label{trumpetrho}
\end{align}
where $E=s^2/2$. We can get disk with one defect with conical angle $2\pi(1-\alpha)$ from a trumpet by analytically continue $b\mapsto 2\pi i\alpha$
\be
\includegraphics[valign=c,width=0.2\textwidth]{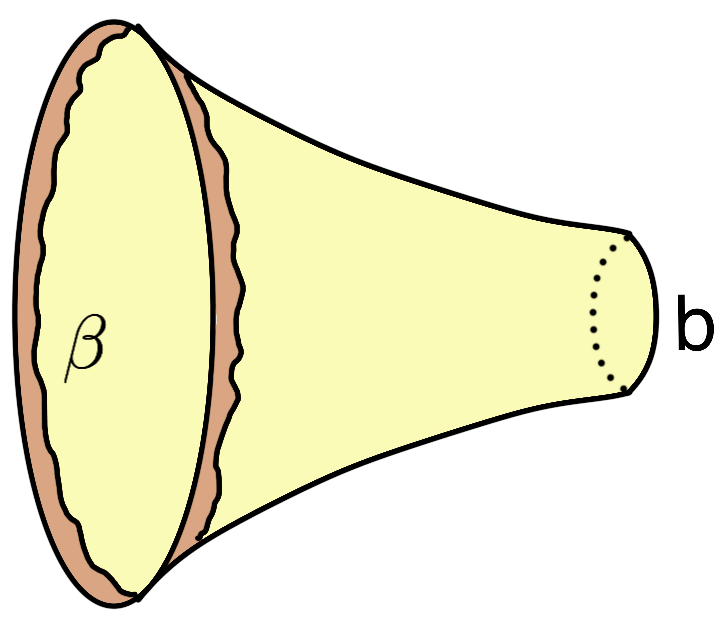}\quad\quad\mapsto\quad\quad \includegraphics[valign=c,width=0.2\textwidth]{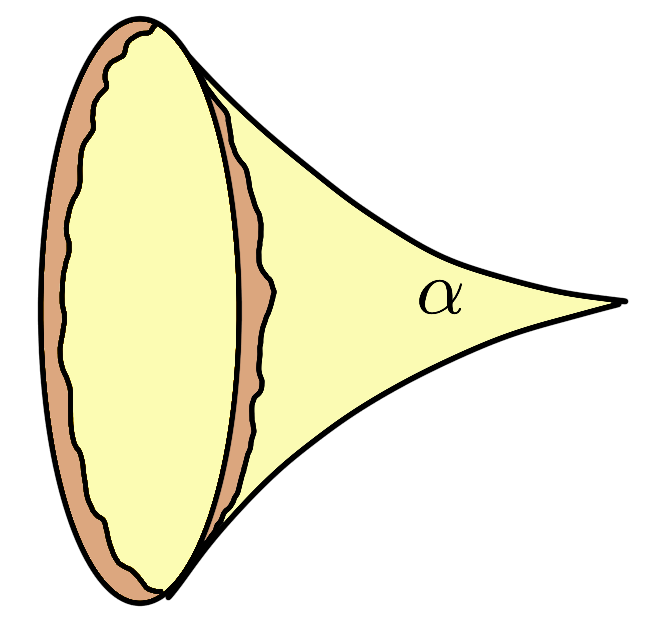}
\ee
After dimensional reduction, BTZ blackhole gets maped to a hyperbolic disk, and $SL(2,\Z)$ blackhole labeled by $(r,d)$ maps to disk with one defect $\alpha=\frac{1}{r}$
\be
\includegraphics[valign=c,width=0.25\textwidth]{figures/btzcircle}\mapsto\includegraphics[valign=c,width=0.2\textwidth]{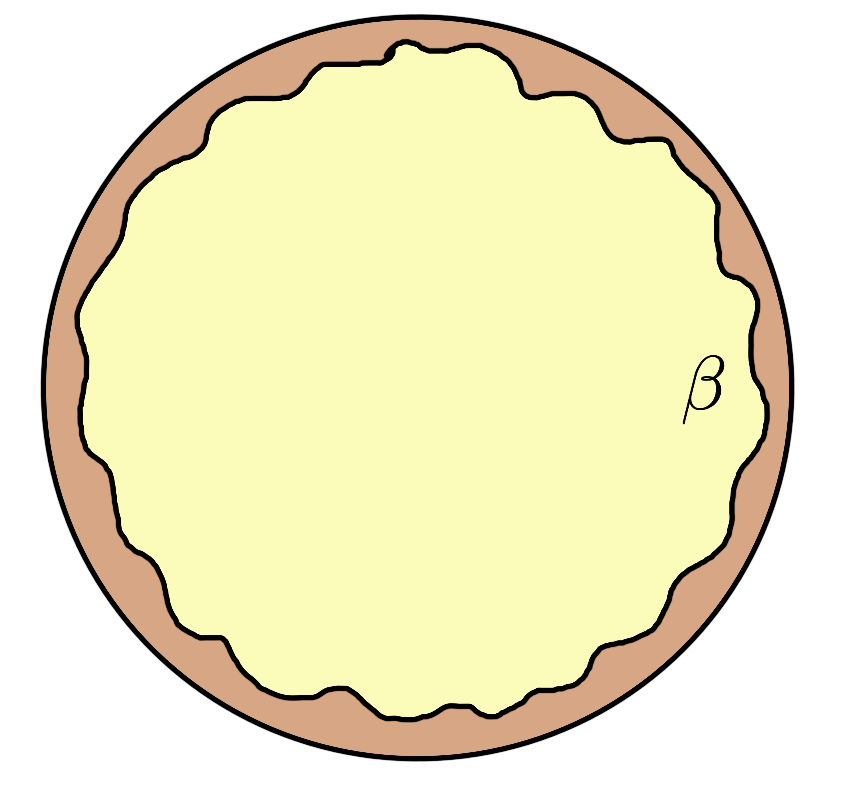}\quad\quad \includegraphics[valign=c,width=0.25\textwidth]{figures/sl2zcircle}\mapsto\includegraphics[valign=c,width=0.2\textwidth]{figures/defect.png}
\ee
In the near extremal limit, the density of states of disk and disk with a defect are given by
\begin{align}
\rho^{2d}_0(s)&=\frac{s \sinh(2\pi s)}{2\pi^2}\approx\frac{s^2}{\pi}\\
\rho^{2d}_1(s,\alpha)&=\frac{\cosh(2\pi\alpha s)}{\pi}\approx\frac{1}{\pi}
\end{align}
And recall that in the near-extremal limit, BTZ blackhole and the leading contribution of SL(2,$Z$) blackhole have density of states
\begin{align}
\rho^{3d}_{BTZ}(P)&\approx32\pi^2e^{S_0(J)}P^2\approx 32\pi^3e^{S_0(J)}\rho^{2d}_0(P)\\
\rho^{3d}_{(2,1)}(P)&\approx4(-1)^Je^{\frac{S_0(J)}{2}}\approx 32\pi^3e^{S_0(J)}\lambda \rho^{2d}_1(P)
\end{align}
where $\lambda=\frac{(-1)^J}{8\pi^2}e^{-\frac{S_0(J)}{2}}$. They are proportional to the disk and disk with one defect respectively, with a defect also associated with a factor of $\lambda$. 

\subsection{Adding Seifert manifolds} 
\label{MTs}

Maxfield-Turiaci proposed that adding Seifert manifolds in addition to BTZ blackhole and SL(2,$\Z$) blackhole would cure the negativity problem. They calculate Seifert manifolds by dimensionally reducing to 2d and calculate disk with defects in JT
\be
\includegraphics[valign=c,width=0.25\textwidth]{figures/seifertn}\quad\mapsto\quad\includegraphics[valign=c,width=0.2\textwidth]{figures/ndefects.png}
\ee
Carve-out and fill-in $n$ tori labeled by $(r_i,d_i)$ inside a solid torus in 3d corresponds to $n$ defects on a disk in 2d with the defect angle $\alpha_i=\frac{1}{r_i}$.

In order to calculate disk with defects in JT we first need to introduce some useful tools.

\paragraph{Useful JT tools} JT gravity can be canonically quantized perturbatively on a disk topology and the resulting Hamiltonian describes a simple quantum mechanical system of a single degree of freedom $\ell$, governed by Hamiltonian 
\be 
H = - \frac{1}{2} \partial^2_\ell + 2 e^{-\ell} .
\ee
This Hamiltonian evolves the states of the left and right asymptotic boundary through $H = H_L$ and $H= H_R$, in pure JT gravity subject to the condition $H_L = H_R$.
From the bulk perspective, $\ell$ captures simply the geodesic length through the bulk between the two asymptotic spatial boundaries. 
The pure gravity perturbative Hilbert space is now simply given by normalizable wavefunctions $\varphi(\ell)$ of the geodesic length $\ket{\ell}$, with the standard inner product
\be 
\mH_{\text{pure gravity}} = L^2(\mathbb R)\,, 
\qquad 
\braket{\varphi|\varphi} = \int d \ell\, |\varphi(\ell)|^2 ,
\ee
with the geodesic length states normalized as $\braket{\ell|\ell'} = \delta(\ell - \ell')$.
The energy eigenstates $\ket{s}$ of the bulk Hamiltonian $H$ in the length basis take the form 
\be 
\ket{s} = \int_{-\infty}^\infty d \ell\, \braket{\ell|s}
,
\qquad
\braket{\ell|s}=2^{3/2}e^{-\ell/2}K_{2is}(4e^{-\ell/2})  , \qquad \text{ where }\quad  E= s^2 / 2\,,
\ee
and the states are normalized as 
\be
\int_{-\infty}^\infty d\ell\,\braket{s|\ell}\braket{\ell|s'}=\frac{\delta(s-s')}{\rho_0(s)}
\ee
The fixed-energy Hartle-Hawking wavefunction is given by
\be
\varphi_s(\ell)=e^{-\ell/2}\braket{\ell|s}
\ee
with normalization
\be
\int_{-\infty}^\infty e^\ell d\ell\,\varphi_s(\ell)\varphi_{s'}(\ell)=\frac{\delta(s-s')}{\rho_0(s)}
\ee
Now let's define the fixed-length Hartle-Hawking wavefunctions
\be
\includegraphics[width=0.2\textwidth,valign=c]{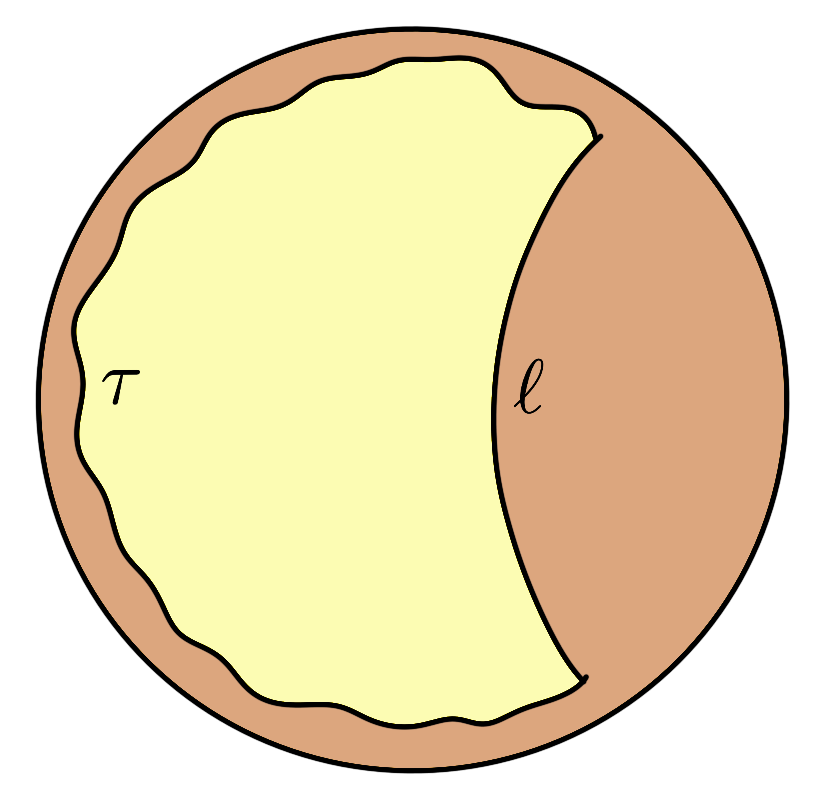}\quad\quad,\quad\quad \includegraphics[width=0.2\textwidth,valign=c]{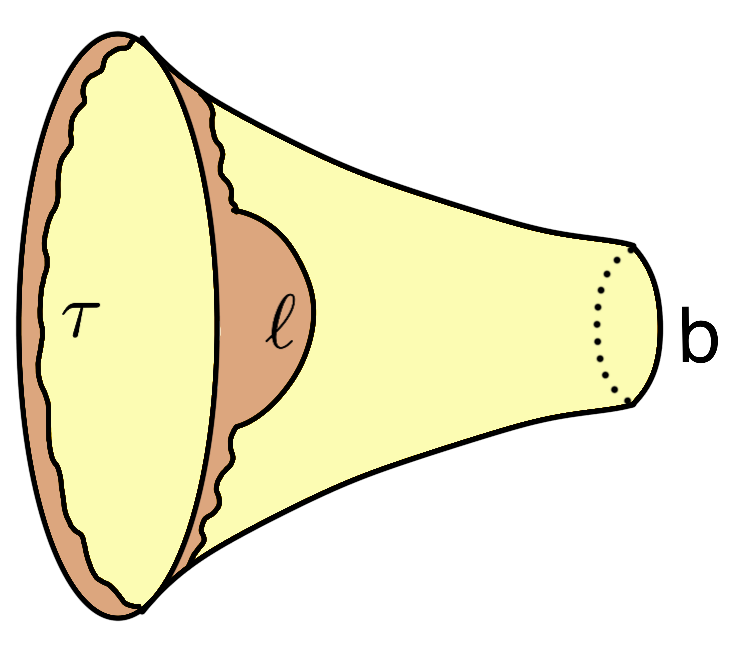}\quad=\quad\includegraphics[width=0.2\textwidth,valign=c]{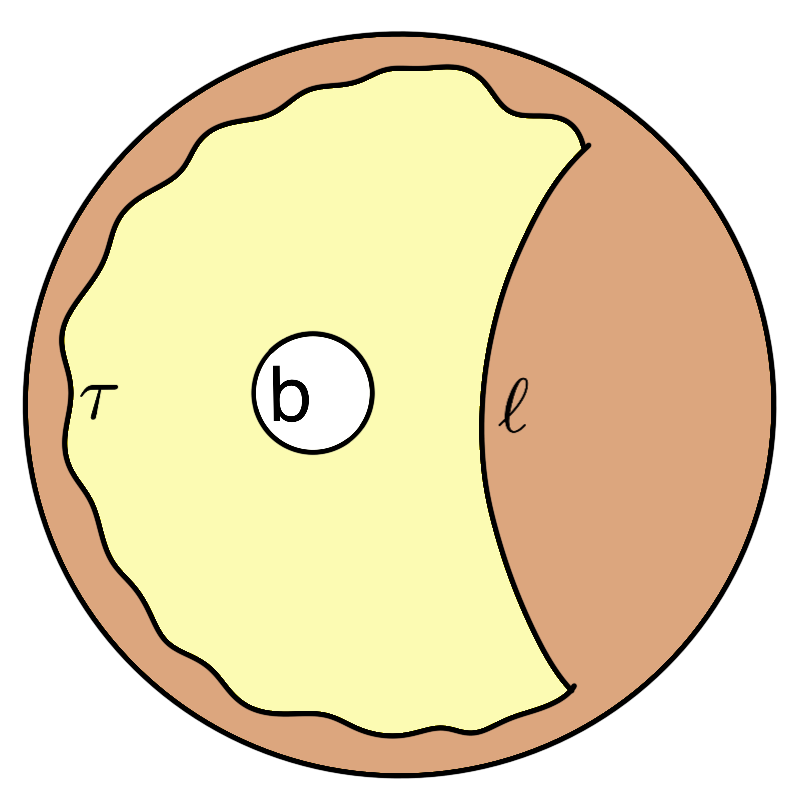}
\ee
\begin{align}
\varphi_{\text{Disk},\tau}(\ell)&=\int_0^\infty ds\,\rho_0(s)e^{-\tau s^2/2}\varphi_s(\ell)\\
\varphi_{\text{Trumpet},\tau}(\ell,b)&=\int_0^\infty ds\,\rho_1(s,b)e^{-\tau s^2/2}\varphi_s(\ell)
\end{align}
We should note that we can get a trumpet HH wavefunction by gluing a disk HH wavefunction to a wedge with a hole
\be
\includegraphics[width=0.08\textwidth]{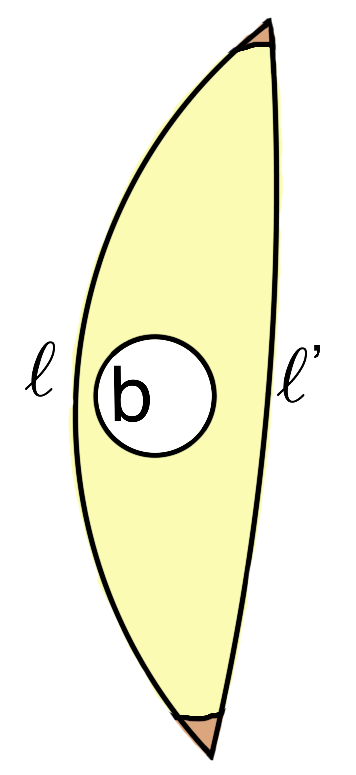}
\ee
which is given by
\be
\braket{\ell,b|\ell'}=\int_0^\infty ds\,\rho_1(s,b)\varphi_s(\ell)\varphi_s(\ell')\label{wedgeeq}
\ee
we can check that gluing this wedge to the disk HH wavefunction gives the trumpet HH wavefunction
\be
\int d\ell e^{\ell}\,\varphi_{\text{Disk}}(\ell)\braket{\ell,b|\ell'}=\varphi_{\text{Trumpet}}(\ell',b)
\ee
We can also check that gluing HH wavefunctions give back the partition functions. For a disk we have
\begin{align}
Z_{\text{Disk}}(\beta)&=\includegraphics[valign=c,width=0.2\textwidth]{figures/diskcopy.png}\\
&=\int\,e^\ell d\ell\includegraphics[valign=c,width=0.2\textwidth]{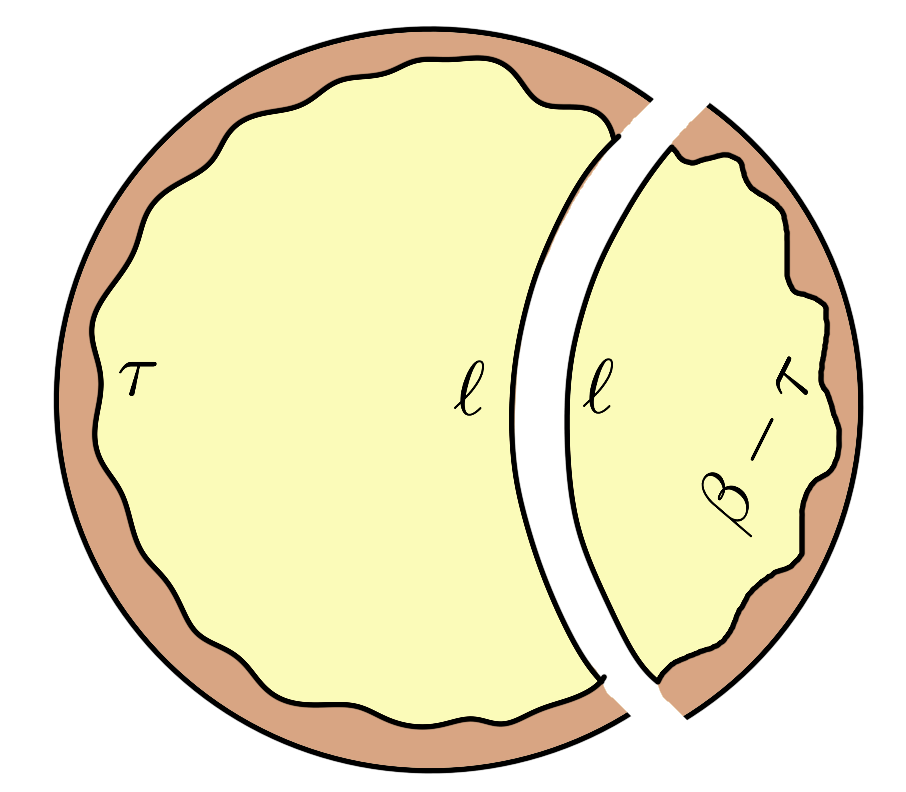}\\
&=e^{S_0}\int\,e^\ell d\ell\,\varphi_{\text{Disk},\tau}(\ell)\varphi_{\text{Disk},\beta-\tau}(\ell)
\end{align}
And for a trumpet, we have
\begin{align}
Z_{\text{Trumpet}}(\beta,b)&=\includegraphics[valign=c,width=0.2\textwidth]{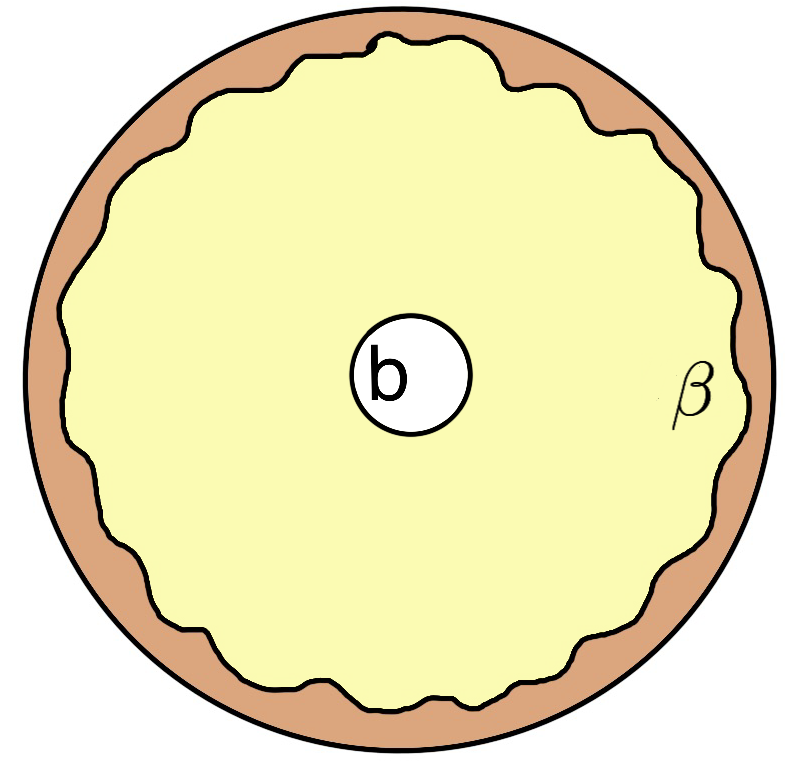}\\
&=\int\,e^\ell d\ell\includegraphics[valign=c,width=0.2\textwidth]{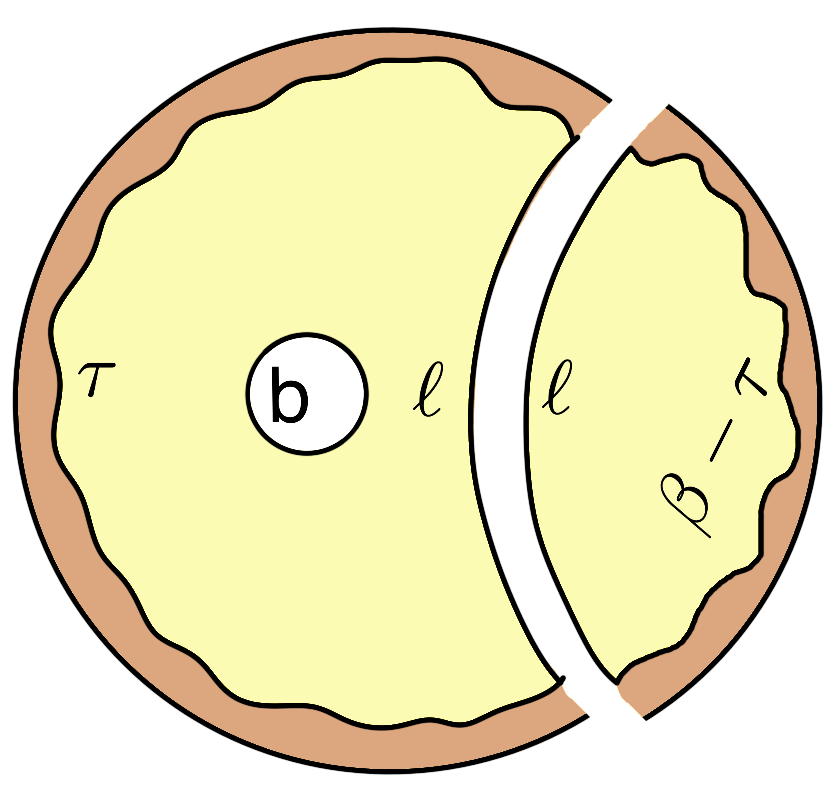}\\
&=\int\,e^\ell d\ell\,\varphi_{\text{Trumpet},\tau}(\ell,b)\varphi_{\text{Disk},\beta-\tau}(\ell)
\end{align}

\paragraph{Disk with n defects} Now let's compute disk with $n$ defects. It turns out in the near-extremal limit, this is the same as disk with $n$ holes with the boundary length of the $n$ holes small. 

We start with calculating a disk with two holes.
\begin{align}
Z_2(\beta,b_1,b_2)&=\includegraphics[valign=c,width=0.2\textwidth]{figures/2holedisk.png}\\
&=\int_0^\infty bdb\,\includegraphics[valign=c,width=0.3\textwidth]{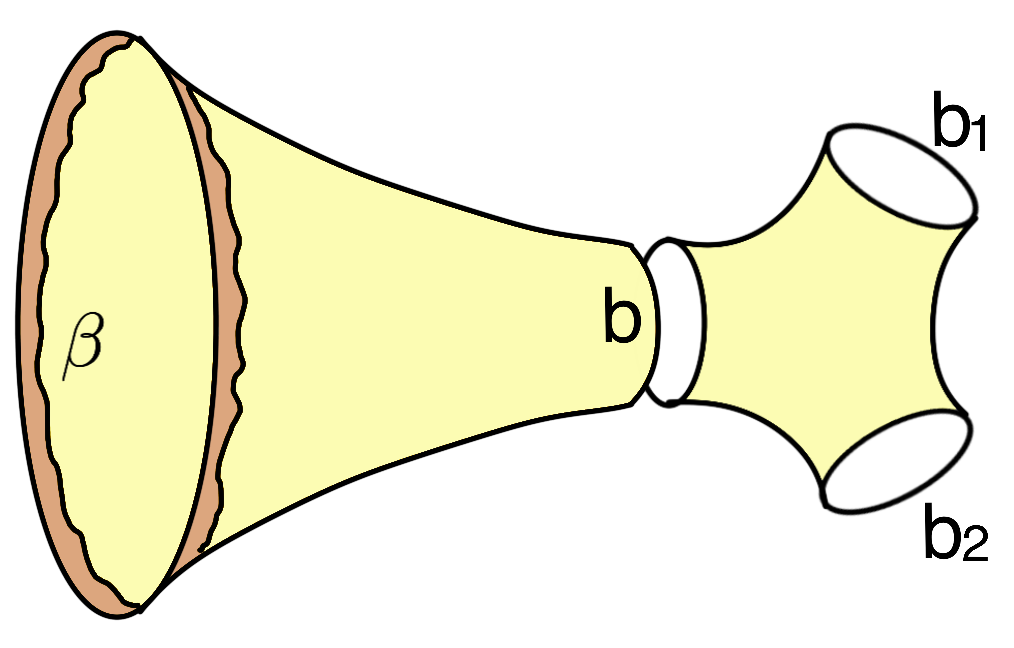}\label{trumpetandvolume}\\
&=\int_0^\infty bdb\,Z_{trumpet}(\beta,b)V_{0,3}(b,b_1,b_2)\\
&=\int_0^\infty ds\,\int_0^\infty bdb\,\rho_1(s,b)e^{-\beta s^2/2}\label{2holedisko}
\end{align}
where $V_{0,3}$ is the volume of three-hole sphere
\be
V_{0,3}(b,b_1,b_2)=1
\ee
This comes from mapping class group. The density of states is manifestly divergent, but after regularization
\begin{align}
\rho_2(s)&=\int_0^\infty bdb\,\rho_1(s,b)\\
&\rightarrow -\frac{1}{\pi\sqrt{\gamma}s^2}\label{2hole}
\end{align}
In general, we can compute disk with $n$ holes by cutting it into a trumpet and a $n+1$-hole sphere
\begin{align}
Z_n(\beta,b_1,\cdots, b_n)&=\includegraphics[valign=c,width=0.2\textwidth]{figures/nholedisk.png}\\
&=\int_0^\infty bdb\,\includegraphics[valign=c,width=0.3\textwidth]{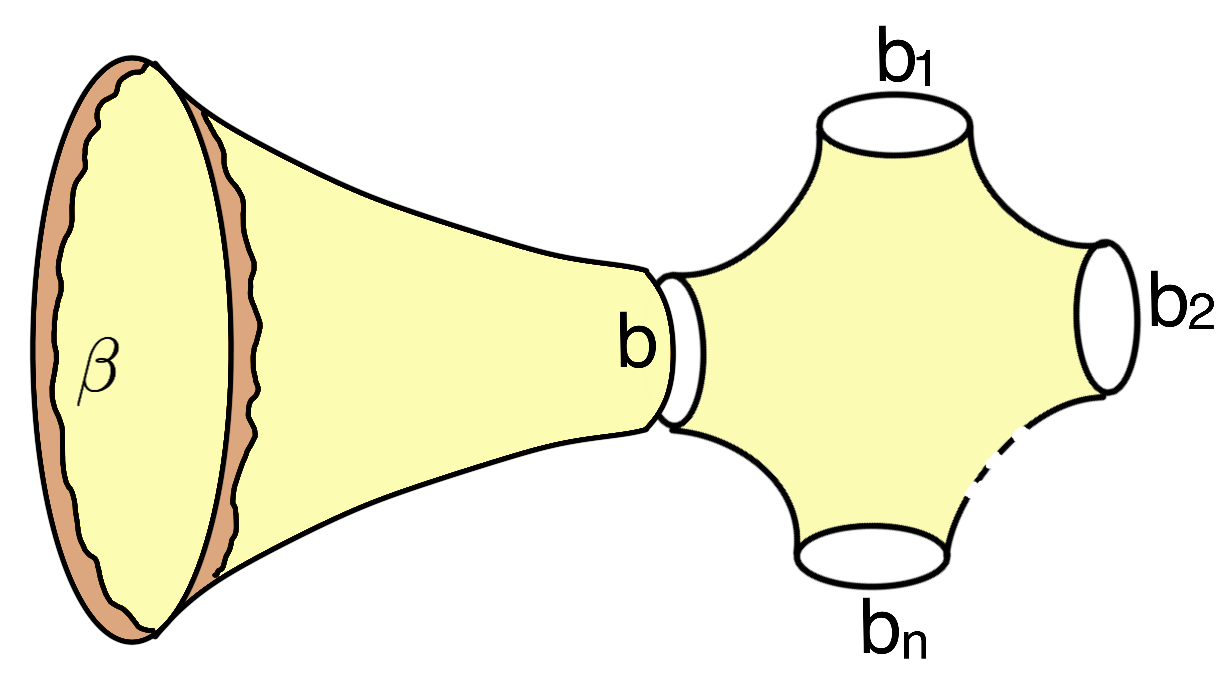}\label{trumpetandvolumen}\\
&=e^{-(n-1)S_0}\int_0^\infty bdb\,Z_{trumpet}(\beta,b)V_{0,n+1}(b,b_1,b_2,\cdots,b_n)\\
&\approx e^{-(n-1)S_0}\int_0^s ds\,\int_0^\infty bdb\,\frac{\rho_1(s,b)}{(n-2)!}\left(\frac{b^2}{2}\right)^{n-2}e^{-\beta s^2/2}\label{MT}
\end{align}
where we have used the fact that in the limit $b\rightarrow\infty$ and $b_i$ small
\be
V_{0,n+1}(b,b_1,b_2,\cdots,b_n)=\frac{1}{(n-2)!}\left(\frac{b^2}{2}\right)^{n-2}+\mathcal{O}(b^{2n-6})
\ee
so the density of states is given by
\begin{align}
\rho_n(s)&\approx \int_0^\infty bdb\,\frac{\rho_1(s,b)}{(n-2)!}\left(\frac{b^2}{2}\right)^{n-2}\\
&=(-1)^{n+1}\frac{2^{2-n}(2n-3)!}{(n-2)!}\frac{1}{\pi s^{2n-2}}\\
&=2^nn!\binom{\frac{1}{2}}{n}\frac{1}{\pi s^{2n-2}}
\end{align}
We should note in particular that $\rho_0(s)$ and $\rho_1(s)$ also satisfy the formulas as $s\rightarrow0$. 

Now let's sum over different defects taking into account the dimensional reduction $\lambda=\frac{(-1)^J}{8\pi^2}e^{-\frac{S_0(J)}{2}}$
\begin{align}
\rho^{3d}(P,\bar{P})&=32\pi^3 e^{S_0(J)}\sum_{n=0}^\infty \frac{\lambda^n}{n!}\rho^{2d}_n\\
&=32\pi^3 e^{S_0(J)}\sum_{n=0}^\infty \binom{\frac{1}{2}}{n}\frac{(2\lambda)^n}{\pi P^{2n-2}}\\
&=\rho^{3d}_{BTZ}\sqrt{1+\frac{2\lambda}{P^2}}
\end{align}
This solves the negativity problem by non-perturbative shift of $P$.

\section{Review: Torus Wormhole}
\label{toruswormhole}
In this section, we review off-shell torus wormhole computation by Cotler-Jensen \cite{Cotler:2020ugk} and compare it to VTQFT on-shell computation in the limit mass insertions go to zero. These computations give different answers and we do not know how to go between them. To gain some intuition, before torus wormholes in \ref{32}, we show a similar situation for 2d cylinders in \ref{22} following \cite{Yan:2023rjh}. Because in JT gravity, we not only know confidently how to compute both the off-shell cylinder and the on-shell cylinder, we can also recover on-shell result by adding winding to an off-shell computation.

\subsection{2d Cylinder}
\label{22}
In this section, we first give the partition function of a cylinder which is off-shell, then we compute a cylinder with one matter insertion of mass $\Delta$ on each boundary with the limit $\Delta\rightarrow0$, and finally we show that by combining the off-shell computation with a winding term we can recover the on-shell computation. This section is a review based on Appendix A of \cite{Yan:2023rjh}.

Before we proceed to compute the partition functions, we need two more useful JT gravity tools. First there is the identity
\be
\int_{-\infty}^\infty e^\ell d\ell\,\varphi_s(\ell)\varphi_{s'}(\ell)e^{-\Delta\ell}=|V_{s,s'}|^2=\frac{\Gamma(\Delta\pm is\pm is')}{2^{2\Delta-1}\Gamma(2\Delta)}
\ee
In particular we take two special limits. When $s=s'$ and $\Delta\rightarrow0$
\be
|V_{s,s}|^2\approx\frac{1}{\Delta\pi\rho_0(s)}
\ee
When $s,s'\rightarrow0$ and $\Delta\rightarrow0$
\be
|V_{s,s'}|^2\approx\frac{4\Delta}{(\Delta^2+(s+s')^2)(\Delta^2+(s-s')^2)}
\ee
Second, in addition to partition functions and Hartle-Hawking states, we review what a propagator evaluates to 
\be
P_{\text{Disk}}(\beta_1,\beta_2,\ell,\ell')=\includegraphics[valign=c,width=0.2\textwidth]{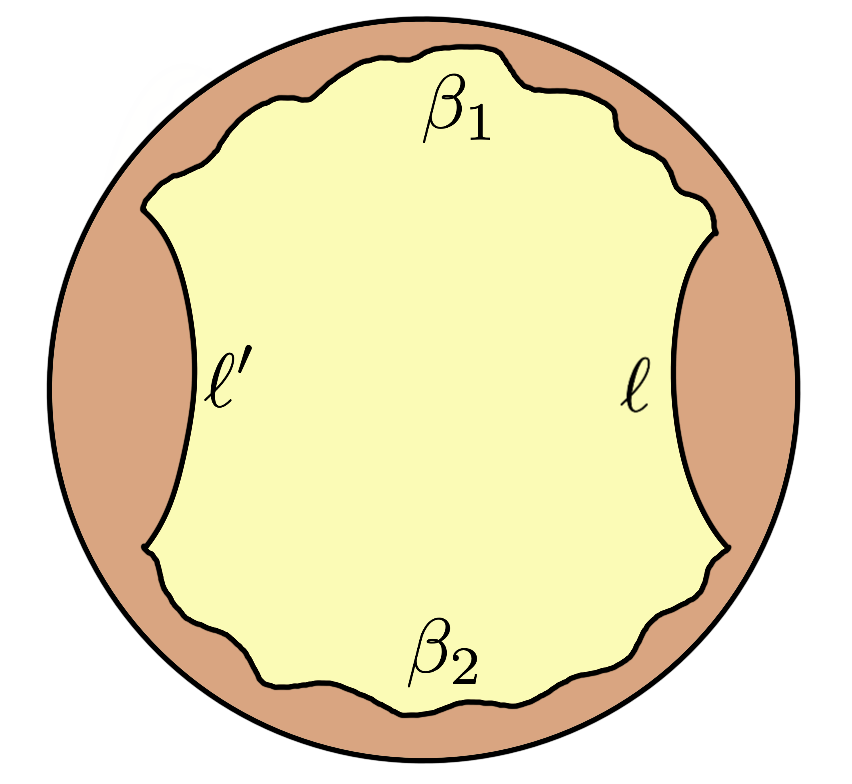}
\ee
A propagator is a time evolution operator of a Hartle-Hawking wavefunction such that
\be
\varphi_{\text{Disk},\beta+\beta_1+\beta_2}(\ell)=\int e^{\ell'}d\ell'\,P_{\text{Disk}}(\beta_1,\beta_2,\ell,\ell')\varphi_{\text{Disk},\beta}(\ell')
\ee
we can check \cite{Saadsingleauthor} that the above relations are solved by
\be
P_{\text{Disk}}(\beta_1,\beta_2,\ell,\ell')=\int ds\,\rho_0(s)e^{-(\beta_1+\beta_2)s^2/2}\varphi_s(\ell)\varphi_s(\ell')\label{diskpropagator}
\ee

\paragraph{On-shell} We can compute a cylinder with one insertion on each boundary by cutting through the geodesics connecting the two insertions and the resulting configuration is a propagator. We evaluate it in the limit mass of the insertions go to zero $\Delta\rightarrow0$.
\begin{align}
Z^{\text{on-shell}}_{\text{cylinder}}&=\includegraphics[valign=c,width=0.2\textwidth]{figures/2doneptpic.png}\\
&=\int e^\ell d\ell\,P_{\text{Disk}}(\beta_1,\beta_2,\ell,\ell)e^{-\Delta\ell}\\
&=\int ds\,\rho_0(s) e^{-(\beta_1+\beta_2)s^2/2}|V_{s,s}|^2\\
&\approx\frac{1}{\pi\Delta}\int ds\,e^{-(\beta_1+\beta_2)s^2/2}\\
&=\frac{1}{\pi\Delta}\sqrt{\frac{\pi}{2(\beta_1+\beta_2)}}
\end{align}

\paragraph{Off-shell} We can compute the cylinder partition function by gluing two trumpets together taking into account the relative twist between them
\begin{align}
Z^{\text{off-shell}}_{\text{cylinder}}&=\includegraphics[valign=c,width=0.2\textwidth]{figures/cylindercom.png}\\
&=\int db\,b\includegraphics[valign=c,width=0.25\textwidth]{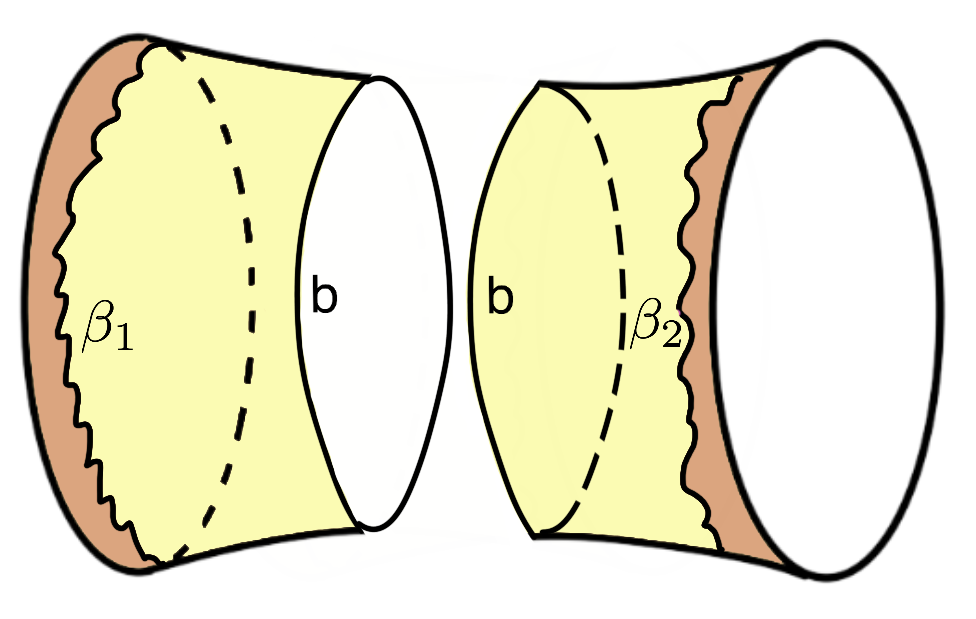}\\
&=\int db\,b\,Z_{\text{Trumpet}}(\beta_1,b)Z_{\text{Trumpet}}(\beta_2,b)\\
&=\frac{\sqrt{\beta_1\beta_2}}{2\pi(\beta_1+\beta_2)}
\end{align}
Note that this gives different answer than the on-shell computation. But we can recover it by include winding in the off-shell computation. Winding term is a sum $e^{-\Delta\ell}$ over all possible ways a geodesic can wind around the cylinder
\be
\sum_{n=-\infty}^\infty e^{-\Delta b |n|}\approx\frac{2}{b\Delta}
\ee
Then the modified off-shell computation becomes
\begin{align}
\tilde{Z}^{\text{off-shell}}_{\text{cylinder}}&=\int db\,b\,\sum_{n=-\infty}^\infty e^{-\Delta b |n|}\,Z_{\text{Trumpet}}(\beta_1,b)Z_{\text{Trumpet}}(\beta_2,b)\\
&=\frac{2}{\Delta}\int db\,Z_{\text{Trumpet}}(\beta,b)Z_{\text{Trumpet}}(\beta',b)\\
&=\frac{1}{\pi\Delta}\sqrt{\frac{\pi}{2(\beta_1+\beta_2)}}\\
&=Z^{\text{on-shell}}_{\text{cylinder}}
\end{align}
We can summarize the result as
\be
\text{on-shell}=\text{off-shell}+\text{winding}
\ee
However, we should note that this relation only works in one direction, i.e. if we know the off-shell computation and the winding we can recover the on-shell computation. However, if we only know the on-shell computation and the winding, because the on-shell computation is lacking the information about which parameter $b$ we want to integrate over, we cannot recover the off-shell computation. This conceptually makes our life difficult, because in 3 dimensions, an on-shell computation is easier, and off-shell computation is what we want to know ultimately. 

\subsection{3d torus wormhole}
\label{32}
A Maldacena-Maoz wormhole \cite{Maldacena:2004rf} $\Sigma\times I$ has two asymptotic boundaries with identical topology and constant moduli. The boundaries are Riemann surfaces. One simple option is for both boundaries to be tori and we get a torus-wormhole
\be
\includegraphics[width=0.2\textwidth]{figures/toruswormhole.png}
\ee
In order for the torus-wormhole to be on-shell, we need at least one insertion on each boundary. 
\be
\includegraphics[width=0.2\textwidth]{figures/twh1pt.png}
\ee
If we take the operator dimension to zero for a torus wormhole, we can connect back to \cite{Cotler:2020ugk}, which was done in \cite{Collier:2023fwi} and \cite{Yan:2023rjh}. The on-shell result is given by
\be
Z^{\text{on-shell}}_{T^2\times I}(\tau_1,\bar{\tau}_1,\tau_2,\bar{\tau}_2)=\sum_{\gamma\in\mathrm{SL}(2,\Z)}Z^{\text{on-shell}}(\tau_1,\bar{\tau}_1,\gamma\tau_2,\gamma\bar{\tau}_2)
\ee
where
\be
Z^{\text{on-shell}}(\tau_1,\bar{\tau}_1,\tau_2,\bar{\tau}_2)\propto\frac{1}{\Delta^2}\int dPd\bar{P}\chi_P(\tau_1)\chi_P(\tau_2)\bar{\chi}_{\bar{P}}(\bar{\tau}_1)\bar{\chi}_{\bar{P}}(\bar{\tau}_2)
\ee
where the character
\be
\chi_P(\tau)=\frac{q^{P^2}}{\eta(\tau)}\quad\quad q=e^{2\pi i\tau}
\ee
On the other hand, the off-shell result is given by
\be
Z^{\text{off-shell}}_{T^2\times I}(\tau_1,\bar{\tau}_1,\tau_2,\bar{\tau}_2)=\sum_{\gamma\in\mathrm{SL}(2,\Z)}Z^{\text{off-shell}}(\tau_1,\bar{\tau}_1,\gamma\tau_2,\gamma\bar{\tau}_2)
\ee
where
\be
Z^{\text{off-shell}}(\tau_1,\bar{\tau}_1,\tau_2,\bar{\tau}_2)\propto\sqrt{\mathrm{Im}(\tau_1)\mathrm{Im}(\tau_2)}\int_0^\infty dPd\bar{P}\chi_P(\tau_1)\chi_{\bar{P}}(\tau_2)\bar{\chi}_{\bar{P}}(\bar{\tau}_1)\bar{\chi}_{\bar{P}}(\bar{\tau}_2)P\bar{P}
\ee
Their difference is reminiscent to 2d cylinder. 

\subsubsection{Off-shell} 
\label{offshell32}
As shown in Cotler-Jensen \cite{Cotler:2020ugk}, we can calculate the path integral of a torus-wormhole by quantizing the phase space of a cylinder $\times$ time, which is two copies of this quantum mechanics, one right-moving and one left-moving.

Recall that for one particle with position $x(t)$ and momentum $p(t)$, ordinary path-integral from $A$ to $B$ is given by
\be
U(B;A)=\int_{x(t_A)=x_A}^{x(t_B)=x_B}\mathcal{D}x(t)\,e^{i\int_{t_A}^{t_B}dt\,L(x(t),\dot{x}(t))}
\ee
while the phase-space path-integral is given by
\be
U(B;A)=\int_{x(t_A)=x_A}^{x(t_B)=x_B}\mathcal{D}x(t)\mathcal{D}p(t)\,e^{i\int_{t_A}^{t_B}dt\,\dot{x}(t)p(t)-i \int_{t_A}^{t_B}dt\, H(x(t),p(t))}
\ee
where on the exponential, the first term is the symplectic measure and the second term is the energy. 

We parametrize the left-moving cylinder by $(b,s)$ and the right-moving cylinder by $(\bar{b},\bar{s})$. To illustrate the quantization procedure, below we just focus on the left-moving cylinder. A cylinder can be viewed as two trumpets glued together. Near the region where two trumpets are glued to each other, set coordinate $\rho$ to measure the distance to the joining locus ($\rho<0$ is the left tube, and $\rho>0$ is the other). The metric is 
\be
ds^2=d\rho^2+\cosh^2\rho\,\left(b\,dx+s\delta(\rho)\,d\rho\right)^2\quad\quad x\sim x+1
\ee
where $\delta(\rho)$ implement the twist by distance $s$ in the $x$ direction. The symplectic form is given by
\be
\Omega_{Cyl}=\alpha\, \delta b\wedge \delta s
\ee

On the left-moving cylinder, there are two boundaries, each boundary would contribute both to the symplectic form and to the energy. Now let's just focus on one boundary, say the one that belongs to the trumpet on the left, and find its symplectic measure and energy. This was done in \cite{Saad:2019lba}.

The boundary condition we impose are 
\be
\left.g_{\theta\theta}\right|=\frac{1}{\epsilon^2}\quad\quad\left.\phi\right|_{bdy}=\frac{\gamma}{\epsilon}\quad\quad\epsilon\rightarrow0
\ee
where $\theta\in[0,2\pi)$ is rescaled proper length along the boundary, so that the total length of the boundary is $2\pi/\epsilon$. 

With this boundary condition, the boundary action is given by the Schwarzian 
\be
I_{bdy}=-\gamma\int_0^{2\pi} d\theta\,\Sch(\theta)
\ee
A trumpet can be obtained from a piece of hyperbolic space
\be
ds^2=d\rho^2+\cosh^2\rho\,d\sigma^2\quad\quad\sigma\sim\sigma+b
\ee
The wiggly boundary is described by a function $\sigma(\theta)$. The boundary action becomes
\be
I_{bdy}=-\gamma\int_0^{2\pi} d\theta\,\Sch(e^{-\sigma(\theta)},\theta)
\ee
The saddle point is given by
\be
\sigma(\theta)=\frac{b}{2\pi}\theta
\ee
with the corresponding Schwarzian given by
\be
\Sch(e^{-\frac{b}{2\pi}\theta},\theta)=-\frac{b^2}{2(2\pi)^2}
\ee
And we consider small fluctuations about the saddle point
\be
\sigma(\theta)=\frac{b}{2\pi}(\theta+\varepsilon(\theta))
\ee
Expand in Fourier modes
\be
\varepsilon(\theta)=\sum_{|n|\geq 1}e^{-in\theta}(\varepsilon_n^{(R)}+i\varepsilon_n^{(I)})
\ee
where we removed the $\varepsilon=1$ mode because of $U(1)$ symmetry of the trumpet. And to keep $\varepsilon(\theta)$ real we need $\varepsilon_n^{(R)}=\varepsilon_{-n}^{(R)}$ and $\varepsilon_n^{(I)}=-\varepsilon_{-n}^{(I)}$, so the independent variables are $\varepsilon_n^{(R)}$ and $\varepsilon_n^{(I)}$ for $n\geq1$. 

\noindent The symplectic measure is given by
\begin{align}
\Omega_{bdy}&=\frac{\alpha}{2}\int_0^{2\pi} d\theta\,\left(d\varepsilon'(\theta)\wedge d\varepsilon''(\theta)-2\Sch(\theta)d\varepsilon(\theta)\wedge d\varepsilon'(\theta)\right)\\
&=\frac{\alpha}{2}\int_0^{2\pi} d\theta\,\left(d\varepsilon'(\theta)\wedge d\varepsilon''(\theta)+\frac{b^2}{(2\pi)^2}d\varepsilon(\theta)\wedge d\varepsilon'(\theta)\right)\\
&=4 \pi\alpha\sum_{n\geq1}(n^3+\frac{b^2}{(2\pi)^2}n)d\varepsilon_n^{(R)}\wedge d\varepsilon_n^{(I)}
\end{align}
and the energy is given by
\begin{align}
\bar{L}_0&=\frac{c}{24}-\gamma\int_0^{2\pi} d\theta\,\Sch(\theta)\\
&=\frac{c}{24}+\gamma\frac{b^2}{4\pi}+\frac{\gamma}{2}\int_0^{2\pi} d\theta\,\left(\varepsilon''(\theta)^2+\frac{b^2}{(2\pi)^2}\varepsilon'(\theta)^2\right)\\
&=\frac{c}{24}+\gamma\frac{b^2}{4\pi}+2\pi\gamma\sum_{n\geq1}(n^4+\frac{b^2}{(2\pi)^2}n^2)\left((\varepsilon_n^{(R)})^2+(\varepsilon_n^{(I)})^2\right)
\end{align}
Now let's set 
\be
\alpha=\frac{c-1}{48\pi}\quad\quad \gamma=\frac{c-1}{24\pi}
\ee
then
\begin{align}
\Omega_{bdy}&=\frac{c-1}{48}\int_0^{2\pi} \frac{d\theta}{2\pi}\,\left(d\varepsilon'(\theta)\wedge d\varepsilon''(\theta)+\frac{b^2}{(2\pi)^2}d\varepsilon(\theta)\wedge d\varepsilon'(\theta)\right)\\
\bar{L}_0&=\frac{c}{24}+\frac{c-1}{24}\frac{b^2}{(2\pi)^2}+\frac{c}{24}\int_0^{2\pi} \frac{d\theta}{2\pi}\,\left(\varepsilon''(\theta)^2+\frac{b^2}{(2\pi)^2}\varepsilon'(\theta)^2\right)
\end{align}
To do the phase-space path integral and get $\Tr(e^{-2\pi i \bar{\tau} (\bar{L}_0-\frac{c}{24})})$, we need to integrate over imaginary time $u$ from $0$ to $2\pi i \bar{\tau}$. Then the exponent looks like
\begin{align}
S(\bar{\tau},b)&= \int_0^{2\pi i \bar{\tau}}du\,\left(i\Omega_{bdy}-( \bar{L}_0-\frac{c}{24})\right)\\
&=-\frac{c-1}{24}\int_0^{2\pi i \bar{\tau}}du\,\frac{b^2}{(2\pi)^2}-\frac{c-1}{24}\int_0^{2\pi i\bar{\tau}} du \int_0^{2\pi}\frac{d\theta}{2\pi}\left(-i(\dot{\varepsilon}'\varepsilon''+\frac{b^2}{(2\pi)^2}\dot{\varepsilon}\varepsilon')+ (\varepsilon''^2+\frac{b^2}{(2\pi)^2}\varepsilon'^2)\right)\\
&=-2\pi i \bar{\tau} \frac{c-1}{24}\frac{b^2}{(2\pi)^2}-\frac{c-1}{24}\int_0^{2\pi i\bar{\tau}}du\int_0^{2\pi}\frac{d\theta}{2\pi}\left(\partial_-\varepsilon'\varepsilon''+\frac{b^2}{(2\pi)^2}\partial_-\varepsilon\varepsilon'\right)
\end{align}
where 
\be
\partial_-=\partial_\theta-i\partial_u
\ee
Now using results from Alekseev-Shatashvili \cite{Alekseev:1988ce} we can do path-integral over $\varepsilon$ for one boundary of the left-moving cylinder and get
\begin{align}
    \int\mathcal{D}\varepsilon\, e^{S(\bar{\tau}_1,b)}&=\prod_{n\geq1}\int d\varepsilon_n^{(R)}d\varepsilon_n^{(I)}\,e^{S(\bar{\tau}_1,b)}\\
    &=\bar{\chi}_{\sqrt{\frac{c-1}{24}}\frac{b}{2\pi}}(\bar{\tau}_1)
\end{align}
And similarly, one boundary on the right-moving cylinder contribute to the path-integral by
\begin{align}
    \int\mathcal{D}\bar{\varepsilon}\, e^{\bar{S}({\tau}_1,\bar{b})}&=\prod_{n\geq1}\int d\bar{\varepsilon}_n^{(R)}d\bar{\varepsilon}_n^{(I)}\,e^{\bar{S}(\tau_1,\bar{b})}\\
    &=\chi_{\sqrt{\frac{c-1}{24}}\frac{\bar{b}}{2\pi}}(\tau_1)
\end{align}
where the characters 
\be
\chi_P(\tau)=\frac{e^{2\pi i\tau P^2}}{\eta(\tau)}
\ee
There are two boundaries (we denote by boundary 1 and boundary 2) each with left-moving and right-moving modes with modular parameters $(\tau_1,\bar{\tau}_1)$ and $(\tau_2,\bar{\tau}_2)$, these give four characters. There are also two symplectic measures from the left-moving and right-moving cylinders themselves. In total the action is given by
\begin{multline}
Z^{\text{off-shell}}(\tau_1,\bar{\tau}_1,\tau_2,\bar{\tau}_2)=\int dbdsd\bar{b}d\bar{s}\, e^{i\int du\,\frac{c}{24\pi}(\dot{b}s+\dot{\bar{b}}\bar{s})}\\
\times\chi_{\sqrt{\frac{c-1}{24}}\frac{\bar{b}}{2\pi}}(\tau_1)\chi_{\sqrt{\frac{c-1}{24}}\frac{\bar{b}}{2\pi}}(\tau_2)\bar{\chi}_{\sqrt{\frac{c-1}{24}}\frac{b}{2\pi}}(\bar{\tau}_1)\bar{\chi}_{\sqrt{\frac{c-1}{24}}\frac{b}{2\pi}}(\bar{\tau}_2)
\end{multline}
If we conjecture that there are two identifications made with $(b,\bar{b},s,\bar{s})$, $s+\bar{s}$ is known to be compact, and \cite{Cotler:2020ugk} made a conjecture $s-\bar{s}$ is also compact. Integrating out these compact modes give an additional factor $b\bar{b}$. The partition function can then be simplified to 
\be
Z(\tau_1,\bar{\tau}_1,\tau_2,\bar{\tau}_2)\propto\sqrt{\mathrm{Im}(\tau_1)\mathrm{Im}(\tau_2)}\int_0^\infty dbd\bar{b}\,b\bar{b}\,\chi_{\sqrt{\frac{c-1}{24}}\frac{\bar{b}}{2\pi}}(\tau_1)\chi_{\sqrt{\frac{c-1}{24}}\frac{\bar{b}}{2\pi}}(\tau_2)\bar{\chi}_{\sqrt{\frac{c-1}{24}}\frac{b}{2\pi}}(\bar{\tau}_1)\bar{\chi}_{\sqrt{\frac{c-1}{24}}\frac{b}{2\pi}}(\bar{\tau}_2)
\ee
And finally to get the torus-wormhole partition function, we need to sum over $\gamma\in SL(2,\Z)$ and get
\be
Z^{\text{off-shell}}_{T^2\times I}(\tau_1,\bar{\tau}_1,\tau_2,\bar{\tau}_2)=\sum_{\gamma\in SL(2,\Z)}Z^{\text{off-shell}}(\tau_1,\bar{\tau}_1,\gamma\tau_2,\gamma\bar{\tau}_2)
\ee

\subsubsection{On-shell}
\label{onshell32}
In remaining part of this section, we review the calculation of on-shell torus-wormhole using VTQFT \cite{Collier:2023fwi} and get and expression of bulk torus-wormhole. 
 
Virasoro TQFT \cite{Collier:2023fwi, Collier:2024mgv} \footnote{Also see \cite{Bhattacharyya:2024vnw, Takahashi:2024ukk}.} starts from SL(2,$\R$)$^2$ Chern-Simons theory on $\Sigma\times\R$
and quantize hyperbolic metrics on $\Sigma$. VTQFT Hilbert space $\mathcal{H}(\Sigma)$ is the space of conformal blocks on $\Sigma$. And the full Hilbert space is given by a tensor product of a chiral part and an anti-chiral part $\mathcal{H}_{gravity}=\mathcal{H}(\Sigma)\otimes \overline{\mathcal{H}}(\Sigma)$.
This only works on-shell.

Maldacena-Maoz wormholes are special in VTQFT, they represent the identity on the Hilbert space $\mathcal{H}_\Sigma$
\be
Z_{Liouville}(\Sigma\times I)=\sum_{\text{conformal blocks on }\Sigma}\frac{\mathcal{F}_\Sigma\overline{\mathcal{F}}_\Sigma}{\braket{\mathcal{F}_\Sigma|\mathcal{F}_\Sigma}}
\ee
Then the gravitational partition function is given by summing over mapping class group
\be
Z_{grav}(\Sigma\times I)=\sum_{\gamma\in MCG(\Sigma)}|Z_{Liouville}(\Sigma;m_1,\gamma\cdot m_2)|^2
\ee
For a torus-wormhole we get
\be
Z^{\text{on-shell}}_{T^2\times I}(\tau_1,\bar{\tau}_1,\tau_2,\bar{\tau}_2)=\sum_{\gamma\in\mathrm{SL}(2,\Z)}Z^{\text{on-shell}}(\tau_1,\bar{\tau}_1,\gamma\tau_2,\gamma\bar{\tau}_2)\label{twsum}
\ee
Now let us just compute one element in the summand using VTQFT. One way of getting $Z^{\text{on-shell}}$ is by starting with an on-shell torus wormhole with one insertion of mass $\Delta$ on each boundary and then taking the limit $\Delta\rightarrow0$.
\begin{align}
Z_{vir}(\includegraphics[valign=c,width=0.15\textwidth]{figures/twh1pt.png})
&=\int dP_idP_j\,\frac{\delta(P_i-P_j)}{\rho_0(P_i)C_0(P_k,P_i,P_i)}\nonumber\\
&\quad\times\rho_0(P_i)\rho_0(P_j)C_0(P_k,P_i,P_i)C_0(P_k,P_j,P_j)\underbrace{\includegraphics[width=0.15\textwidth, valign=c]{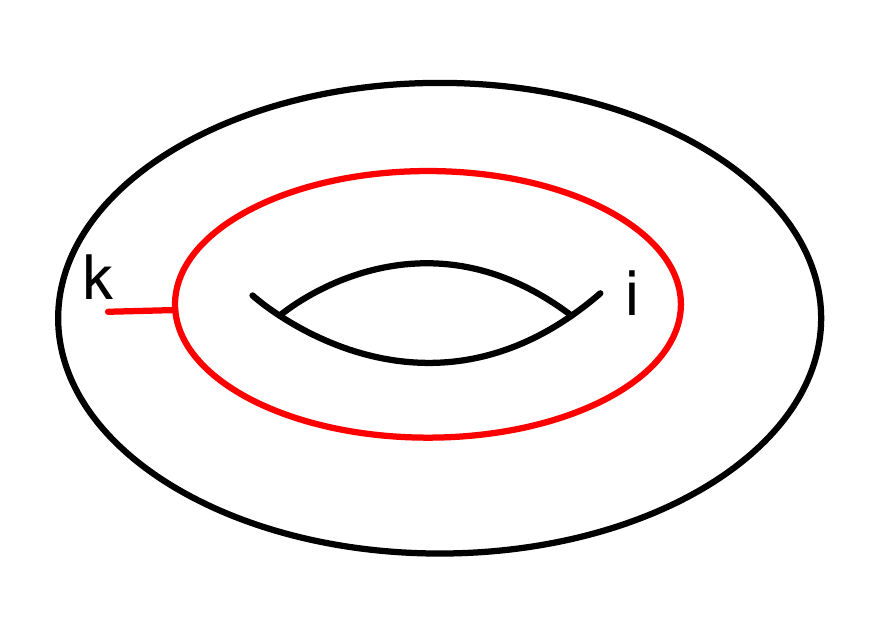}\includegraphics[width=0.15\textwidth, valign=c]{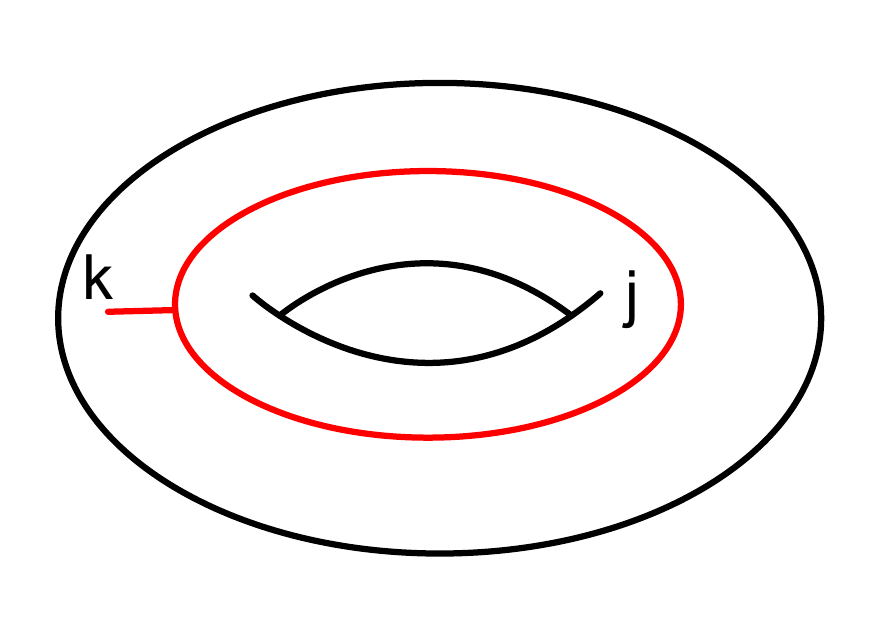}}_{\text{conformal blocks}}\\
&=\frac{1}{\Delta}\int dP_idP_j\,\delta(P_i-P_j)\chi_{P_i}(\tau_1)\chi_{P_j}(\tau_2)
\end{align}
where we used the fact that in the limit $\Delta\rightarrow0$
\be
C_0(P_k,P_i,P_i)\propto \frac{1}{\Delta\rho_0(P_k)}
\ee
and the conformal blocks become characters. Note that each torus boundary is associated with $\rho_0(P_i)C_0(P_k,P_i,P_i)$ times a conformal block, which in the limit $\Delta\rightarrow0$ is proportional to $\frac{1}{\Delta}\chi_{P_i}(\tau_i)$. Therefore, combing the chiral and anti-chiral parts, we have
\be
Z^{\text{on-shell}}(\tau_1,\bar{\tau}_1,\tau_2,\bar{\tau}_2)\propto\frac{1}{\Delta^2}\int dPd\bar{P}\chi_P(\tau_1)\chi_P(\tau_2)\bar{\chi}_{\bar{P}}(\bar{\tau}_1)\bar{\chi}_{\bar{P}}(\bar{\tau}_2)
\ee
We can think of this result as obtained from gluing two 3d trumpets to a bulk torus-wormhole. A 3d trumpet is a solid torus with a Wilson loop labeled by $(P,\bar{P})$ running through the middle \cite{Mertens:2022ujr, Jafferis:2024jkb} which is given by 
\be
Z^{3d}_{\text{trumpet}}=\chi_P(-1/\tau)\chi_{\bar{P}}(-1/\bar{\tau})
\ee
Thus our result of the on-shell torus-wormhole implies that the chiral part of the bulk torus-wormhole is proportional to $\Delta \delta(P_i-P_j)$ and we need to multiply that by $\frac{1}{\Delta^2}$ because we need to glue to two asymptotic 3d trumpets, and in total get $\frac{\delta(P_i-P_j)}{\Delta}$. Now suppose we have a bulk torus-wormhole, and we are only gluing it to one asymptotic 3d trumpet, then the bulk torus-wormhole contributes
\be
\delta(P_i-P_i)\delta(\bar{P}_i-\bar{P}_j)\label{tw1bd}
\ee

Another component of (\ref{twsum}) is where we do an S-tranform on one of the asymptotic torus boundary. Agian we first have an insertion on each boundary and then take the limit $\Delta\rightarrow0$.
\begin{align}
Z_{vir}(\includegraphics[width=0.2\textwidth, valign=c]{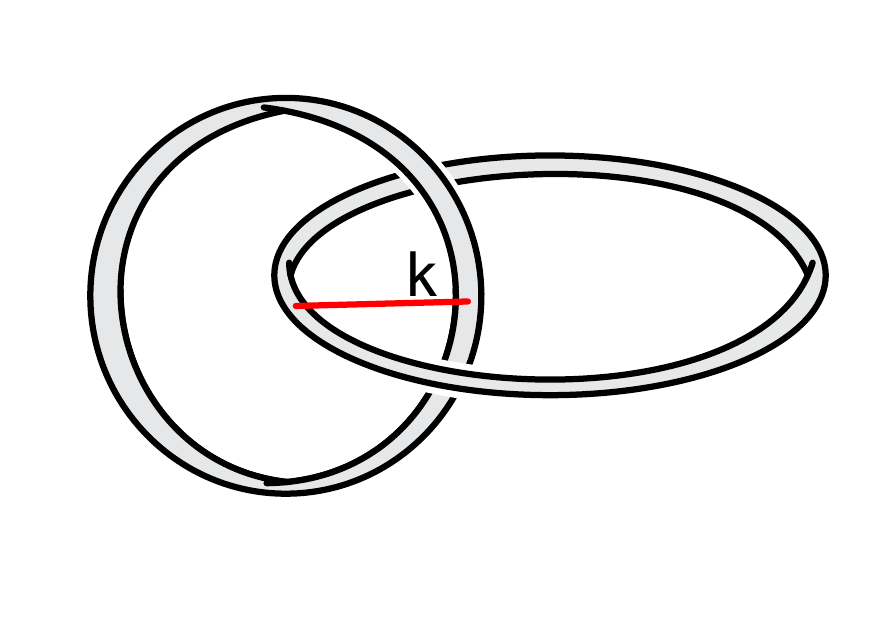})&=\int dP_idP_j\,\frac{e^{-\pi i\Delta_k}\mathbb{S}_{P_i,P_j}[P_k]}{\rho_0(P_i)C_0(P_k,P_i,P_i)}\nonumber\\
&\quad\times\rho_0(P_i)\rho_0(P_j)C_0(P_k,P_i,P_i)C_0(P_k,P_j,P_j)\underbrace{\includegraphics[width=0.15\textwidth, valign=c]{figures/opi}\includegraphics[width=0.15\textwidth, valign=c]{figures/opj}}_{\text{conformal blocks}}\\
&=\frac{1}{\Delta}\int dP_idP_j\,\mathbb{S}_{P_i,P_j}[\one]\chi_{P_i}(\tau_1)\chi_{P_j}(\tau_2)
\end{align}
Note that in the diagram the boundaries are asymptotic boundaries. Again we used the fact that each torus boundary is associated with $\rho_0(P_i)C_0(P_k,P_i,P_i)$ times a conformal block, which in the limit $\Delta\rightarrow0$ is proportional to $\frac{1}{\Delta}\chi_{P_i}(\tau_i)$. We can interpret torus wormhole with one of the boundaries S-transformed as gluing two 3d trumpets to a Hopf link.
\be
\adjustbox{valign=c}{\begin{tikzpicture}
        \draw[very thick, black] (-1/2,0) circle (1);
        \draw[draw=white, fill=white] (0,.85) circle (.08);
        \draw[very thick, blue] (1/2,0) circle (1);
        \draw[draw=white, fill=white] (0,-.85) circle (.08);
        \draw[very thick, black] (-.07,-.91) to (.08,-.82);
        \node[scale=.9,left](P) at (-3/2,0) {$P_i$};
        \node[scale=.9,left,blue](P1) at (3/2,0) {$P_j$};
    \end{tikzpicture}}\label{thlink}
\ee
The chiral part of a Hopf link by itself is proportional to $\Delta \mathbb{S}_{P_i,P_j}[\one]$ and we need to multiply that by $\frac{1}{\Delta^2}$ because we need to glue to two asymptotic 3d trumpets, and in total get $\frac{\mathbb{S}_{P_i,P_j}[\one]}{\Delta}$. Now suppose we have a Hopf link, and we are only gluing it to one asymptotic 3d trumpet, then the Hopf link contributes 
\be
\mathbb{S}_{P_i,P_j}[\one]\mathbb{S}_{\bar{P}_i,\bar{P}_j}[\one]\label{thlink1bd}
\ee

\section{Links in VTQFT}
\label{link}

In this section, we compute the Hopf link and more complicated links in the framework of VTQFT using fusion moves 
\be
\adjustbox{valign=c}{\begin{tikzpicture}
        \draw[very thick,red] (-0.5,0) to (0.5,0);
        \draw[very thick,red] (-0.5,0) to (-1,1);
        \draw[very thick,red] (-0.5,0) to (-1,-1);
        \draw[very thick,red] (0.5,0) to (1,1);
        \draw[very thick,red] (0.5,0) to (1,-1);
        \node[scale=.9,left](P3) at (-1,1) {$P_3$};
        \node[scale=.9,right](P2) at (1,1) {$P_2$};
        \node[scale=.9,left](P4) at (-1,-1) {$P_4$};
        \node[scale=.9,right](P1) at (1,-1) {$P_1$};
        \node[scale=.9](Ps) at (0,0.2) {$P_s$};
    \end{tikzpicture}}=\int dP_t\,\,\mathbb{F}_{st}\begin{bmatrix}P_3&P_2\\P_4&P_1\end{bmatrix}\adjustbox{valign=c}{\begin{tikzpicture}
        \draw[very thick,red] (0,-0.5) to (0,0.5);
        \draw[very thick,red] (0,-0.5) to (1,-1);
        \draw[very thick,red] (0,-0.5) to (-1,-1);
        \draw[very thick,red] (0,0.5) to (1,1);
        \draw[very thick,red] (0,0.5) to (-1,1);
        \node[scale=.9,left](P3) at (-1,1) {$P_3$};
        \node[scale=.9,right](P2) at (1,1) {$P_2$};
        \node[scale=.9,left](P4) at (-1,-1) {$P_4$};
        \node[scale=.9,right](P1) at (1,-1) {$P_1$};
        \node[scale=.9](Pt) at (0.2,0) {$P_t$};
    \end{tikzpicture}}\label{fusionmove}
\ee
and braiding moves
\be
\adjustbox{valign=c}{\begin{tikzpicture}
        \draw[very thick] (0,0) arc (-90:45:0.5);
        \draw[very thick,blue] (0,0) arc (-90:-225:0.5);
        \draw[very thick] (0.35,0.85) to (-0.8,1.8);
        \draw[draw=white, fill=white] (0,1.15) circle (.08);
        \draw[very thick, blue] (-0.35,0.85) to (0.8,1.8);
        \draw[very thick, red] (0,0) to (0,-0.6);
        \node[scale=.9,left](P1) at (-1,2) {$P_1$};
        \node[scale=.9,right,blue](P2) at (1,2) {$P_2$};
        \node[scale=.9,left,red](P3) at (0,-0.5) {$P_3$};
    \end{tikzpicture}}=\bB_{12}^3\adjustbox{valign=c}{\begin{tikzpicture}
        \draw[very thick] (0,0.8) to (-0.8,1.8);
        \draw[very thick, blue] (0,0.8) to (0.8,1.8);
        \draw[very thick, red] (0,0.8) to (0,-0.4);
        \node[scale=.9,left](P1) at (-0.8,1.8) {$P_1$};
        \node[scale=.9,right,blue](P2) at (0.8,1.8) {$P_2$};
        \node[scale=.9,left,red](P3) at (0,-0.5) {$P_3$};
    \end{tikzpicture}}
\ee
where
\be
\bB_{12}^3=\pm e^{\pi i(h_3-h_1-h_2)}
\ee

\subsection{Hopf link}
\label{linkthlink}

A Hopf link is 
\be
\adjustbox{valign=c}{\begin{tikzpicture}
        \draw[very thick, black] (-1/2,0) circle (1);
        \draw[draw=white, fill=white] (0,.85) circle (.08);
        \draw[very thick, blue] (1/2,0) circle (1);
        \draw[draw=white, fill=white] (0,-.85) circle (.08);
        \draw[very thick, black] (-.07,-.91) to (.08,-.82);
        \node[scale=.9,left](P) at (-3/2,0) {$P$};
        \node[scale=.9,left,blue](P1) at (3/2,0) {$P_1$};
    \end{tikzpicture}}
\ee
we compute this by first connect the two loops by a matter and then take the mass $\Delta\rightarrow0$ limit.
\begin{align}
&\adjustbox{valign=c}{\begin{tikzpicture}
        \draw[very thick, black] (-1/2,0) circle (1);
        \draw[draw=white, fill=white] (0,.85) circle (.08);
        \draw[very thick, blue] (1/2,0) circle (1);
        \draw[draw=white, fill=white] (0,-.85) circle (.08);
        \draw[very thick, black] (-.07,-.91) to (.08,-.82);
        \node[scale=.9,right](P) at (-3/2,0) {$P$};
        \node[scale=.9,left,blue](P1) at (3/2,0) {$P_1$};
        \draw[very thick, vert] (-1/2,0) to (1/2,0);
        \node[scale=.9,above,vert] at (0,0) {$P'$};
    \end{tikzpicture}}=\int_0^\infty dP_1'\,\bF_{\one P_1'}\begin{bmatrix}P&P_1\\P&P_1\end{bmatrix}(\bB_{PP_1}^{P_1'})^2\adjustbox{valign=c}{\begin{tikzpicture}
        \draw[very thick] ({{cos(130)}}, {sin(130)}) arc (50:310:1);
        \draw[very thick,blue] ({{cos(130)}}, {sin(130)}) arc (130:-130:1);
        \draw[very thick, blue] ({{cos(130)}}, {sin(130)}) to ({{cos(130)}},-0.8);
        \draw[very thick, vert] (-1.2,-1) to (0,-1);
        \node[scale=.9,right](P) at (-2.3,0) {$P$};
        \node[scale=.9,left,blue](P1p) at (-0.7,0) {$P_1'$};
        \node[scale=.9,left,blue](P1) at (1.1,0) {$P_1$};
        \node[scale=.9,vert](Pp) at (-0.5,-1.3) {$P'$};
    \end{tikzpicture}}\\
&=\int_0^\infty dP_1'dP_1''\,\bF_{\one P_1'}\begin{bmatrix}P&P_1\\P&P_1\end{bmatrix}(\bB_{PP_1}^{P_1'})^2\bF_{P_1'P_1''}\begin{bmatrix}P_1&P_1\\P&P\end{bmatrix}\adjustbox{valign=c}{\begin{tikzpicture}
        \draw[very thick, black] (-1,0) circle (0.6);
        \draw[very thick, blue] (1,0) circle (0.6);
        \draw[very thick, blue] (-0.45,0.2) to (0.45,0.2);
        \draw[very thick, vert] (-0.45,-0.2) to (0.45,-0.2);
        \node[scale=.9,right](P) at (-1.5,0) {$P$};
        \node[scale=.9,left,blue](P1) at (3/2,0) {$P_1$};
        \node[scale=.9,blue](P1pp) at (0,0.4) {$P_1''$};
        \node[scale=.9,vert](Pp) at (0,-0.4) {$P'$};
    \end{tikzpicture}}\\
&=\int_0^\infty dP_1''\,\frac{\bF_{\one P_1''}\begin{bmatrix}P_1&P_1\\P_1&P_1\end{bmatrix}\bS_{P_1P}[P_1'']}{\rho_0(P)}\adjustbox{valign=c}{\begin{tikzpicture}
        \draw[very thick, black] (-1,0) circle (0.6);
        \draw[very thick, blue] (1,0) circle (0.6);
        \draw[very thick, blue] (-0.45,0.2) to (0.45,0.2);
        \draw[very thick, vert] (-0.45,-0.2) to (0.45,-0.2);
        \node[scale=.9,right](P) at (-1.5,0) {$P$};
        \node[scale=.9,left,blue](P1) at (3/2,0) {$P_1$};
        \node[scale=.9,blue](P1pp) at (0,0.4) {$P_1''$};
        \node[scale=.9,vert](Pp) at (0,-0.4) {$P'$};
    \end{tikzpicture}}\\
&=\frac{\bS_{P_1P}[P']}{\rho_0(P)}\adjustbox{valign=c}{\begin{tikzpicture}
        \draw[very thick] ({{cos(130)}}, {sin(130)}) arc (90:270:1);
        \draw[very thick] ({{cos(130)}}, {sin(130)}) arc (90:-90:1);
        \draw[very thick, vert] ({{cos(130)}}, {sin(130)}) to ({{cos(130)}}, -1.25);
        \node[scale=.9,left](P) at (-1.8,-0.2) {$P$};
        \node[scale=.9,left,vert](P1) at (-0.6,-0.2) {$P'$};
        \node[scale=.9,left](P1) at (0.4,-0.2) {$P$};
    \end{tikzpicture}}\\
&=\frac{\bS_{P_1P}[P']}{\rho_0(P)C_0(P',P,P)}
\end{align}
In the limit $\Delta\rightarrow0$, we get
\be
\adjustbox{valign=c}{\begin{tikzpicture}
        \draw[very thick, black] (-1/2,0) circle (1);
        \draw[draw=white, fill=white] (0,.85) circle (.08);
        \draw[very thick, blue] (1/2,0) circle (1);
        \draw[draw=white, fill=white] (0,-.85) circle (.08);
        \draw[very thick, black] (-.07,-.91) to (.08,-.82);
        \node[scale=.9,left](P) at (-3/2,0) {$P$};
        \node[scale=.9,left,blue](P1) at (3/2,0) {$P_1$};
    \end{tikzpicture}}\quad\propto \Delta \bS_{P_1P}[\one]
\ee
This is consistent with our analysis below (\ref{thlink}). In particular note that if we are gluing one of the loops of a Hopf link to one asymptotic 3d trumpet, there is an additional factor of $\frac{1}{\Delta}$, and in total the Hopf link contributes $\bS_{P_1P}[\one]$.

\subsection{More Loops added to Hopf link}
\label{link3more}
Let us first consider three loops linked together.
\begin{align}
&\adjustbox{valign=c}{\begin{tikzpicture}
        \draw[very thick, black] (-1/2,0) circle (1);
        \draw[draw=white, fill=white] (0,.85) circle (.08);
        \draw[draw=white, fill=white] (0.35,.6) circle (.08);
        \draw[very thick, red] (1/2,0) circle (1);
        \draw[draw=white, fill=white] (0,-.85) circle (.08);
        \draw[very thick, black] (-.07,-.91) to (.08,-.82);
        \draw[very thick, blue] (0.5,0) circle (0.6);
        \draw[draw=white, fill=white] (0.35,-.6) circle (.08);
        \draw[very thick, black] (0.27,-.65) to (0.37,-.5);
        \node[scale=.9,left](P) at (-3/2,0) {$P$};
        \node[scale=.9,left,blue](P1) at (1.2,0) {$P_1$};
        \node[scale=.9,right,red](P2) at (3/2,0) {$P_2$};
    \end{tikzpicture}}=\int_0^\infty dP_1'\,\bF_{\one P_1'}\begin{bmatrix}P&P_1\\P&P_1\end{bmatrix}(\bB_{PP_1}^{P_1'})^2\adjustbox{valign=c}{\begin{tikzpicture}
        \draw[very thick, black] (-1/2,0) circle (1);
        \draw[draw=white, fill=white] (0,.85) circle (.08);
        \draw[very thick, red] (1/2,0) circle (1);
        \draw[draw=white, fill=white] (0,-.85) circle (.08);
        \draw[very thick, black] (-.07,-.91) to (.08,-.82);
        \draw[very thick, blue, fill=white] (0.5,0) circle (0.6);
        \node[scale=.9,left](P) at (-3/2,0) {$P$};
        \node[scale=.9,left,blue](P1) at (1.2,0) {$P_1$};
        \node[scale=.9,left,blue](P1p) at (0.6,0) {$P_1'$};
        \node[scale=.9,right,red](P2) at (3/2,0) {$P_2$};
    \end{tikzpicture}}\\
&=\int_0^\infty dP_1'dP_1''\,\bF_{\one P_1'}\begin{bmatrix}P&P_1\\P&P_1\end{bmatrix}(\bB_{PP_1}^{P_1'})^2\bF_{P_1'P_1''}\begin{bmatrix}P_1&P_1\\P&P\end{bmatrix}\adjustbox{valign=c}{\begin{tikzpicture}
        \draw[very thick, black] (-1/2,0) circle (1);
        \draw[draw=white, fill=white] (0.25,0.65) circle (.08);
        \draw[very thick, red] (1,0) circle (1);
        \draw[draw=white, fill=white] (0.25,-.65) circle (.08);
        \draw[very thick, black] (0.18,-0.73) to (0.32,-0.58);
        \draw[very thick, blue, fill=white] (1.5,0) circle (0.4);
        \draw[very thick, blue] (0.5,0) to (1.1,0);
        \node[scale=.9,left](P) at (-3/2,0) {$P$};
        \node[scale=.9,left,blue](P1) at (2,0) {$P_1$};
        \node[scale=.9,blue](P1pp) at (0.8,0.2) {$P_1''$};
        \node[scale=.9,right,red](P2) at (2,0) {$P_2$};
    \end{tikzpicture}}\\
&=\int_0^\infty dP_1''\,\frac{\bF_{\one P_1''}\begin{bmatrix}P_1&P_1\\P_1&P_1\end{bmatrix}\bS_{P_1P}[P_1'']}{\rho_0(P)}\adjustbox{valign=c}{\begin{tikzpicture}
        \draw[very thick, black] (-1/2,0) circle (1);
        \draw[draw=white, fill=white] (0.25,0.65) circle (.08);
        \draw[very thick, red] (1,0) circle (1);
        \draw[draw=white, fill=white] (0.25,-.65) circle (.08);
        \draw[very thick, black] (0.18,-0.73) to (0.32,-0.58);
        \draw[very thick, blue, fill=white] (1.5,0) circle (0.4);
        \draw[very thick, blue] (0.5,0) to (1.1,0);
        \node[scale=.9,left](P) at (-3/2,0) {$P$};
        \node[scale=.9,left,blue](P1) at (2,0) {$P_1$};
        \node[scale=.9,blue](P1pp) at (0.8,0.2) {$P_1''$};
        \node[scale=.9,right,red](P2) at (2,0) {$P_2$};
    \end{tikzpicture}}\\
&=\frac{\bS_{P_1P}[\one]}{\rho_0(P)}\adjustbox{valign=c}{\begin{tikzpicture}
        \draw[very thick, black] (-1/2,0) circle (1);
        \draw[draw=white, fill=white] (0,.85) circle (.08);
        \draw[very thick, red] (1/2,0) circle (1);
        \draw[draw=white, fill=white] (0,-.85) circle (.08);
        \draw[very thick, black] (-.07,-.91) to (.08,-.82);
        \node[scale=.9,left](P) at (-3/2,0) {$P$};
        \node[scale=.9,left,red](P2) at (3/2,0) {$P_2$};
    \end{tikzpicture}}
\end{align}
We see that the additional loop $P_1$ gives a multiplicative factor $\frac{\bS_{P_1P}[\one]}{\rho_0(P)}$ to the Hopf link. Now assume we glue one asymptotic 3d trumpet to the black loop, so the Hopf link contribute $\bS_{P_1P}[\one]$. By induction, we can add any number of loops and get
\be
\includegraphics[valign=c,width=0.3\textwidth]{figures/linkpn}=\frac{\prod_{i=1}^n\bS_{P_iP}[\one]}{\bS_{P\one}[\one]^{n-1}}\phantom{=\frac{\bS_{P_2P}[\one]}{\bS_{P\one}[\one]}\includegraphics[valign=c,width=0.3\textwidth]{figures/linkp1}}
\ee
the independence of $\Delta$ is provided we only glue one of the loops to one asymptotic 3d trumpet. 

\subsection{3 Loops with matter}
\label{3m}
The following configuration is useful in our analysis in section \ref{second}.
\begin{align}
&\adjustbox{valign=c}{\begin{tikzpicture}
        \draw[very thick, black] (-1/2,0) circle (1);
        \draw[draw=white, fill=white] (0,.85) circle (.08);
        \draw[draw=white, fill=white] (0.35,.6) circle (.08);
        \draw[very thick, red] (1/2,0) circle (1);
        \draw[draw=white, fill=white] (0,-.85) circle (.08);
        \draw[very thick, black] (-.07,-.91) to (.08,-.82);
        \draw[very thick, blue] (0.5,0) circle (0.6);
        \draw[draw=white, fill=white] (0.35,-.6) circle (.08);
        \draw[very thick, black] (0.27,-.65) to (0.37,-.5);
        \draw[very thick, vert] (0.7,-0.55) to (1.2,-0.75);
        \node[scale=.9,left](P) at (-3/2,0) {$P$};
        \node[scale=.9,left,blue](P1) at (1.2,0) {$P_1$};
        \node[scale=.9,right,red](P2) at (3/2,0) {$P_2$};
        \node[scale=.9,left,vert](Pp) at (1.5,-0.5) {$P'$};
    \end{tikzpicture}}=\int_0^\infty dP_1'\,\bF_{\one P_1'}\begin{bmatrix}P&P_1\\P&P_1\end{bmatrix}(\bB_{PP_1}^{P_1'})^2\adjustbox{valign=c}{\begin{tikzpicture}
        \draw[very thick, black] (-1/2,0) circle (1);
        \draw[draw=white, fill=white] (0,.85) circle (.08);
        \draw[very thick, red] (1/2,0) circle (1);
        \draw[draw=white, fill=white] (0,-.85) circle (.08);
        \draw[very thick, black] (-.07,-.91) to (.08,-.82);
        \draw[very thick, blue, fill=white] (0.5,0) circle (0.6);
        \draw[very thick, vert] (0.7,-0.55) to (1.2,-0.75);
        \node[scale=.9,left](P) at (-3/2,0) {$P$};
        \node[scale=.9,left,blue](P1) at (1.2,0) {$P_1$};
        \node[scale=.9,left,blue](P1p) at (0.6,0) {$P_1'$};
        \node[scale=.9,right,red](P2) at (3/2,0) {$P_2$};
        \node[scale=.9,left,vert](Pp) at (1.5,-0.5) {$P'$};
    \end{tikzpicture}}\\
&=\int_0^\infty dP_1'dP_1''\,\bF_{\one P_1'}\begin{bmatrix}P&P_1\\P&P_1\end{bmatrix}(\bB_{PP_1}^{P_1'})^2\bF_{P_1'P_1''}\begin{bmatrix}P_1&P_1\\P&P\end{bmatrix}\adjustbox{valign=c}{\begin{tikzpicture}
        \draw[very thick, black] (-1/2,0) circle (1);
        \draw[draw=white, fill=white] (0.25,0.65) circle (.08);
        \draw[very thick, red] (1,0) circle (1);
        \draw[draw=white, fill=white] (0.25,-.65) circle (.08);
        \draw[very thick, black] (0.18,-0.73) to (0.32,-0.58);
        \draw[very thick, blue, fill=white] (1.5,0) circle (0.4);
        \draw[very thick, blue] (0.5,0) to (1.1,0);
        \draw[very thick, vert] (1.5,-0.4) to (1.5,-0.9);
        \node[scale=.9,left](P) at (-3/2,0) {$P$};
        \node[scale=.9,left,blue](P1) at (2,0) {$P_1$};
        \node[scale=.9,blue](P1pp) at (0.8,0.2) {$P_1''$};
        \node[scale=.9,right,red](P2) at (2,0) {$P_2$};
        \node[scale=.9,left,vert](Pp) at (1.5,-0.7) {$P'$};
    \end{tikzpicture}}\\
&=\int_0^\infty dP_1''\,\frac{\bF_{\one P_1''}\begin{bmatrix}P_1&P_1\\P_1&P_1\end{bmatrix}\bS_{P_1P}[P_1'']}{\rho_0(P)}\adjustbox{valign=c}{\begin{tikzpicture}
        \draw[very thick, black] (-1/2,0) circle (1);
        \draw[draw=white, fill=white] (0.25,0.65) circle (.08);
        \draw[very thick, red] (1,0) circle (1);
        \draw[draw=white, fill=white] (0.25,-.65) circle (.08);
        \draw[very thick, black] (0.18,-0.73) to (0.32,-0.58);
        \draw[very thick, blue, fill=white] (1.5,0) circle (0.4);
        \draw[very thick, blue] (0.5,0) to (1.1,0);
        \draw[very thick, vert] (1.5,-0.4) to (1.5,-0.9);
        \node[scale=.9,left](P) at (-3/2,0) {$P$};
        \node[scale=.9,left,blue](P1) at (2,0) {$P_1$};
        \node[scale=.9,blue](P1pp) at (0.8,0.2) {$P_1''$};
        \node[scale=.9,right,red](P2) at (2,0) {$P_2$};
        \node[scale=.9,left,vert](Pp) at (1.5,-0.7) {$P'$};
    \end{tikzpicture}}\\
&=\frac{\bS_{P_1P}[P']}{\rho_0(P)}\adjustbox{valign=c}{\begin{tikzpicture}
        \draw[very thick, black] (-1/2,0) circle (1);
        \draw[draw=white, fill=white] (0,.85) circle (.08);
        \draw[very thick, red] (1/2,0) circle (1);
        \draw[draw=white, fill=white] (0,-.85) circle (.08);
        \draw[very thick, black] (-.07,-.91) to (.08,-.82);
        \node[scale=.9,right](P) at (-3/2,0) {$P$};
        \node[scale=.9,left,red](P2) at (3/2,0) {$P_2$};
        \draw[very thick, vert] (-1/2,0) to (1/2,0);
        \node[scale=.9,above,vert] at (0,0) {$P'$};
    \end{tikzpicture}}\\
&=\frac{\bS_{P_1P}[P']\bS_{PP_2}[P']}{\rho_0(P)\rho_0(P_2)C_0(P',P_2,P_2)}
\end{align}

\section{3-Boundary Wormhole in 2D}
\label{23}

In this section we consider a slightly more complicated geometry: a 3-boundary wormhole, because it will provide us with important intuition when we go up to 3 dimensions. We first compute a 3-boundary wormhole with two matter insertions of mass $\Delta$ on each boundary with the limit $\Delta\rightarrow0$ which is on-shell, we then compute the off-shell 3-boundary wormhole directly. Finally, we show that by combining the off-shell computation with three winding terms we can recover the on-shell computation. 

\paragraph{On-shell} We insert 6 matter operators, i.e. 2 insertions on each boundary. In the limit $s, s'\rightarrow$ and $\Delta\rightarrow0$ we have the following
\begin{align}
Z^{\text{on-shell}}&=\includegraphics[valign=c,width=0.25\textwidth]{figures/3holematter3}\\
&=e^{-S_0}\int dsds'\, \rho_0(s)\rho_0(s')|V_{s,s'}(\Delta)|^6 e^{-\sum \beta_i \frac{s^2}{2}}\\
&\approx e^{-S_0}\int dsds'\,\frac{s^2}{\pi}\frac{s'^2}{\pi}\left(\frac{4\Delta}{(\Delta^2+(s+s')^2)(\Delta^2+(s-s')^2)}\right)^3\\
&\approx \frac{e^{-S_0}}{8\Delta^3}
\end{align}
where we recalled that in the $s\rightarrow0$ limit
\be
\rho_0(s)=\frac{s\,\sinh(2\pi s)}{2\pi^2}\approx\frac{s^2}{\pi}
\ee

\paragraph{Off-shell} On the other hand, without matter, we can cut the 3-boundary wormhole into 3 trumpets and a 3-hole sphere
\begin{align}
Z^{\text{off-shell}}&=e^{-S_0}\includegraphics[valign=c,width=0.25\textwidth]{figures/3holenomatter}\\
&=e^{-S_0}\int_0^\infty b_1db_1\,b_2db_2\,b_3db_3\,\includegraphics[valign=c,width=0.3\textwidth]{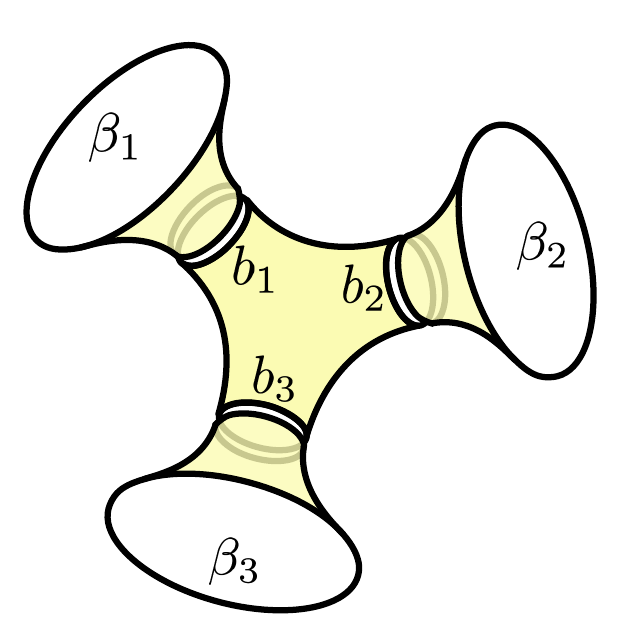}\\
&=e^{-S_0}\int_0^\infty b_1db_1\,b_2db_2\,b_3db_3\,Z_{trumpet}(\beta_1,b_1)Z_{trumpet}(\beta_2,b_2)Z_{trumpet}(\beta_3,b_3)\nonumber\\
&\hspace{6cm} \times V_{0,3}(b_1,b_2,b_3)\\
&=\sqrt{\frac{\beta_1\beta_2\beta_3}{(2\pi)^3}}
\end{align}
where
\be
Z_{trumpet}(\beta,b)=\frac{e^{-\frac{b^2}{2\beta}}}{\sqrt{2\pi\beta}}
\ee
Let's now analyze the windings we need to include. $\ell_2$ and $\ell_3$ together could wind around $b_1$, $\ell_1$ and $\ell_3$ together could wind around $b_2$, and also $\ell_1$ and $\ell_2$ together could wind around $\beta_3$. Thus we need to include three winding factors multiplied together each with effective mass $2\Delta$
\be
\frac{2}{b_12\Delta}\frac{2}{b_22\Delta}\frac{2}{b_32\Delta}
\ee
and the off-shell computation get modified to 
\begin{align}
\tilde{Z}^{\text{off-shell}}&=e^{-S_0}\int_0^\infty b_1db_1\,b_2db_2\,b_3db_3\,\frac{1}{b_1b_2b_3\Delta^3}Z_{trumpet}(\beta_1,b_1)Z_{trumpet}(\beta_2,b_2)Z_{trumpet}(\beta_3,b_3)\nonumber\\
&\hspace{7.5cm}\times V_{0,3}(b_1,b_2,b_3)\\
&=\frac{e^{-S_0}}{8\Delta^3}\\
&=Z^{\text{on-shell}}
\end{align}
so we get back our on-shell answer.

\section{Heegard splitting details}
\label{heegard}
\subsection{First method}
\label{heegard1}
In this section, we compute the following Heegard splitting
\begin{multline}
   \includegraphics[width=0.25\textwidth, valign=c]{figures/3T1}\\
   \rightarrow\quad  M_1=\includegraphics[width=0.25\textwidth, valign=c]{figures/3Tm1} \quad \text{and}\quad M_2=\includegraphics[width=0.25\textwidth, valign=c]{figures/3Tm2} 
\end{multline}
then
\be
Z_{vir}(M)=\braket{Z_{vir}(M_1)|Z_{vir}(M_2)}
\ee
Now, how do we compute $\ket{Z_{vir}(M_1)}$ and $\ket{Z_{vir}(M_2)}$? For each original boundary of $M_1$, we insert a complete set of states (i.e. conformal blocks) on its Hilbert space. In addition to those states, there is a state in the Hilbert space of the splitting surface $\Sigma_g$ (in this case, the blue torus). It is obtained by gluing the handle body (in this case a solid torus) with Wilson lines corresponding to the conformal blocks appearing in the complete set of states in the original boundary Hilbert spaces.
\begin{multline}
\ket{Z_{vir}(M_1)}=\int dP_adP_b\,\rho_0(P_a)\rho_0(P_b)C_0(P_1,P_a,P_b)C_0(P_3,P_a,P_b)\\
\times\underbrace{\includegraphics[width=0.2\textwidth, valign=c]{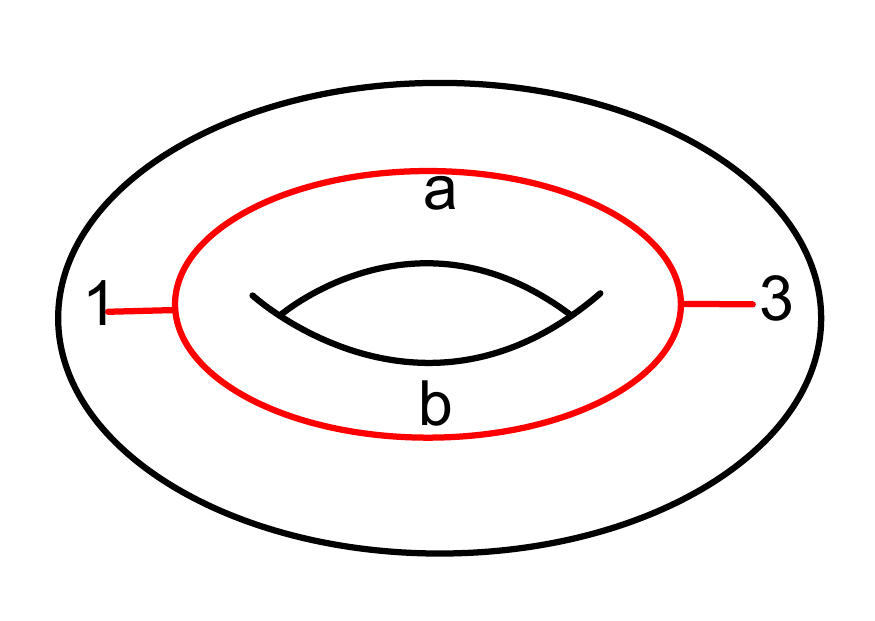}}_{\text{original bdy conformal block}}\quad \quad \underbrace{\ket{\includegraphics[width=0.2\textwidth, valign=c]{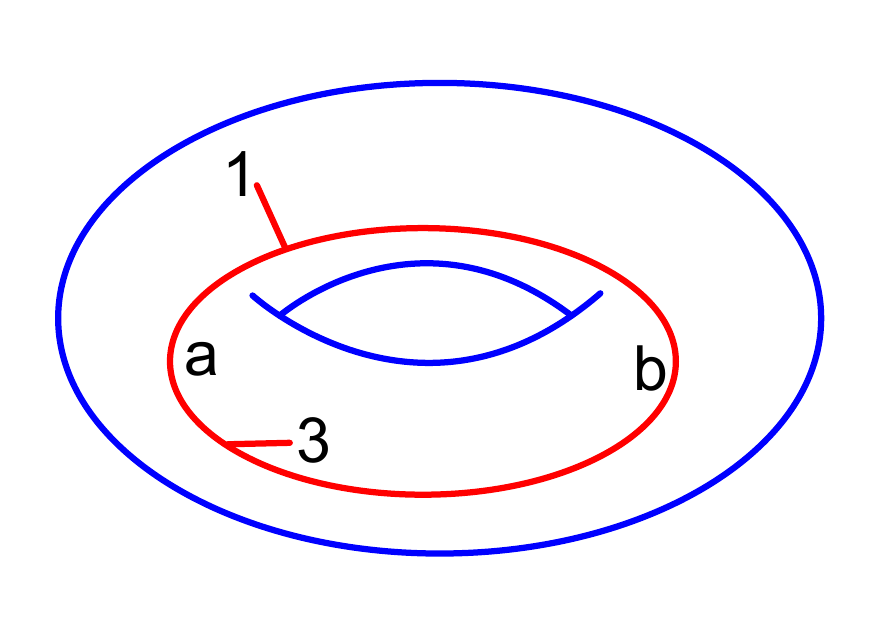}}}_{\Sigma_g\text{ conformal block}}
\end{multline}
and similarly
\begin{multline}
\ket{Z_{vir}(M_2)}=\int dP_pdP_qdP_rdP_s\,\rho_0(P_p)\rho_0(P_q)\rho_0(P_r)\rho_0(P_s)\\
\times C_0(P_1,P_p,P_q)C_0(P_2,P_p,P_q)C_0(P_2,P_r,P_s)C_0(P_3,P_r,P_s)\\
\times\underbrace{\includegraphics[width=0.2\textwidth, valign=c]{figures/diag7}}_{\text{original bdy conformal block}}\quad \underbrace{\includegraphics[width=0.2\textwidth, valign=c]{figures/diag8}}_{\text{original bdy conformal block}}\quad  \underbrace{\ket{\includegraphics[width=0.2\textwidth, valign=c]{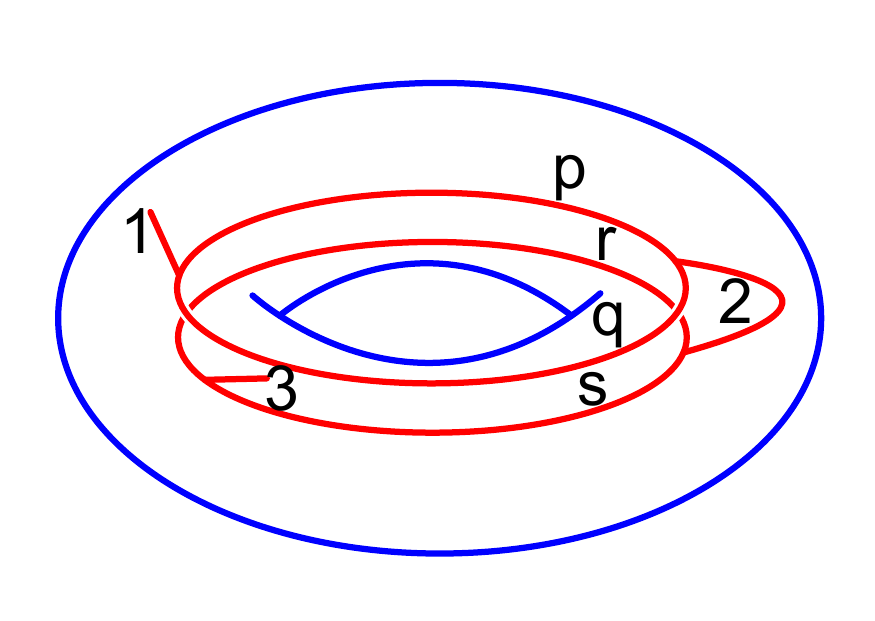}}}_{\Sigma_g\text{ conformal block}}
\end{multline}
We now need to compute the inner product between $\ket{Z_{vir}(M_1)}$ and $\ket{Z_{vir}(M_2)}$. Before doing that, we simplify the $\Sigma_g$ conformal block of $M_2$ using the fusion move (\ref{fusionmove}) and get
\begin{align}
&\ket{\includegraphics[width=0.2\textwidth, valign=c]{figures/diag1}}=\int dP_t\, \mathbb{F}_{2t}\begin{bmatrix}P_p&P_r\\P_q&P_s\end{bmatrix}\ket{\includegraphics[width=0.2\textwidth, valign=c]{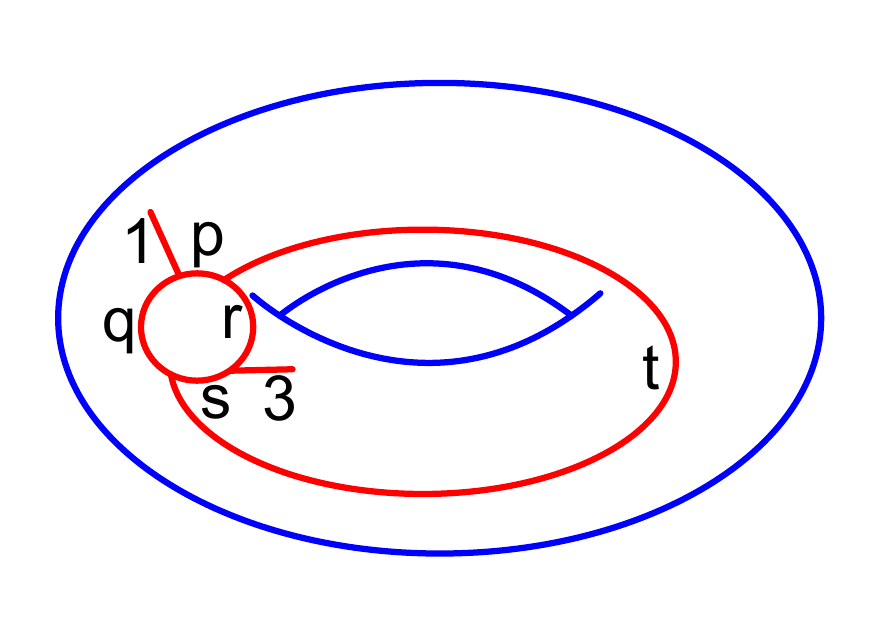}}\\
&=\int dP_tdP_udP_v\, \mathbb{F}_{2t}\begin{bmatrix}P_p&P_r\\P_q&P_s\end{bmatrix}\mathbb{F}_{pu}\begin{bmatrix}P_1&P_t\\P_q&P_r\end{bmatrix}\mathbb{F}_{sv}\begin{bmatrix}P_3&P_t\\P_r&P_q\end{bmatrix}\ket{\includegraphics[width=0.2\textwidth, valign=c]{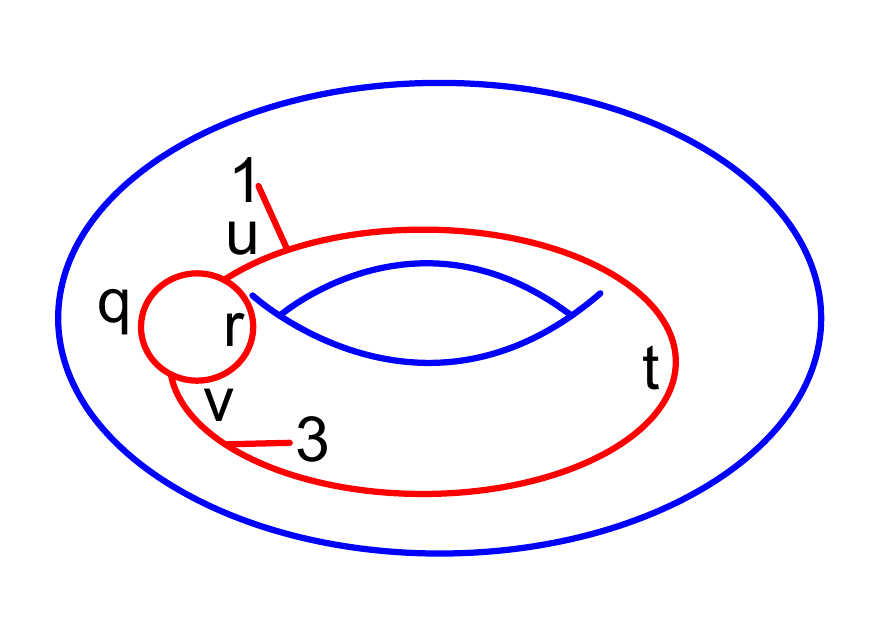}}\\
&=\int dP_tdP_u\, \frac{\mathbb{F}_{2t}\begin{bmatrix}P_p&P_r\\P_q&P_s\end{bmatrix}\mathbb{F}_{pu}\begin{bmatrix}P_1&P_t\\P_q&P_r\end{bmatrix}\mathbb{F}_{su}\begin{bmatrix}P_3&P_t\\P_r&P_q\end{bmatrix} }{\rho_0(P_u)C_0(P_q,P_r,P_u)}\ket{\includegraphics[width=0.2\textwidth, valign=c]{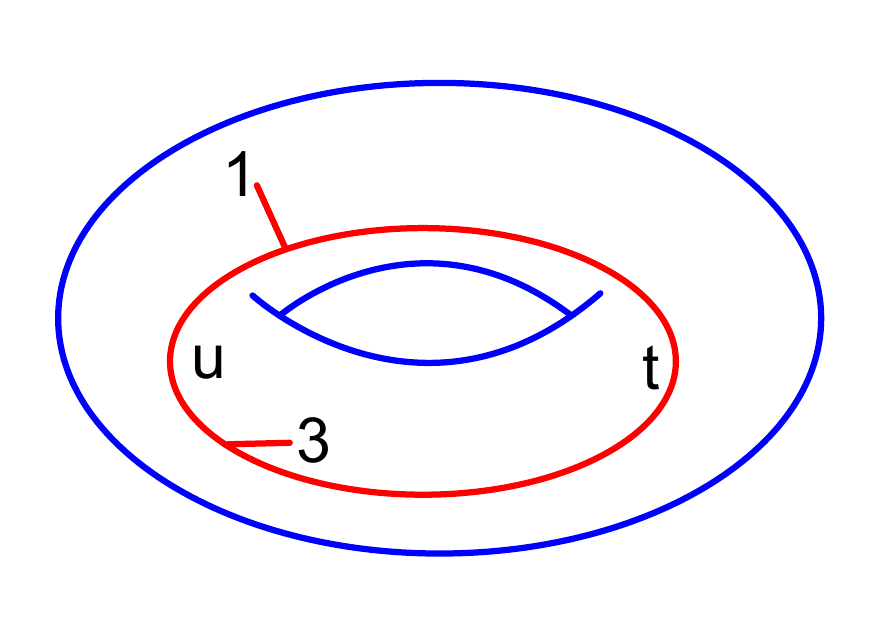}}
\end{align}
where we also used
\be
\adjustbox{valign=c}{\begin{tikzpicture}
        \draw[very thick, red] (0,0) circle (0.5);
        \draw[very thick,red] (-1.5,0) to (-0.5,0);
        \draw[very thick,red] (0.5,0) to (1.5,0);
        \node[scale=.9](P3) at (0,0.8) {$P_3$};
        \node[scale=.9](P2) at (1,0.2) {$P_2$};
        \node[scale=.9](P4) at (0,-0.8) {$P_4$};
        \node[scale=.9](P1) at (-1,0.2) {$P_1$};
    \end{tikzpicture}}=\frac{\delta(P_1-P_2)}{\rho_0(P_a)C_0(P_1,P_3,P_4)}\adjustbox{valign=c}{\begin{tikzpicture}
        \draw[very thick,red] (-1.2,0) to (1.2,0);
        \node[scale=.9](P1) at (0,0.2) {$P_1$};
        \node[scale=.9,white](P2) at (0,-0.2) {$P_2$};
    \end{tikzpicture}}
\ee
How do we compute inner product between conformal blocks of $\Sigma_g$. The claim of \cite{Cotler:2020ugk} is
\be
\braket{\includegraphics[width=0.2\textwidth, valign=c]{figures/diag5}|\includegraphics[width=0.2\textwidth, valign=c]{figures/diag4}}
=\frac{\delta(P_a-P_u)\delta(P_b-P_t)}{\rho_0(P_t)\rho_0(P_u)C_0(P_1,P_t,P_u)C_0(P_3,P_t,P_u)}
\ee
Then putting everything together
\begin{multline}
Z_{vir}(M)=\int dP_p dP_q dP_r dP_s dP_t dP_u\,\frac{\rho_0(P_p)\rho_0(P_q)\rho_0(P_r)\rho_0(P_s)}{\rho_0(P_u)}\\\times\frac{C_0(P_1,P_p,P_q)C_0(P_2,P_p,P_q)C_0(P_2,P_r,P_s)C_0(P_3,P_r,P_s)}{C_0(P_q,P_r,P_u)}\mathbb{F}_{2t}\begin{bmatrix}P_p&P_r\\P_q&P_s\end{bmatrix}\mathbb{F}_{pu}\begin{bmatrix}P_1&P_t\\P_q&P_r\end{bmatrix}\mathbb{F}_{su}\begin{bmatrix}P_3&P_t\\P_r&P_q\end{bmatrix}\\
\times\underbrace{\includegraphics[width=0.2\textwidth, valign=c]{figures/diag7}\quad \includegraphics[width=0.2\textwidth, valign=c]{figures/diag8}\quad \includegraphics[width=0.2\textwidth, valign=c]{figures/diag9}}_{\text{conformal blocks}}
\end{multline}
We observe that the above expression doesn't look symmetric between $1\leftrightarrow2\leftrightarrow3$ but it should be. In order to make the process of taking the limit $\Delta\rightarrow0$ easier, we do the following 3 steps. First, we rewrite the F-blocks in terms of Virasoro 6j symbols \cite{Eberhardt:2023mrq}. Then, we express $Z_{vir}(M)$ in terms of Virasoro 6j symbols and rearrange the letters within the 6j symbols according to 6j identities. And finally, we reexpress the new Virasoro 6j symbols in terms of F-blocks. The F-block and Virasoro 6j symbol conversion is given by the following formula. 
\be
\mathbb{F}_{2t}\begin{bmatrix}P_p&P_r\\P_q&P_s\end{bmatrix}=\begin{Bmatrix}P_p&P_r&P_t\\P_s&P_q&P_2\end{Bmatrix} \rho_0(P_t)\sqrt{\frac{C_0(P_p,P_r,P_t)C_0(P_s,P_q,P_t)}{C_0(P_p,P_q,P_2)C_0(P_s,P_r,P_2)}}\label{6jformula}
\ee
so rewriting $Z_{vir}(M)$ in terms of Virasoro 6j symbols give
\begin{multline}
Z_{vir}(M)=\int dP_p dP_q dP_r dP_s dP_t dP_u\,\rho_0(P_p)\rho_0(P_q)\rho_0(P_r)\rho_0(P_s)\rho_0(P_t)\rho_0(P_u)\\\times\sqrt{C_0(P_1,P_p,P_q)C_0(P_1,P_t,P_u)C_0(P_2,P_p,P_q)C_0(P_2,P_r,P_s)C_0(P_3,P_r,P_s)C_0(P_3,P_t,P_u)}\\
\times\begin{Bmatrix}P_p&P_r&P_t\\P_s&P_q&P_2\end{Bmatrix}\begin{Bmatrix}P_p&P_t&P_r\\P_u&P_q&P_1\end{Bmatrix}\begin{Bmatrix}P_s&P_t&P_q\\P_u&P_r&P_3\end{Bmatrix}\\
\times\includegraphics[width=0.2\textwidth, valign=c]{figures/diag7}\quad \includegraphics[width=0.2\textwidth, valign=c]{figures/diag8}\quad \includegraphics[width=0.2\textwidth, valign=c]{figures/diag9}
\end{multline}
and then again back in terms of F blocks
\begin{multline}
Z_{vir}(M)=\int dP_p dP_q dP_r dP_s dP_t dP_u\,\rho_0(P_p)\rho_0(P_s)\rho_0(P_u)\\
\times \frac{C_0(P_1,P_p,P_q)C_0(P_1,P_t,P_u)C_0(P_2,P_p,P_q)C_0(P_2,P_r,P_s)C_0(P_3,P_r,P_s)C_0(P_3,P_t,P_u)} {C_0(P_p,P_r,P_t)C_0(P_q,P_r,P_u)C_0(P_t,P_q,P_s)}\\
\times\mathbb{F}_{2t}\begin{bmatrix}P_p&P_r\\P_q&P_s\end{bmatrix}\mathbb{F}_{1r}\begin{bmatrix}P_p&P_t\\P_q&P_u\end{bmatrix}\mathbb{F}_{3q}\begin{bmatrix}P_s&P_t\\P_r&P_u\end{bmatrix}\\
\times\includegraphics[width=0.2\textwidth, valign=c]{figures/diag7}\quad \includegraphics[width=0.2\textwidth, valign=c]{figures/diag8}\quad \includegraphics[width=0.2\textwidth, valign=c]{figures/diag9}\label{method1b}
\end{multline}

\subsection{Second method}
\label{heegard2}
In this section, we use a different Heegaard splitting with a genus-2 surface
\be
\includegraphics[width=0.25\textwidth, valign=c]{figures/3T2}
\ee
and this splits $M$ into two parts
\be
M_1=\includegraphics[width=0.25\textwidth, valign=c]{figures/3Tm1p}\quad\text{and}\quad M_2=\includegraphics[width=0.25\textwidth, valign=c]{figures/3Tm2p}
\ee
Then we calculate the Virasoro partition function of $M_1$ and $M_2$ respectively
\begin{multline}
\ket{Z_{vir}(M_1)}=\int dP_pdP_qdP_rdP_s\,\rho_0(P_p)\rho_0(P_q)\rho_0(P_r)\rho_0(P_s)\\
\times C_0(P_1,P_p,P_q)C_0(P_2,P_p,P_q)C_0(P_2,P_r,P_s)C_0(P_3,P_r,P_s)\\
\times\includegraphics[width=0.2\textwidth, valign=c]{figures/diag7}\quad \includegraphics[width=0.2\textwidth, valign=c]{figures/diag8}\quad \ket{\includegraphics[width=0.2\textwidth, valign=c]{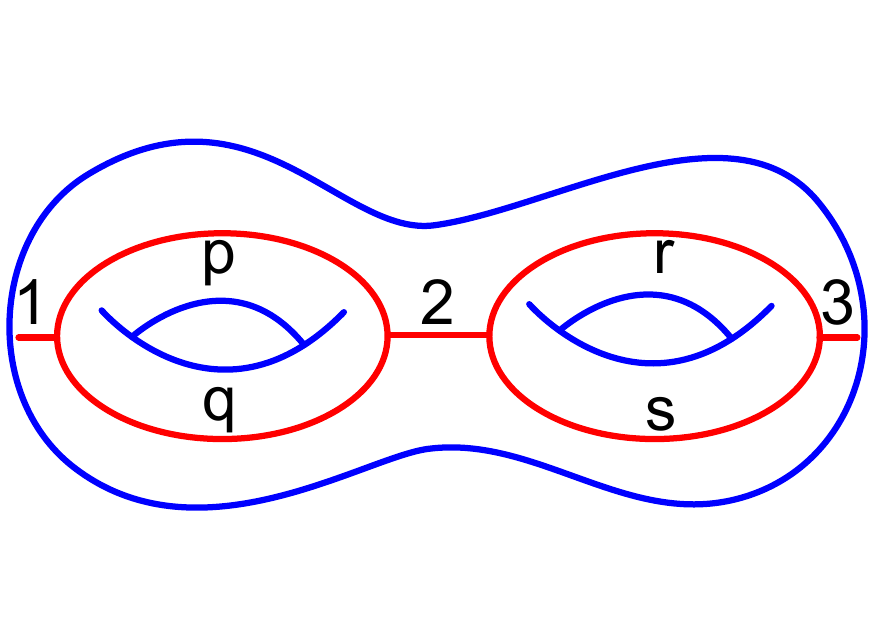}}
\end{multline}
and
\begin{multline}
\ket{Z_{vir}(M_2)}=\int dP_tdP_u\,\rho_0(P_t)\rho_0(P_u)\,C_0(P_1,P_t,P_u)C_0(P_3,P_t,P_u)\\
\times\includegraphics[width=0.2\textwidth, valign=c]{figures/diag9}\quad  \ket{\includegraphics[width=0.2\textwidth, valign=c]{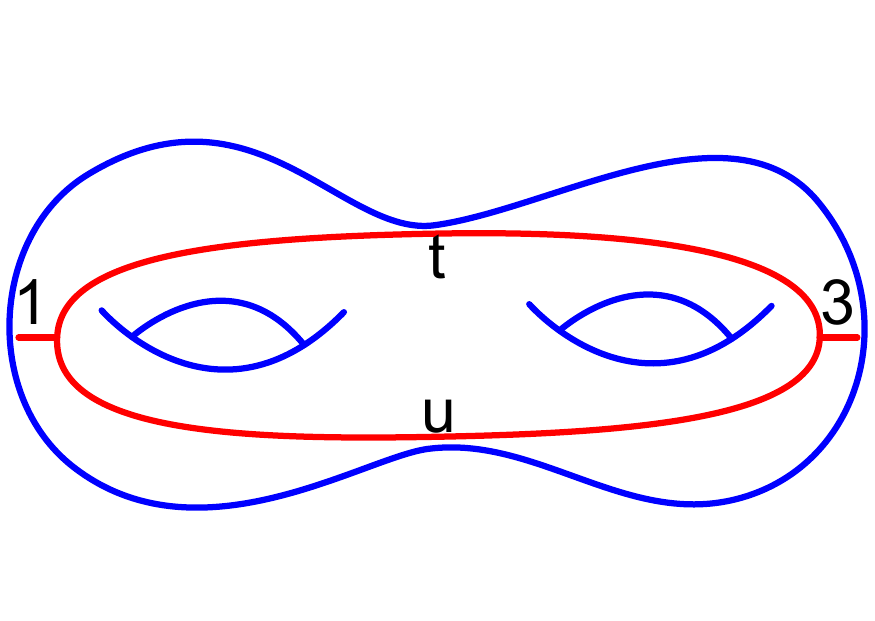}}
\end{multline}
We could simplify the inner product by using a fusion move
\begin{align}
\ket{\includegraphics[width=0.2\textwidth, valign=c]{figures/diag11}}&=\ket{\includegraphics[width=0.2\textwidth, valign=c]{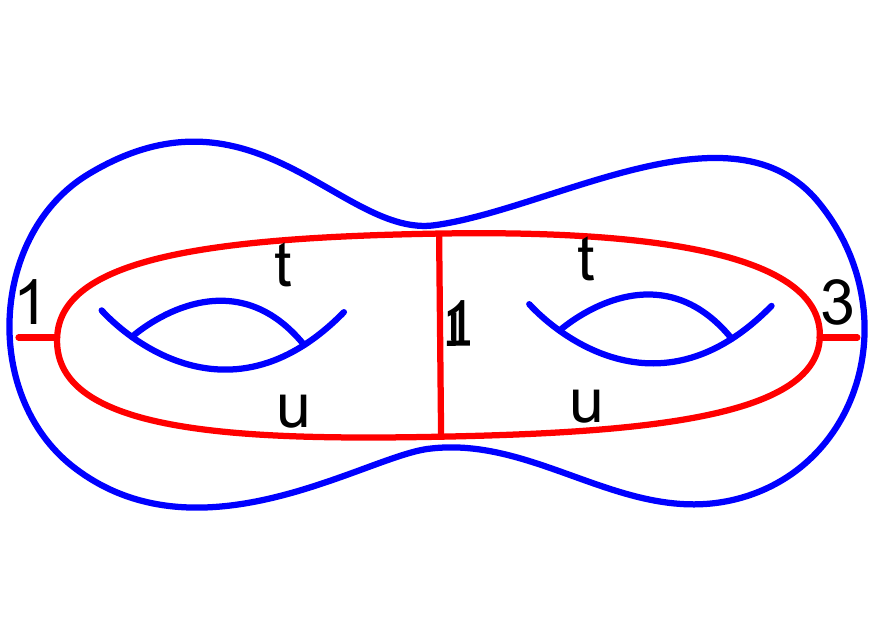}}\\
=&\int dP_v\, \mathbb{F}_{\mathbb{1}P_v}\begin{bmatrix}P_t&P_u\\P_t&P_u\end{bmatrix}\ket{\includegraphics[width=0.2\textwidth, valign=c]{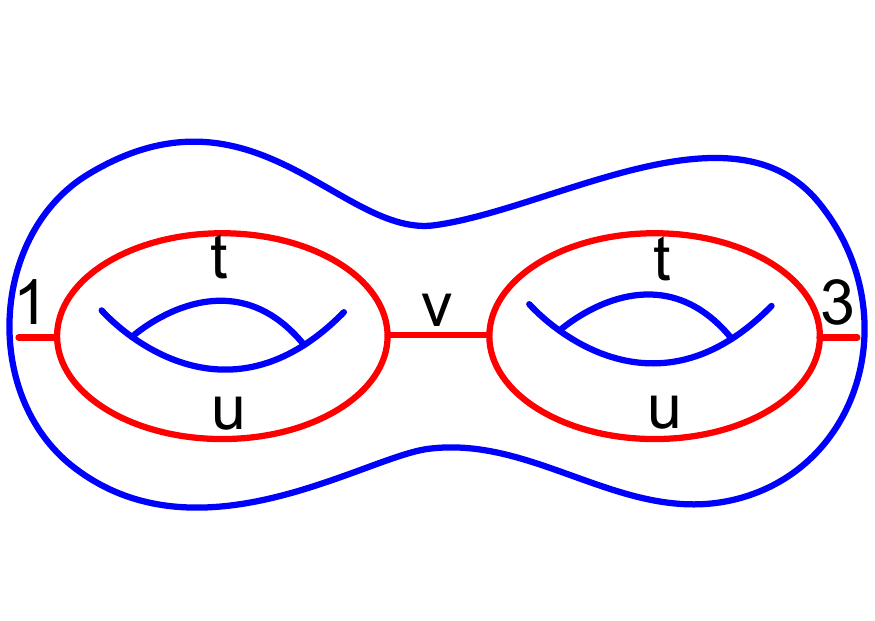}}
\end{align}
and remembering the identity
\be
\mathbb{F}_{P_t\mathbb{1}}\begin{bmatrix}P_2&P_1\\P_2&P_1\end{bmatrix}=\rho_0(P_t)C_0(P_1,P_2,P_t)
\ee
Recall that it was conjectured \cite{Collier:2023cyw} that the inner product between conformal blocks is given by the following
\begin{multline}
\Braket{\includegraphics[width=0.2\textwidth, valign=c]{figures/diag10}|\includegraphics[width=0.2\textwidth, valign=c]{figures/diag13}}\\
=\frac{\delta(P_p-P_t)\delta(P_q-P_u)\delta(P_r-P_t)\delta(P_s-P_u)\delta(P_2-P_v)}{\rho_0(P_p)\rho_0(P_q)\rho_0(P_r)\rho_0(P_s)\rho_0(P_2)C_0(P_1,P_p,P_q)C_0(P_2,P_p,P_q)C_0(P_2,P_r,P_s)C_0(P_3,P_r,P_s)}
\end{multline}
so in the end the chiral part of the partition function is given by
\begin{multline}
Z_{vir}(M)=\int dP_u dP_t\,\rho_0(P_u)\rho_0(P_t)\, C_0(P_1,P_t,P_u)C_0(P_2,P_t,P_u)C_0(P_3,P_t,P_u)\\
\times\underbrace{\includegraphics[width=0.2\textwidth, valign=c]{figures/diag7pp}\quad \includegraphics[width=0.2\textwidth, valign=c]{figures/diag8pp}\includegraphics[width=0.2\textwidth, valign=c]{figures/diag9}}_{\text{conformal blocks}}\label{method2}
\end{multline}

\section{Mapping Class Group of Some 3d Manifolds}
\label{MCG}
In this section, we compute the mapping class group of a genus $g$ surface times $S^1$ for $g\geq2$, and the mapping class group of $n$-boundary torus wormhole for $n\geq3$.

\subsection{Mapping Class Group $\Mod(\Sigma_g\times S^1)$ for $\Sigma_g$ without Boundaries}
We show that 
\be
\Mod(\Sigma_g\times S^1)=\Z^{2g}\rtimes\Mod(\Sigma_g)\quad\quad\text{for}\quad\quad g\geq2
\ee
following \cite{Chen}.
\subsubsection{Fundamental Group}
Recall that genus $g$ Riemann surface $\Sigma_g$ has fundamental group
\be
\pi_1(\Sigma_g)=\langle A_i, B_i|\prod_{i=1}^g[A_i,B_i]=1\rangle
\ee
In particular, we should note that it only has one constraint. Now for a general $S^1$ bundle over $\Sigma_g$ denoted by $M^3$
\be
S^1\rightarrow M^3\xrightarrow{p} \Sigma_g
\ee
the fundamental group looks like
\be
\pi_1(M^3)=\langle\hat{A}_i,\hat{B}_i,t|[\hat{A}_i,t]=1, [\hat{B}_i,t]=1\quad \forall i \cdots\text{ and more}\rangle
\ee
which has more constraints, because $A_i$ ($B_i$) and the circle $S^1$ forms a torus. In particular
\be
p\left(\prod_{i=1}^g[\hat{A}_i,\hat{B}_i]\right)=\prod_{i=1}^g[A_i,B_i]=1
\ee
This means 
\be
\prod_{i=1}^g[\hat{A}_i,\hat{B}_i]=t^e
\ee
where $e$ is called Euler number. We now show that $e$ is well-defined. Take
\be
\hat{A}'_1=t^i\hat{A}_1\quad\quad \hat{B}'_1=t^j\hat{B}_1
\ee
then 
\be
[\hat{A}'_1,\hat{B}'_1]=[t^i\hat{A}_1,t^j\hat{B}_1]=t^i\hat{A}_1t^j\hat{B}_1\hat{A}_1^{-1}t^{-i}\hat{B}_1^{-1}t^{-j}=[\hat{A}_1,\hat{B}_1]
\ee
so $e$ is well-defined. For the bundle $M^3=\Sigma_g\times S^1$, we have $e=0$.

We have the exact sequence
\be
\begin{tikzcd}
 1 \arrow[r] & \pi_1(S^1) \ar[draw=none]{d}[sloped,auto=false,  near end]{\approx} \arrow[r] & \pi_1(M^3) \arrow[r, "p"] & \pi_1(\Sigma_g) \ar[r] & 1\\
 &\Z&&&
\end{tikzcd}
\ee
This sequence splits if there exists a section 
\be
s:\pi_1(\Sigma)\hookrightarrow \pi_1(M^3)
\ee
This does exist for our case $e=0$, because we can write any point of the bundle $M^3$ as $(x,z_0)\in \Sigma_g\times S^1$. 

\subsubsection{Mapping class group of $\Sigma_g\times S^1$ for $g\geq 2$}
Recall that the definition
\be
\Mod(M^3)=\pi_0(\Homeo\, M^3)
\ee
Given a homeomorphism of $M^3$ that fixes a point $m_0\in M^3$, there is an automorphism of $\pi_1 M^3$. But we also notice that $\pi_1 M^3$ is $K(\pi, 1)$, so basepoint-preserving homotopy equivalences (HEQ) correspond to $\Aut\,\pi_1M^3$. 

Waldhausen proved a theorem: every homotopy equivalence (HEQ) of $M^3$ correspond to a homeomorphism of $M^3$ \cite{Waldhausen}. Therefore, we get the equivalence
\be
\text{HEQ}\quad\leftrightarrow\quad \Aut\,\pi_1M^3\quad\leftrightarrow\quad \pi_0\,\Homeo\, M^3
\ee
In general, we have the following diagram 
\be
\begin{tikzcd}
  1 \arrow[r] &\pi_1(S^1) \arrow[d, "id"] \arrow[r] & \pi_1(M^3) \arrow[d, "\hat{\varphi}\in\Aut\,\pi_1M^3"] \arrow[r, "p"] & \pi_1(\Sigma_g) \arrow[d, "\varphi\in\Aut\,\pi_1\Sigma_g"] \arrow[r] & 1 \\
  1 \arrow[r] & \pi_1(S^1) \arrow[r] & \pi_1(M^3) \arrow[r, "p"] & \pi_1(\Sigma_g) \ar[r] & 1
\end{tikzcd}
\ee
Then we have the short exact sequence
\be
\begin{tikzcd}
 1 \arrow[r] & \Hom(\pi_1\Sigma_g,\Z) \ar[draw=none]{d}[sloped,auto=false,  near end]{\approx} \arrow[r] & \Mod(M^3) \ar[draw=none]{d}[sloped,auto=false,  near end]{\ni} \arrow[r] & \Mod(\Sigma_g) \ar[draw=none]{d}[sloped,auto=false,  near end]{\ni} \ar[r] & 1\\
 &H^1(\Sigma_g,\Z)&\hat{\varphi}&\varphi&
\end{tikzcd}\label{sequence2}
\ee
In general this sequence wouldn't split because given an automorphism $\hat{\varphi}$ of $\pi_1 M^3$ that corresponds to the identity in $\pi_1\Sigma_g$, we could have
\be
\hat{\varphi}(\hat{A}_i)=\hat{A}_it^{\alpha_i}\quad\quad\alpha_i\in\Z
\ee
which is illustrated in the following picture. 
\be
\includegraphics[width=0.55\textwidth,valign=c]{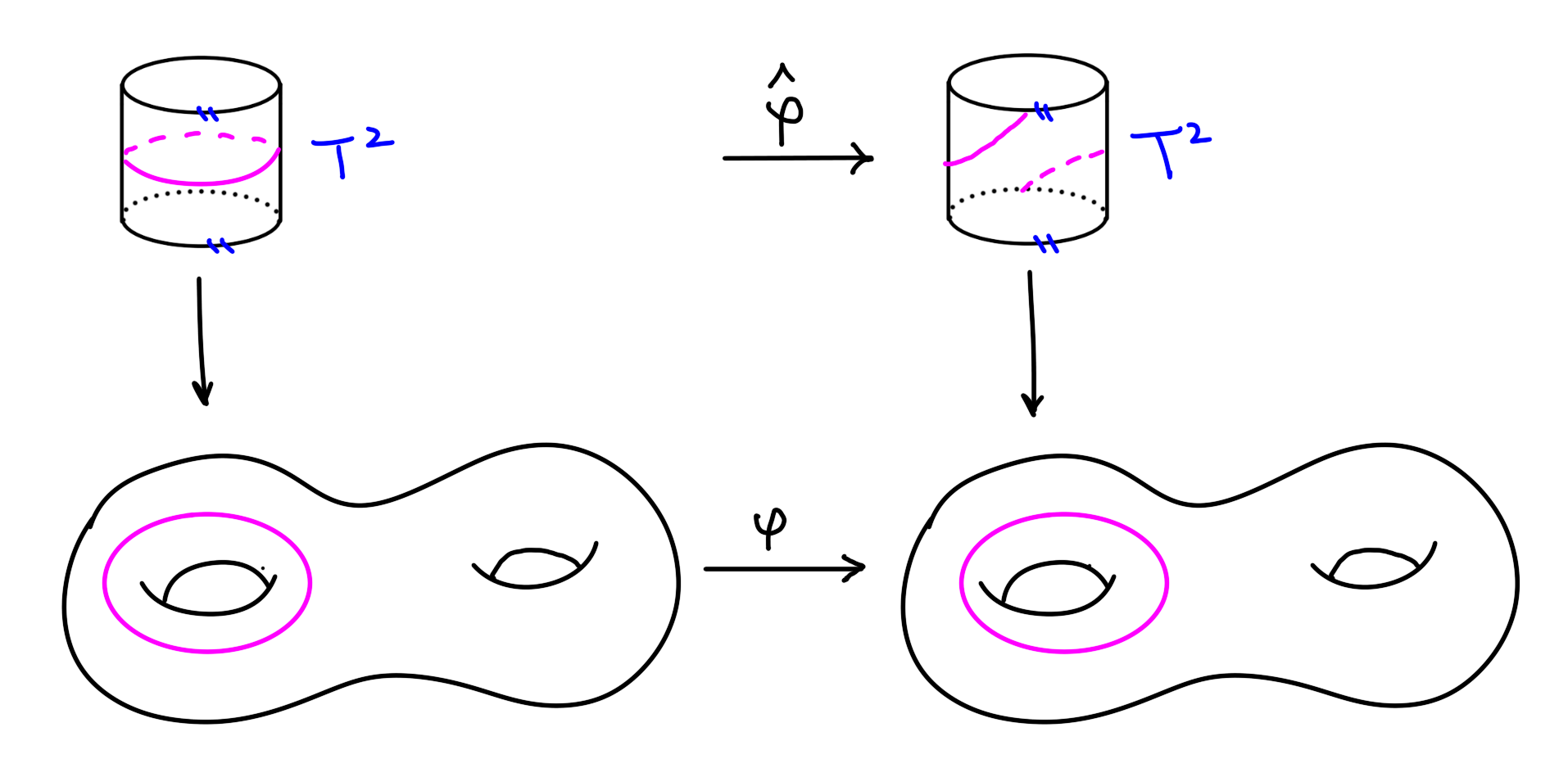}
\ee
But in our particular case where $e=0$, the short exact sequence (\ref{sequence2}) does split, because there exists
\be
s: \Mod(\Sigma_g)\hookrightarrow \Mod(M^3)
\ee
Therefore, we have 
\be
\Mod(\Sigma_g\times S^1)=H^1(\Sigma_g,\Z)\rtimes\Mod(\Sigma_g)=\Z^{2g}\rtimes\Mod(\Sigma_g)\quad\quad\text{for}\quad\quad g\geq2
\ee

\subsection{Mapping Class Group of $n$-boundary torus wormhole}
We denote an $n$-hole sphere by $\Sigma^n$, so we are interested in the mapping class group $\Mod(\Sigma^n\times S^1)$. Here we only consider $n\geq3$.

If $n\geq 3$, the surface $\Sigma^n$ has no center in the fundamental group
\be
\Sigma^n\simeq\underbrace{S^1\vee\cdots \vee S^1}_{n-1}=\vee_{i=1}^{n-1} S^1_i\quad\quad n\geq 3
\ee
so
\be
\pi_1(\Sigma^n)=\pi_1(\vee_{i=1}^{n-1} S^1_i)=*_{i=1}^{n-1}\pi_1(S^1)=\underbrace{\Z*\cdots*\Z}_{n-1}
\ee
where we used the fact the wedge of geometries has fundamental group a free product of each fundamental groups of the components. Thus this has no center.

In this case we can still use the short exact sequence which splits
\be
\begin{tikzcd}
 1 \arrow[r] & \Hom(\pi_1\Sigma^n,\Z) \ar[draw=none]{d}[sloped,auto=false,  near end]{\approx} \arrow[r] & \Mod(M^3) \ar[draw=none]{d}[sloped,auto=false,  near end]{\ni} \arrow[r] & \Mod(\Sigma^n) \ar[draw=none]{d}[sloped,auto=false,  near end]{\ni} \ar[r] & 1\\
 &H^1(\Sigma^n,\Z)&\hat{\varphi}&\varphi&
\end{tikzcd}
\ee
and we know
\be
H^1(\Sigma^n,\Z)=H^1(\vee_{i=1}^{n-1} S^1_i,\Z)=\Z^{n-1}
\ee
so
\be
\Mod(\Sigma^n\times S^1)=H^1(\Sigma^n,\Z)\rtimes \Mod(\Sigma^n)=\Z^{n-1}\rtimes\Mod(\Sigma^n)\quad\quad\text{for}\quad\quad n\geq3
\ee

\bibliographystyle{utphys}
\bibliography{references}
\end{document}